\documentclass[12pt,a4paper,twoside]{amsart}
\usepackage{amsmath,bbm,amssymb,amsxtra}
\allowdisplaybreaks
\usepackage{epsfig}
\usepackage{ae}
\numberwithin{equation}{section}
\theoremstyle{definition}

\newtheorem{lemma}{Lemma}[section]
\newtheorem{proposition}[lemma]{Proposition}

\newtheorem{theorem}[lemma]{Theorem}
\newtheorem{corollary}[lemma]{Corollary}
\newtheorem{remark}[lemma]{Remark}
\newtheorem{definition}[lemma]{Definition}

\newcommand{\prop}[1]{\begin{proposition}\label{#1}
\sl }
\newcommand{\eprop}{\end{proposition}}
\newcommand{\thm}[1]{\begin{theorem}\label{#1}
\ƒ }
\newcommand{\ethm}{\end{theorem}}
\newcommand{\lem}[1]{\begin{lemma}\label{#1}
\sl }
\newcommand{\elem}{\end{lemma}}

\newcommand{\defin}[1]{\begin{definition}\label{#1}
\sl }
\newcommand{\edefin}{\end{definition}}

\newcommand{\beqno}{\begin{eqnarray*}}
\newcommand{\eeqno}{\end{eqnarray*}}
\newcommand{\beqla}[1] {\begin {eqnarray}\label{#1}}
\def\eeq {\end {eqnarray}}
\newcommand{\beq}{\begin {eqnarray}}

\newcommand{\dist}{{{\rm dist}\,}}
\def\ga{\gamma}         

\def\al{\alpha}
\def\ep{\epsilon}
\def\la{\lambda}        
\def\de{\delta}         
\def\om{\omega}         \def\Om{\Omega}

\def\CA{{\mathcal A}}              \def\CC{{\mathcal C}}
\def\CD{{\mathcal D}}              
       \def\CH{{\mathcal H}}       
\def\CJ{{\mathcal J}}              \def\CL{{\mathcal L}}
\def\CM{{\mathcal M}}       \def\CN{{\mathcal N}}       \def\CO{{\mathcal O}}
\def\CP{{\mathcal P}}       \def\CQ{{\mathcal Q}}       \def\CR{{\mathcal R}}
\def\CS{{\mathcal S}}       \def\CT{{\mathcal T}}       \def\CU{{\mathcal U}}
       \def\CW{{\mathcal W}}       \def\CX{{\mathcal X}}
\def\CY{{\mathcal Y}}       
\def\CH{{\mathcal H}}
\def\CX{{\mathcal X}}
\def\CY{{\mathcal Y}}
\def\CW{{\mathcal W}}
\def\qq{ \begin{eqnarray} }
\def\qqq{ \end{eqnarray} }
\def\rr{ \begin{equation} }
\def\rrr{ \end{equation} }
\def\non{ \nonumber }
\newcommand{\NJ}{{{\bf J}}}
\newcommand{\Nv}{{{\bf v}}}
\newcommand{\Nw}{{{\bf w}}}
\newcommand{\Nu}{{{\bf u}}}

\newcommand{\Nz}{{{\bf z}}}

\newcommand{\Z}{{{\mathbb{Z}}}}
\newcommand{\R}{{{\mathbb{R}}}}

\newcommand{\E}{{{\mathbb{E}}}}
\newcommand{\T}{{{\mathbb{T}}}}
\newcommand{\N}{{{\mathbb{N}}}}
\def\qq{ \begin{eqnarray} }
\def\qqq{ \end{eqnarray} }
\def\non{ \nonumber }
\newcommand{\no}{\noindent}

\newcommand{\hf}{{_1\over^2}}
\newcommand{\rea}{{\rm Re}}
\newcommand{\ima}{{\rm Im}}

\newcommand{\BbbZ}{\mathbb{Z}}
\newcommand{\BbbR}{\mathbb{R}}

\newcommand{\BbbT}{\mathbb{T}}

         \def\halmos{{\ \vbox{\hrule\hbox{\vrule
height1.3ex\hskip0.8ex\vrule}\hrule}}
            \par\medskip}

\begin{document}
\title[Diffusion in Energy Conserving  Coupled Maps]
{Diffusion in Energy Conserving Coupled Maps }

\author[Jean Bricmont]{Jean Bricmont}
\thanks{Partially supported  by the Belgian IAP program P6/02.}
\address{UCLouvain, IRMP, chemin du Cyclotron 2,
B-1348  Louvain-la-Neuve, Belgium}
\email{Jean.Bricmont@uclouvain.be}

%\thanks{The first author was  supported
%by ??, project \#}

\author[Antti Kupiainen]{Antti Kupiainen}
\address{University of Helsinki, Department of Mathematics and Statistics, P.O. Box 68, FIN-00014 University of Helsinki, Finland}
\email{antti.kupiainen@helsinki.fi}
\thanks{Supported by the Academy of Finland and European Research Council}
%    General info
%\subjclass[2000]{???}
\date{28.10.10}
%\dedicatory{This paper is dedicated }
%\keywords{Random welding, quasi-conformal maps, SLE}

%%%%%%%%%%%%%%%%%%%%%%%%%%%%%%%%%%%%%%%%%%%%%%%%%%%%%%%%%%%%%%%%%%%%%%%%%%%%%%%%

\begin{abstract}
We consider a dynamical system consisting of subsystems indexed by a lattice. Each subsystem
has one conserved degree of freedom (``energy") the rest being uniformly hyperbolic. The subsystems are weakly coupled together so that the sum of the subsystem energies remains conserved.
We prove that the subsystem energies satisfy the diffusion equation in a suitable scaling limit.
\end{abstract}

\maketitle

%%%%%%%%%%%%%%%%%%%%%%%%%%%%%%%%%%%%%%%%%%%%%%%%%%%%%%%%%%%%%%%%%%%%%%%%%%%%%%%%%

\section{Coupled Maps with a Conservation Law}

\subsection{Diffusion in Hamiltonian Dynamics} One of the fundamental problems in deterministic dynamics is to understand the
microscopic origin of diffusion. On a microscopic level, a physical system such as
a fluid or a crystal can be modeled by Schr\"odinger or Hamiltonian dynamics, with a macroscopic
number of degrees of freedom. Although the microscopic dynamics is not dissipative,
dissipation should emerge in large spatial and temporal scales e.g. in the form
of diffusion of heat or of concentration of particles. 

Dynamically, diffusion is related to the existence in the system of conserved quantities
such as the energy which are {\it extensive} i.e. sums (or integrals) of local contributions that are `almost
conserved". Thus, if the system has a microscopic energy density $E(t,x)$, $x\in \BbbR^d$ the total energy
$E_{\rm tot}=\int E(t,x)dx$ is a constant of motion but the energy density is, in general, not conserved since
the dynamics redistributes it:
\qq
\dot E(t,x)=\nabla\cdot{\bf J(t,x)}.
\label{conser}
\qqq
The divergence acting on the energy current $J$ guarantees conservation of the total energy.
One would like to show that the conservative dynamics (\ref{conser}) turns, in a suitable scaling limit, to a
diffusive one. Such a limit involves
diffusive scaling of space and time, and taking typical initial conditions with respect to
the Liouville measure with prescribed initial energy profile. The resulting macroscopic energy density
should then satisfy
 a nonlinear diffusion equation of the type
\qq
\partial_t E=\nabla\cdot (\kappa(E)\nabla E)
\label{hydro}
\qqq
where $\kappa(E)$ is the conductivity function. 

There has been a lot of numerical and theoretical work in recent years around these questions
in the context of {\it coupled dynamics}. One considers a  dynamical system
consisting of a large number of elementary systems indexed
 by a subset $V$ of
a $d$-dimensional lattice $\mathbb{Z}^d$. The total
energy $E$ of the system is a sum $\sum_{x\in V}E(x)$
of energies $E(x)$ which involve the dynamical
variables of the system at lattice site $x$ and
nearby sites and describe the energy of the system
at $x$ and its interaction energy with  its neighbors.

In particular, two classes of models have been discussed. The first consists
of Hamiltonian dynamics of coupled {\it weakly anharmonic} oscillators. In the
weak anharmonicity scaling  limit, one may use kinetic theory to compute the conductivity $\kappa$.
Rigorous justification of the kinetic limit (see \cite{LS}), let alone the case of  fixed (small) coefficient of the
anharmonic term (see  \cite{BK1}), is still lacking. 
%For progress in the former case, see \cite{LS}, and for
%an attempt to study the latter one, see \cite{BK1}.

A second class of models deals with a complementary situation of {\it weakly coupled
chaotic} systems. Hamiltonian systems of this type are obtained by putting at
each lattice site a chaotic system, e.g. a billiard, and coupling them
weakly to each other \cite{buni}. On might hope that the strong mixing
properties of the billiard dynamics could help in proving diffusion. 
Rigorous results on such Hamiltonian systems are rare: in  \cite{buni}
ergodicity is proved for the one dimensional case. Numerically, diffusion comes
out cleanly   \cite{GG}.

In this paper we will consider these issues in the framework of
Coupled Map Lattices (CML) which are  discrete time 
models for extended dynamical systems. We introduce
a class of Coupled Map Lattices which satisfy a discrete
space and time version of the conservation law (\ref{conser}). 
These systems have, apart from the conserved ``energy" variables,
chaotic variables that are coupled to the energies.
We formulate for such systems
a general framework for the derivation of the diffusive dynamics (\ref{hydro})
and prove diffusion for a class of such CML's. 

Our approach is to view
the fast chaotic dynamics as a noise acting on the slow dynamics.
We show that under quite general assumptions the slow dynamics becomes
nonlinear diffusion in a random environment.  We then prove diffusive scaling limit
under the assumption of weak nonlinearity and weak randomness. The proof is based
on a multi scale Renormalization Group (RG) method. In RG jargon, we show that
both the noise and the nonlinearity are irrelevant. We want to stress that
the systems we consider are far from realistic Hamiltonian systems like the
ones mentioned above. The most radical simplification we make is to
assume that the slow dynamics cannot slow down the fast one. This will not
hold in  Hamiltonian systems. However, such slowing down can in principle
be studied using the RG and
we view the paper as a first step in such studies.

\subsection{Coupled Map Lattices} A CML is a dynamical system built out of subsystems $(M_x,f_x)$ indexed by
the lattice $ \mathbb{Z}^d$. The dynamical systems $f_x:M_x\to M_x$
are copies of a fixed system $(M,f)$. 
Let $\Lambda \subset \BbbZ^d$.
The phase space
of the CML in volume $\Lambda$  is ${\mathcal M}_{\Lambda}=\times_{x\in\Lambda}M_x$ and the CML dynamics  ${\mathcal F}:{\mathcal M}_{\Lambda}\to
{\mathcal M}_{\Lambda}$ is a perturbation of the uncoupled dynamics $\times_{x\in\Lambda}f_x$.

To motivate our choice of $M$ and ${\mathcal F}$ consider again the coupled billiard case.
The billiard dynamics has very good chaotic properties. There are two
zero Lyapunov exponents corresponding to the conserved energy and 
the time shift ($\nabla H$ and the Hamiltonian
vector field). A discrete time version (given, say, by a Poincare map) would
have one vanishing exponent and the remaining ones nonzero (positive and negative).

The following general class of CML models this situation.
We let the subsystems have $M= \mathbb{R}_+\times N$
and denote the elements of $M$ by 
$(E,\theta)$. Hence at lattice site $x\in\BbbZ^d$ we have $(E(x),\theta(x))$.
We call energy the non-negative variables $E(x)$. $\theta$ represent the chaotic
variables in the billiard case and they will take values in some manifold $N$.

The uncoupled dynamics is simply defined to be
\qq
(E(x),\theta(x))\to (E(x),g(\theta(x),E(x)))
\label{uc}
\qqq
for each $x\in\mathbb{Z}^d$. 
The energies at each lattice site are thus conserved (as in the
Hamiltonian case).
The variables $\theta(x)$ are the fast, chaotic ones. In the billiard case, 
the dynamical system $\theta\to g(\theta, E)$ is uniformly hyperbolic for any fixed E.
We will model this situation by taking $g(\theta, E)=g(\theta)$
a fixed chaotic map, independent of $E$. 
Examples are given by
$N=\BbbT^1=\mathbb{R}/\mathbb{ Z}$ and $g$  an
expansive circle map, e.g. $g(\theta)=2\theta$ or 
$N=\BbbT^2=\mathbb{R}^2/\mathbb{ Z}^2$ and $g$  a
hyperbolic toral automorphism.  In general we will need good chaotic properties of the
$\theta$ dynamics, in particular an invariant Sinai-Ruelle-Bowen (SRB) measure 
of Gibbsian type (see below).

We want to stress that in a realistic Hamiltonian system, such as the billiard, the $E$ dependence
of $g$ can not be ignored. Indeed, it is obvious that, as $E\to 0$, the Lyapunov exponents
of $g(\cdot,E)$ also tend to zero since $E$ sets the time scale. 

\subsection{Conservative Coupling} The CML dynamics
is now written as
\qq
{\mathcal F}(x,E,\theta)= (E(x)+\phi(x,E,\theta),g(\theta(x))+\psi(x,{\theta}))
\label{uc1}
\qqq
where $\phi$ and $\psi$ are the perturbation of the local
dynamics and will be taken to be small local functions of $(E,\theta)$  (for $\phi$) or of   $\theta$  (for $\psi$)
i.e. to depend weakly on  
 $(E(y),\theta(y))$ for $|x-y|$ large.
 
 We will consider perturbations such that the {\it total energy} $$\sum_{x}E(x)$$
 is conserved. This follows if, formally,
 $$
\sum_{x}\phi(x,E,\theta)=0
$$
 for all $E,\theta$. A natural way to guarantee this is to 
 consider a ``vector field" $$\NJ(x)=\{ J^\mu(x)\}_{\mu=1,\dots, d}$$  and take
  \qq
\phi(x,E,\theta)=(\nabla\cdot \NJ)(x,E,\theta):=\sum_\mu (J^\mu(x+e_\mu, E,\theta)-J^\mu(x,E,\theta)),
\label{current}
\qqq
where $(e_\mu)_{\mu= 1}^d$ is the canonical basis in $\mathbb{ Z}^d$.
With these definitions, we arrive at the following class of dynamical
systems which are natural discrete space time  versions of  (\ref{conser}):
\qq
E(t+1,x)&=&E(t,x)+\nabla\cdot \NJ(x,E(t),\theta(t)):= F(x,E(t),\theta(t))
\label{Edyn}\\
\theta(t+1,x)&=&g(\theta(t,x))+\psi(x,\theta(t)):= h(x,\theta(t)).\label{thdyn}
\label{dyn}
\qqq
Examples of the dynamics (\ref{Edyn}) will be discussed in Remark 2.4 below.

The purpose of this paper is to prove that the deterministic, conservative dynamical system (\ref{Edyn}),  (\ref{thdyn})
gives rise to diffusion under suitable assumptions on the functions $\bf J$ and $h$ .

\section{The Result}
\subsection{Random Environment}
We would like to inquire under what conditions and in what sense  the dynamics (\ref{Edyn}) of the energies is diffusive.
Let us start by specifying in what sense we want to prove diffusion. The strongest and most natural result would be
to prove diffusion 
 {\it for almost all initial values of $\theta$}.
By "almost all" we mean the following. 

Suppose first $\psi=0$.  The
$\theta$ dynamics is then local and the $g$ dynamics has an invariant Sinai-Ruelle-Bowen measure $\nu_0$ on $N$. In this case, it would be natural 
to prove that the $E$ -dynamics is diffusive (in a sense to be specified below)  a.s. in $\theta(0,\cdot)$ with respect to the measure  $\nu=\nu_0^{\Z^d}$.  

For  $\psi\neq 0$, if the $\theta$-dependence of $\psi$
is local and smooth
 and $\psi$ small in a suitable sense \cite{BS, J, JP, PS, BKL, BFG, BGG, CAM, BK5} the $\theta$ dynamics still has an
 invariant SRB measure  $\nu$ defined on the
 cylinder sets of $N^{\Z^d}$. Then we want to prove diffusion a.s. with respect to $\nu$.

Sampling  $\theta(0,\cdot)$ with the probability measure $\nu$ makes 
  $\theta(t,x)$  random variables on the same probability space. 
   Thus the
  $\theta$ dynamics provides a space time {\it random environment} for the
  $E$-dynamics:
   $$
 {\bf J}_t(x,E):={\bf J}(x,E,\theta(t))
 $$
 are random functions of $E$. 
  For $h$ as above, the
  $\theta$ dynamics is exponentially mixing in time and space. 
 If the   $\theta$-dependence of  ${\bf J}$ is suitably local, then ${\bf J}_t(x,E)$ are 
 weakly correlated random variables.  
 We will next spell out these facts precisely.
 
\subsection{Gibbs states} 
We assume $g$ is a uniformly expansive (i.e. $|g'|\geq\lambda>1$)
  $C^{1+\alpha}$ circle map or a $C^{1+\alpha}$
  perturbation of a linear toral hyperbolic automorphism. For the coupling
  assume $\psi$ is $C^2$ (actually $C^{1+\alpha}$ is enough, see  \cite{BK5})
  and local, namely
  \qq
  \label{hass}
  \|{\partial^2 \psi(x,\theta)\over\partial \theta_y\partial \theta_z}\|_\infty \leq \kappa e^{-m(|x-y|+|x-z|)}
  \qqq
 for some $m>0$ and $\kappa $ small enough. Here 
  $
   \| f \|_\infty= \sup_{\theta \in N^{\Z^d}} |f(\theta)|
   $.
   
  Then the following  hold (see Appendix):
 
First,  there is a map  $\Gamma: \Omega\to N^{\Z^d}$
where
$$ \Omega=\times_{(x,t)\in {\Z}^{d+1}}\Omega_{x,t}$$ 
(or a subset thereof, see Appendix) 
where $\Omega_{x,t}$ is a copy of  a given finite
set (whose elements will be called ``spins") such that $\Gamma$ conjugates the
dynamical system  $(N^{\Z^d}, h)$ in (\ref{dyn}) to  a time shift $\tau$ in a space of symbol sequences
$$
h\circ\Gamma=\Gamma\circ\tau.
$$

Secondly  $(N^{\Z^d}, h)$
has an invariant SRB measure $\nu$ which is the image under $\Gamma$
 of a Gibbs measure $\mu$ on  the $\sigma$-algebra $\Sigma$ generated by the cylinder sets
 of $\Omega$:
   \qq
  \label{hass1}
 \int F(\theta) \nu(d\theta)= \int (F\circ\Gamma) (\omega) \mu(d\omega).
\qqq
This Gibbs state is local in the following sense. For $A\subset\Z^{d+1}$
a finite set
let $\Omega_A=\times_{(x,t)\in A}\Omega_{x,t}$. We say $F : \Omega_A \to {\R}$
is supported in $A$. Then

\vskip 2mm

\no (a) 
There exists $m>0, C<\infty$, such that, for
all $F_1$, $F_2$, with $F_i$ supported in $A_i$
% : \Omega_{A_i} \to {\R}$:%, $F_2 : \Omega_B \to {\R}$: 
\qq
| \E (F_1F_2) - \E (F_1) \E (F_2)
 |  \leq C \min (|A_1|,|A_2|) \|F_1\|_\infty\|F_2\|_\infty e^{-md({A_1},{A_2})},
\label{decay} 
\qqq 
where $d({A_1},{A_2})$ is the
distance between the sets $A_1$ and $A_2$ and $\E F =\int F d\mu$.

\vskip 2mm

\no (b) 
There exists $m>0, C<\infty$ such that, if $F_i \geq 0$ are supported on $A_i$, $i=1,\dots,k$
\rr
\mathbb{E} (\prod^k_{i=1} F_i) \leq \prod^k_{i=1}( \mathbb{E} (F_i) \exp (C |A_i| e^{-mR}))
\label{decay1}
\rrr
where $R=\min_{i\neq j}{\rm dist}(A_i,A_j)$.

\vskip 2mm

Property (a) is familiar  for Gibbs states of spin systems with weak interactions. Property (b) is less familiar, but is proven in the Appendix
for the class of Gibbs states that correspond to the SRB measures.

Thirdly, $\Gamma$ maps local functions of $\theta$ to local functions of $\omega$ as follows.
 We say that $w:N^{\Z^{d}}\to\R$ is   {\it local} if $w$ is $C^2$  with, $ \forall z \in \Z^d$,
\qq
  \| \frac{\partial w(\theta)}{\partial \theta_z}\|_\infty \leq C(w)  \exp (-m |z|)
\label{decay2}
\qqq
for some $C(w)<\infty$, $m>0$. Then one can localize  $w\circ \Gamma$ as
\qq
(w\circ \Gamma)(\omega)= \sum_{A\subset\Z^{d+1}}w_A(\om)
\label{locality}
\qqq
where $w_A$ is supported in $A$ and
\rr
\sum_{A\subset\Z^{d+1}}|w_A|e^{\lambda d(A\cup \{0\})}<C (w)
\label{1norm1}
\rrr
for some $\lambda>0$,  and where $d(B)$ is the diameter of the set $B$. This means that $w(\theta)$ becomes
under the conjugation a local function of the
spins near the space time point $0$.
In our problem we will have $$\NJ(x,E,\theta(t))=\NJ(0, \tau_xE, \tau_xh^t(\theta))$$
where $\tau_x$ is the translation by $x \in \Z^d$.  Since $h^t\circ\Gamma=\Gamma\circ\tau^t$, we have
\qq
\tau_xh^t(\theta)= \Gamma( \tau_x \circ\tau^t (\omega))
\label{covar}
\qqq
where, on both sides, $\tau_x$ denotes the natural action of lattice translations by $x\in \Z^d$,
and thus
$\NJ(0, \tau_xE, \tau_xh^t\circ\Gamma(\omega))$ has an expansion like in (\ref{locality})
where in the estimate (\ref{1norm1}) the origin is replaced by the point $(x,t)$ i.e.
it is a local function
of the spins around $(x,t)$.

\subsection{Quenched Diffusion} We may now rephrase the problem of deriving diffusion in deterministic dynamics
 as that of {\it almost sure, or quenched, diffusion in a random dynamics}. Consider the dynamics
 \qq
E(t+1,x)&=&E(t,x)+\nabla\cdot \NJ_t(x,E(t),\omega):=f_t(x,E(t),\omega)\label{rdyn}
\qqq
 where $\NJ_t$ is a random field defined on the probability space $(\Omega, \Sigma, \mu)$ described above where $\mu$ satisfies 
 properties (a) and (b).
 %(we usually suppress the $\omega$ dependence). 
 Let us stress that, in our main Theorem \ref{main}
 below, we assume only those properties for $\mu$, plus some symmetries (see (iii) in Section 2.5 below)  and nothing else;  $\mu$
 {\it does not have to correspond to an SRB measure or even to be a Gibbs state}. The results for the deterministic dynamical system  (\ref{Edyn}-\ref{dyn}),
 stated in Corollary \ref{corol_main} below,
 will be consequences of the results on the random dynamics and of (\ref{locality}, \ref{1norm1}).

 We want to prove that
 the function $E(t)$ diffuses {\bf almost surely} in $\omega$.

Consider first the 
{\it annealed} problem, i.e. the  averaged equation (\ref{rdyn}): 
$$
E(t+1,x)-E(t, x)=\nabla\cdot\E [\NJ_t(x,E(t),\omega)]
:=\nabla\cdot {\mathcal J}(x,E(t)).
$$
where, since we will assume below that $\mu$ is invariant under lattice translations (which corresponds for the time direction to the  stationarity of $\nu$), ${\mathcal J}$ is time independent.
Supposing that $h$ and $\NJ$ have natural symmetries under lattice
translations and rotations (see assumption (iii) in section 2.5 below), we infer that ${\mathcal J}$ vanishes at constant $E$
and then, locality assumptions (\ref{hass}) that we assumed for $h$ 
imply
$${\mathcal J}(x,E)=\sum_y\kappa (x,y,E)\nabla^\dagger  E(y),$$
where  $\nabla_\mu^\dagger$ is the adjoint of $\nabla_\mu$ in $l^2(\Z^d)$, and the matrix $\kappa (x,y,E)$
is a rapidly decaying function of $|x-y|$.
Hence, the annealed dynamics is a {\it discrete nonlinear diffusion}
\qq
\label{anneal}
E(t+1)-E(t)=\nabla\cdot \kappa( E(t))\nabla^\dagger E(t)
\qqq
provided the diffusion matrix $ \kappa( E(t))$ is positive. 

Let now
$$\beta_t (x,E(t))=\NJ_t (x,E(t))- {\mathcal J}(x,E(t))$$
be the fluctuating part. Then, the  $E$-dynamics  (\ref{rdyn}) becomes
\qq
\label{slow}
E(t+1)-E(t)=\nabla\cdot \kappa( E(t))\nabla^\dagger E(t)
+\nabla \cdot \beta_t (E(t))
\qqq
with
$$\E\ \beta_t(E(t))=0.$$
 
 Let us stress that the equations (\ref{anneal}) and  (\ref{slow}) are
 completely general and require only reasonable assumptions of
 locality and smoothness of the functions $\bf J$ and $h$ such as the ones we are going to make in section 2.5 below.
 In particular, the matrix $\kappa$ is close to diagonal  (exponentially decaying).
 In a physical model, one would expect  $\kappa( E(t))$ to be
{\it positive},  although not necessarily uniformly in $E$. If we assume that $\kappa( E(t))$ is uniformly
positive in $E$ and that $\beta$ is a small perturbation, then one would expect  diffusion to occur almost surely.
%If, furthermore,
 %$\beta$ turned out to be a small perturbation, quenched diffusion might be provable. BUT WE PROVE THAT
% EXCEPT FOR UNIFORMITY.
 In what follows, we will make essentially such assumptions on the functions $\bf J$ and $h$ and prove
 almost sure diffusion.
  
\subsection{Random Walk in a Random Environment} Before stating the assumptions let us make one more
 reduction. It is reasonable to assume that $E=0$ is preserved by the dynamics.
 This then implies  $\beta_t (0)=0$. Let us study the linearization at $E=0$:
  \qq
E(t+1)-E(t)=\nabla\cdot \kappa( 0)\nabla^\dagger E(t)
+\nabla \cdot (D\beta_t(0)E(t))
\label{lin}
\qqq
where, by symmetry, $\kappa( 0)$ is a matrix which is a multiple of the identity (multiple given by $\frac{D_0}{2d}$  in (\ref{bite111}) below);
in other words
 \qq
E(t+1, x)=\sum_{y}p_{xy}(t)E(t, y)
\label{lin2}
\qqq
with 
$$\sum_x p_{xy}(t)=1.$$
Since $E\geq 0$ we have $p_{xy}\geq 0$ i.e. $p_{xy}(t)$ are 
{\it transition probabilities of a random walk}. $p_{xy}(t)$  is space and time dependent
and random i.e. it defines a
{\it  random walk in a random environment}.

\subsection{Assumptions} 
 We prove 
 that under suitable assumptions on $\kappa(0)$ and $D\beta_t(0)$ in (\ref{lin})
 the random dynamical system (\ref{rdyn}) is diffusive $\mu$ a.s. in $\omega$
 and $E(0)$ small enough
 i.e. the deterministic system (\ref{Edyn}),  (\ref{dyn}) is diffusive $\nu$ a.s. in  $\theta(0)$.
 
 To state our  results,  it is convenient to introduce the following notation: we identify
  $f: (L^{-n}\Z)^d \to \R$,   to 
 a function $f: \R^d \to \R$, which takes the constant  value $f(x)$ on
 $L^{-n}$ cubes in $\R^d$ centered at $x \in (L^{-n}\Z)^d$, so that we can write, e.g.:
\qq
\int dx f(x):=L^{-nd} \sum_{x\in (L^{-n}\Z)^d} f(x)
\label{integral}
\qqq
Since the $x$ argument of $E(t,x)$ is in $ \Z^d$, the $x$ argument of $E(t, L^{n} x)$ is in $(L^{-n}\Z)^d$. 

For $f: (L^{-n}\Z)^d \to \R$,   
  we denote by $\|f\|_{1}$ the $L^1(\R^d)$ norm of the associated function $f: \R^d \to \R$. When $n=0$, this coincides with the $\ell^1 (\Z^d)$ norm.

Let $B^+_\delta$ denote the set of $E(x) \geq 0$, $x\in \Z^d$, with $\|E\|_{1} \leq \delta$.
We assume that  $f_t$  in (\ref{rdyn}) is measurable in $\omega$ and $C^2$ in $E$ in $B^+_\delta$
for some $\delta>0$.
Moreover, we assume:

\vskip 2mm

\no (i)  {\it Positivity:} $f_t(x, E,\omega)\geq 0$ for $E\geq 0$, $\forall x \in \Z^d$, and a.s. in $\omega$.

\vskip 2mm

\no (ii) {\it Conservation law:} 
\qq
\sum_{x\in \Z^d}f_t(x,E,\omega)=\sum_{x\in \Z^d}E(x)
\label{conserv}
\qqq
This implies, since $E\geq 0$, that $f(x, 0, \omega)=0$.
\vskip 2mm

It is natural to assume that the randomness is statistically symmetric:

\no (iii) {\it Symmetry:} The law of $f_t(x,E,\omega)$
is invariant under the natural action of translations of $\Z^{d+1}$
and rotations fixing $\Z^d$.
\vskip 2mm

\no (iv) {\it Locality:} 
We assume
\qq
|\partial_{ E(y)}f_t(x,E,\omega)|&\leq& Ce^{-m|x-y|}\non\\
|\partial_{ E(y)}\partial_{ E(z)}f_t(x,E,\omega)|&\leq& Ce^{-m(|x-y|+|x-z|)}
\non
%\label{conserv}
\qqq
%$$ e^{m|x-y|}(|\partial_{ E_y}f_t(x,E,\omega)|+|\partial_{ E_y}\partial_{ E_z}f_t(x,E,\omega)|e^{m|x-z|})\leq C $$
%\vskip 2mm
for some $m>0$,  uniformly in $\omega$ and $E\in B^+_\delta$.

\vskip 2mm

 Let us write
the  average of the derivative of the map at $E=0$ as
\qq
\E (\partial_{E(y)} f_t(x,0,\omega)):= T(x-y)
\label{defT}
\qqq
(i.e. $T=1+\nabla\cdot\kappa(0)\nabla^\dagger$ in (\ref{lin})).
From (i) and (ii), we have  $T(x)\geq 0$, $\sum_{x \in \Z^d} T(x)=1$
and from (iv)
$ T(x) \leq C e^{-m|x|}$.
Let $\hat T(k)=\sum_{x \in \Z^d} T(x) e^{-ikx}$. 
%Assumption (iv) implies that  $\hat T(k)$ is analytic in a neighborhood of $\T^d$.
% I WILL CHANGE THE NUMBERING OF THE ASSUMPTIONS
By assumption (iii), we may  write 
\rr
\hat T(k)=1-\frac{D_0}{2d}k^2+\CO(k^4)
\label{bite111}
\rrr
where $D_0$ is the diffusion constant 
$$
D_0=\sum_u u^2T(u).
$$

We moreover assume: 
\vskip 1mm

\no  (v) 
%{\it Ellipticity:}  
$|\hat T(k)|<1$ for $k\neq 0$. 

This is a convenient assumption: although it is not true for all random walks, in particular for the simple random walk, when all the components of $k$ equal $\pi$, 
 it becomes true when one iterates that walk once.
\vskip 2mm
\no Write
\qq
\partial_{E(y)}  f_t(x,E, \omega)
-T(x-y):=\nabla\cdot b_t(x, y, E ,\omega)
\label{defbeta}
\qqq
 By (iv) $b_t(x,y, 0, \omega)$ 
 decays exponentially  in $|x-y|$.
We assume that  it is local in $\omega$ in the sense of Section 2.2 and small:
\vskip 1mm
\no (vi) {\it Weak randomness.}   Assume that
 \qq
b_t(x,y, 0, \omega)    =\sum_{A\subset {\BbbZ^{d+1}}}b_{t, A}(x, y ,\omega)
\label{bsum}
\qqq
with $b_{t, A}$ supported on $A$ (see Section 2.2) and, for some $\la>0$,
\rr
\sum_{A\subset {\BbbZ^{d+1}}, x \in \Z^d}| b_{t, A
}(x,y,\omega)|e^{\la d(A\cup (x,t) \cup (y,t))}<\epsilon
\label{1norm}
\rrr
%\begin{remark} As explained in Section 2.2.  the representation
 %(\ref{bsum})  satisfying (\ref{1norm1}) follows from a local $b(E,\theta)$ (see (\ref{decay2})).
 %\end{remark} 
 Below we will usually drop the argument $\omega$.
 
Then our main result is: 
\begin{theorem}\label{main}
{\it Consider the dynamical system} (\ref{rdyn}). {\it Assume that the random field $(\Om, \Sigma, \mu)$ satisfies}  (\ref{decay}),  (\ref{decay1}){\it 
and that $f_t$ satisfies the assumptions} (i)-(vi) {\it above.
Assume moreover that $L$ is an integer, that $d \geq 2$   and that $\epsilon$ and $\delta$ are small enough. 

Then
there exists, almost surely in $\omega$, and  for all $E\in B^+_\delta$,
a $D>0$ such that, for all
$ G\in C^1(\BbbR^{d}) $, with $\|G\|_\infty$, $\|\nabla  G\|_\infty$, finite,}
\qq
\lim_{n\to \infty} \int dx G(x) \Big(L^{nd} E(L^{2n} , L^{n} x)-\|E\|_{1}T_D^*(x)\Big) =0
\label{result}
\qqq
where $T_{D}^*(x)=( \frac{d}{2\pi D})^{d/2}e^{-dx^2/2D}$.% and $\|E\|_1=\sum_{x \in \Z^d} E(x)$. 
%and $[\cdot]$ denotes the integer part. NOT A GOOD NOTATION-USED LATER.
\ethm
%WE NEED TO DEFINE THE INTEGRAL NOTATION FOR LATTICE SUMS SOMEWHERE!
\begin{remark} The restriction to $d\geq 2$ is done for convenience. With a more detailed nonlinear analysis 
(see Section 5) one could extend the result to $d=1$.

Moreover, the result is stated for a subsequence of times
of the form $L^{2n}$, but a  general sequence of times 
and even a result for several   times, leading to the convergence of finite dimensional distributions
and  to an invariance principle, could be obtained along the lines of the arguments in \cite{BK08}.

Finally, the convergence is weak, but is uniform over functions $G$ with fixed norms $\|G\|_\infty$ and $\|\nabla  G\|_\infty$. 
 \end{remark} 
 
 Combining this theorem with  (\ref{locality} ), (\ref{1norm1}), we can now formulate a similar result for the
 deterministic  dynamical system (\ref{Edyn}, \ref{dyn}). We assume that $F$ in (\ref{Edyn}) satisfies the same assumptions (i), (ii) (iv), (v), 
stated above for $f_t$ in (\ref{rdyn}), and satisfies, instead
 of (iii),
 \vskip 2mm

 \no (iii') {\it Symmetry:}  $F (x,E(t),\theta (t))$
is invariant under the natural action of translations of $\Z^{d}$
and rotations fixing $\Z^d$.
\vskip 2mm

As in (\ref{defbeta}), we define $T(x-y)= \E( \partial_{E(y)} F(x,0, \theta))$ and
\qq
\partial_{E(y)}  F(x,E, \theta)
-T(x-y):=\nabla\cdot b (x, y, E ,\theta)
\label{defbeta1}
\qqq
By (iii'), $b(x,y,E, \theta)= b(0,y-x, \tau_x E, \tau_x \theta)$. 
\no Instead of (vi), we assume for $b$:
 \vskip 2mm 

 \no (vi') {\it Weak randomness:}
    We assume that $w(\theta):= b(0,x,0,\theta)$ is local in the sense 
 (\ref{decay2}), with $C(w)=\ep' e^{-m|x|}$.
   \vskip 2mm
 
Then, we have the
 \begin{corollary}\label{corol_main}{\it 
 Let g, $\psi$ in (\ref{dyn}) satisfy the assumptions of  Section }2.2, {\it with $\kappa$ small enough, and let  $F$ in} (\ref{Edyn}){\it 
 satisfy}  (i), (ii)  (iii')  (iv), (v), (vi') {\it above. Assume that   $\ep'$ and $\delta$ small enough. Then, the result stated in
 theorem }\ref{main} {\it holds also for the dynamical system} (\ref{dyn}) (\ref{Edyn}), {\it almost surely  in $\theta$ with respect to the SRB measure $\nu$.}
\end{corollary}
 \begin{remark} 
Concrete examples of deterministic or random dynamical systems that satisfy the
 assumptions of Corollary \ref{corol_main} or Theorem  \ref{main} respectively can be
 obtained  by taking in \eqref{Edyn} 
 $$
 {F}(x,E,\theta)=E(x)+\sum_y d(x-y)E(y)+\tilde{F}(x,E,\theta)
 $$
 where $d(x) \geq 0$, for $x\neq 0$, $-\hat d(k)/k^2\geq d_0>0$  and where $\tilde{F}$ is a
 perturbation as in (vi') with $\epsilon'$ small enough.  A generic (small, local and $C^2$) coupling
 function $\bf J$ in  \eqref{Edyn}  would give rise to $d_0$ and $\epsilon'$ comparable and hence not satisfying
 our assumptions. However we expect to be able to handle such cases as well by perturbing
 around a weak coupling limit i.e. considering first a time scale of order $\epsilon^{-1}$
 to produce  `effective" $T$ and $b$ that fit into our assumptions \cite{bkinpreparation}.
  \end{remark} 
 The proofs of Theorem \ref{main} and of Corollary \ref{corol_main} will be given in section 6. In the following section, we will explain the Renormalization Group method used
 in the proofs. In section 4, we discuss the linearized Renormalization Group, and we state and prove the auxiliary Propositions \ref{sell},\ref{prob1},\ref{T}, that are used in Section 6.
 In section 5, we give the estimates on the nonlinear parts of the Renormalization Group transformation, also used in Section 6 (Proposition \ref{sell3}). The Appendix is devoted to properties of Gibbs states and SRB measures that are used in the proof of Corollary \ref{corol_main}.
 
\section{Renormalization group for random coupled maps}
The proof of  Theorem  \ref{main} is based on a Renormalization Group (RG)
method introduced in  \cite{BK91} and   \cite{BK08}. We will next explain this RG
and give an outline of how it yields the Theorem.  
\subsection{Renormalized Dynamics} Let us introduce the
{scaling}  transformation $S_L$:
 \qq
(S_LE)(x)=L^dE(Lx).
\label{scaling}
\qqq
where $L>1$. Note that $S_LE$ is defined on a finer lattice $(L^{-1}\Z)^d$.
Fix $L$ and define, for each $n\in\N$, {\it renormalized energies}
$$
E_n(t, x)=S_{L^{n}}E(L^{2n}t,x),
$$
where $t\in L^{-2n}\N $, $x\in (L^{-n}\Z)^d$.
We can then rephrase the scaling limit inside the integral in  (\ref{result}) as 
\qq
\lim_{n\to\infty}L^{nd}E(L^{2n},{L^n x})=\lim_{n\to\infty}E_n(1,x).
\label{jb4}
\qqq
$E_n(t)$  inherits its dynamics from the one of $E$. We will call this
the {\it renormalized dynamics}:
\qq
E_n(t+1)=f_{n,t}( E_n(t)).
\label{jb5}
\qqq
Explicitely we have:
 $$f_{n,t}=S_{L^{n}}\circ f_{L^{2n}t+L^{2n}-1}\circ\cdot\cdot\cdot\circ f_{L^{2n}t}\circ S_{L^{-n}} .$$

\subsection{Renormalization Group} The dynamics changes with the scale as:
$$f_{n+1}={\mathcal R}f_n,$$ 
where
\qq\label{RG}
{\mathcal R}f_{n,t}%=S_Lf^{\circ L^2}S_{L}^{-1}
=S_L\circ f_{n,t_{L^2}}\circ\cdot\cdot\cdot\circ
f_{n,t_1}\circ S_{L}^{-1}
\qqq
with $t_k=L^2t +k-1$, $k=1, \dots L^2$.
% and $t_{L^2}=L^2(t+1)-1$.
The map $\mathcal R$ is the {\it Renormalization Group} map acting in a suitable space of random
maps $f$ . We may rephrase the diffusive scaling limit as a property of the Renormalization Group flow.
We will prove that, {\it almost surely}, 
the renormalized maps converge, in a suitable sense,
$$
{\mathcal R}^nf\to f^*
$$
where the fixed point is {\it nonrandom and linear}:
$$
 f^*(E)= e^{\frac{D\Delta}{2d}}E.
$$
$\Delta$ is the Laplacean on $\R^d$ (note that $ f^*$ is defined on $E(x)$, $x\in\R^d$) 
and $D>0$.
Moreover, we show that the renormalized energies converge almost surely to the fixed point
$$E_n(1,x) -\|E\|_{1}( \frac{d}{2\pi D})^{d/2}e^{-dx^2/2D}\to 0$$
(in the sense of (\ref{result})),
which is the diffusive scaling limit. 
These results may be summarized by saying that both the randomness
and the nonlinearity are {\it irrelevant} in the RG sense. Let us
sketch the reasons for this.
 
\subsection{Linearization} Let us adopt the convention that objects on scale $n+1$ are
denoted by prime and the unprimed ones are on scale $n$ (and delete the  indices $n$ or $n+1$). As we will see in Section 6,
it will be sufficient to control derivative of $f$ (see (\ref{Df2})). For $f'=\CR f$ 
the iteration (\ref{RG}) implies a recursion for the derivative of $f$: let $Df$ denote
 the kernel  of the derivative map (using definition (\ref{integral}) for the integrals), $Df(x,y)=\partial_{E(y)}f(x, E)$, $x,y\in(L^{-n}\Z)^d$. We get:
\qq
\label{LRG}
Df'_{t'}(E')%=S_Lf^{\circ L^2}S_{L}^{-1}
=L^d{\mathcal S}_L(Df_{t_{L^2}}(E_{L^2})Df_{t_{L^2-1}}(E_{L^2-1})\dots
Df_{t_1}(E_1))
\qqq
where 
$$E_s=f_{t_{s-1}}\circ\dots \circ f_{t_1}(E),\ \  \ E=L^{-d}E'({\cdot/ L}),
$$
$t_k=L^2t +k-1$, $k=1, \dots L^2$.
%$t_1=L^2t'$, $t_{L^2}=L^2(t'+1)-1$, 
with the convention $E_1=E$.
 ${\mathcal S}_L$
is the scaling 
$$({\mathcal S}_LM)(x,y)=M(Lx,Ly).$$
%$Df$  denotes the derivative map.
The factor $L^d$ in (\ref{LRG}) comes from the fact that $Df(x,y)$ is a kernel, the definition (\ref{integral}), and the fact that $Df'$ on the left hand side of (\ref{LRG})
is on scale $n+1$ while $Df$ on the right hand side is on scale $n$.

Let us consider this iteration at $E=0$. Set
$$
p_t:=Df_{t}(0).
$$
Then
\qq
\label{rwreite}
p'_{t'}(x',y') = L^dp_{t_{L^2}}\dots p_{t_1}(Lx',Ly').
\qqq
Next, use  (\ref{defT}), (\ref{defbeta}),  to write $p$ as a sum of its average and fluctuation:
\qq
p_t({x,y})=T(x-y)+\nabla_n\cdot b_t({x,y})
 \label{bdef}
\qqq
where $\nabla_n$ acts on the $x$ variable and, for a function $f : (L^{-n}\Z)^d \to \R$,  $\nabla_{n,\mu}f (x)=L^n (f(x+ L^{-n} e_\mu)-f(x))$,
and where $ b_t({x,y})$ is the value of $ b_t({x,y, E})$ at $E=0$ 
(we will suppress the index $n$ on $\nabla_n $ below, when the scale 
on which it is defined will be obvious from the context).
We have
 \qq
 \label{Tdef}
 \E p_t(x,y):=T(x-y)
 \qqq
  and $\E b_t=0$. 
  
  Then, for
$p'=T'+\nabla_{n+1}\cdot b'$, we get
 \qq
T'(x'-y')=L^dT^{L^2}(Lx'-Ly')+P(x'-y')
\label{tite}
\qqq
where $P$ is the expectation of a polynomial in $b$.

 For the
noise, we get
 \qq
&&
\nabla_{n+1}\cdot b'_{t'}=
\\
&&
L^{d}\sum_{t\in I_{t'}}\int dxdy
T^{L^2-i-1}(Lx'-x)\nabla_n \cdot 
b_{t}(x,y)T^{i}(y-Ly')
+G_{t'},
\label{bite}
\qqq
where 
$$I_{t'}=[L^2t', L^2(t'+1)-1],$$
 $i=t-L^2t'$
and  $G_{t'}$ involves quadratic and higher order polynomials in $b$. Note that, since $T$ is time independent, we can write in the sum
 $T^{L^2-i-1}$, $T^{i}$, where the exponents
 refer to amounts of time spent in the interval $I_{t'}$,
but do not depend on $t'$, unlike the random variable $b_{t}(x,y)$, where $t\in I_{t'}$.

\subsection{Linear RG} Treating the noise $b$ as a perturbation, consider 
 first the iterations (\ref{tite}) and  (\ref{bite})  to leading order in $b$.
 For $b=0$,  we get 
 \qq
\CT_n:= L^{nd}T^{L^{2n}}(L^n\cdot)
\label{bite1}
\qqq
i.e.
 \rr
 \hat \CT_n(k)=\hat T(k/L^n)^{L^{2n}}.
 \label{bite11}
\rrr
 Then, using (\ref{bite111}), we get, as $n\to\infty$:
\qq
\hat \CT_n(k)\to e^{-\frac{D_0}{2d}k^2}:=\hat T^*_{D_0}(k)
\label{jb10}
\qqq
which explains the form of the fixed point.

For the noise, the linear RG map  is gotten from (\ref{bite}) by integrating the $\nabla_n $
by parts (actually, Riemann summing by parts) and commuting with ${\mathcal S}_L$ (which gives rise to the factor $L^{-1}$, 
compared to (\ref{bite})): 
\qq
\label{linearRG}
(\CL b)_{t'}&=&\sum_{t\in I_{t'}}\CL_{i} b_{t},\\
\CL_i b({x',y'})
& :=& L^{d-1}\int dxdy
T^{L^2-i-1}(Lx'-x)
b(x,y)T^{i}(y-Ly'),
\nonumber
\qqq
with $I_{t'}$, $i$ as in (\ref{bite}).
In the next Section we will show  that a suitable norm of the variance of $ b$ contracts under $\CL$.
Let us explain the intuitive reason for this. Take e.g. $x'=y'=0$ and $t'=0$, the estimate
being similar for other $x', y', t'$. For $t$ of order $L^2$, 
$T^t(Lx'-x)\sim L^{-d}e^{-|x'-x/L|}$. Hence the $x$ and the $y$ integrals are
localized in an $L$ cube at the origin.
As it will turn out, $b_t({x,y})$ has exponential decay in $|x-y|$, $\forall t$, so let us put heuristically $x=y$, 
 \qq
(\CL b) ({0,0}) \sim
  L^{-d-1}\sum_{t=0}^{L^2-1}\int_{|x|<L}dx
b_{t}({x,x}).
\label{line}
\qqq
Since correlations
of $b$ decay exponentially in space and time (\ref{line}) is effectively a sum
of $L^{d+2}$ independent random variables of variance $L^{-2d-2}(\E b)^2$
and we conclude $\E(\CL b)^2\sim L^{-d}\E b^2$. The actual contraction factor
depends on the norm on $b$ and will be slightly different.

Taking into account the corrections $P$ and $G$ in (\ref{tite}) and (\ref{bite})
that are smaller than the linear term,
we conclude that we should expect the  variance to contract 
$$\E(b_{n,t})^2\sim \epsilon^2_n\to 0$$
(for any fixed $t$) as $n\to\infty$  and
the iteration of the mean to become
 \qq
T_{n+1}=L^dT_n^{L^2}(L\cdot)+\CO(\epsilon^2_n).
\label{titen}
\qqq
The fixed point is the same but the $\CO(\epsilon^2_n)$ renormalizes the
diffusion constant $D$ at each iteration step (less and less as $n\to
\infty$).

\subsection{Random Drift} There is a problem however once we try to make  this perturbative
analysis rigorous. Deterministically the noise is {\it relevant}: from (\ref{line})
we see that $\|\CL b\|_\infty$
can be as big as 
$
\CO( L)\| b\|_\infty
$. This means that there are unlikely events in the environment where the random
walk develops a drift. Recall that $b_{n,t}(x,y)$ is the fluctuating part of the transition
probability density for random walks in time $L^{2n}$ from $L^nx$ to $L^ny$.  There will
be a (random) region $D_t\subset \Z^d$ where 
we write (see (\ref{f3a})-(\ref{f24}) below for  a precise definition),
 for $u,v\in\Z^d$, and any given $t$,
$$
  \sup_{x\in \Nu, y\in \Nv}  |b_{n, t} (x,y)  | 
\leq  \ep_n e^{N_{n, z}},
$$
with $ \Nu,  \Nv$ the unit cubes in $\R^d$ centered on $u, v$ and $z=(t, u)$.
Then $N_{n, z}$ can be (very) large, but with (very) small probability:
$${\rm Prob}(N_{n, z}>N)\leq e^{-K(n+N)}$$
with $K$ large (see Proposition \ref{prob1} below for a precise bound).

\subsection{Nonlinear corrections} Finally, to control the {\it nonlinear} contributions to $f_n$ we
show that  the second
derivative $D^2_Ef$ is irrelevant in all dimensions due to
the scaling of $E$. Differentiating (\ref{LRG}) we obtain
\qq
\label{LRG1}
D^2f'_{t'}(E')%=S_Lf^{\circ L^2}S_{L}^{-1}
=L^d\sum_{i=1}^{L^2}{\mathcal S}_L(Df^{(L^2)}\dots Df^{(i+1)}D^2f^{(i)}(Df^{(i-1)}\dots Df^{(1)})^{\otimes^2})
\qqq
where we denote $Df^{(i)}=Df_{t_i}(E_i)$, 
%with the convention $Df^{(0)} (E) =E$, and
and where
 we write, for matrices $F$, $G$, 
\qq
D^2f( F \otimes  G) (x,y,z) = \int dudv D^2f(x,u,v) F (u,y)G (v,z).
\label{f10a}
\qqq
Evaluating this at $E'=0$, and to first order in the noise, we arrive at the
 linear RG map 
 %(still denoted by $\CL$) 
 on $D^2f_t(0):=\nabla\cdot \rho_t$: 
\begin{eqnarray}
&&(\tilde \CL \rho)_{t'}( x^\prime,y^\prime,z^\prime) = \nonumber \\
&& L^{d-1}\sum_{t\in I_{t'}}\int dxdydz T^{L^2-i-1} (Lx^\prime-x) \rho_{t}( x,y,z)T^{i} (y-Ly^\prime) T^{i} (z-Lz^\prime).
\label{f41}
\end{eqnarray}
with $I_{t'}$, $i$ as in (\ref{bite}).
We will show in  Section 5 that $\tilde \CL$ is a contraction in a suitable norm. It is not hard
to see why this should be so. We will show that $\rho_t$ is (exponentially) local, so let us set, heuristically, $x=y=z$.
Using the bounds on $T$ (proven in Corollary \ref{ttilde} below), $\|T\|_1=1$ to control the integral over $x=y=z$,
and $\|T^j\|_\infty=\CO(j^{-d/2})$ for the other powers of $T$,
 we conclude that the integral yields
a smallness factor 
$$
\min\{i^{-d},i^{-d/2}(L^2-i)^{-d/2}\} \leq 2^{d/2}L^{-d} i^{-d/2},
$$
whose sum over $i \in [1, L^2]$, multiplied by $L^{d-1}$, is small for $d>1$,
i.e. we expect  $\tilde \CL$ to be contractive for $d>1$.

\section{Random Walk in a Random Environment}

We saw in Section 3.3 that the linearization of our random dynamical system, $p_t=Df_t(0)$, has a closed evolution (\ref{rwreite})
under the RG.
Thus we solve this first and use the result as a driving term for the nonlinear
part of $f_t$. $p_t$ defines a random walk in a random environment that
has exponentially decaying correlations in both space and time. An almost sure 
(or quenched) central limit theorem has been established for this problem
\cite{DL, DKL,BK08}, but we need a very detailed knowledge of it in order to control
the nonlinear flow. The proof given in this section is a (much simplified)
version of the (much more difficult) proof \cite{BK91} for environments that
are time independent.

\subsection{Localization}
Recall from Section 2.5 that the fluctuating part of the map $f_t$ has the expansion (\ref{bsum})
in terms of localized terms that are (almost) independent for disjoint $A$'s. As explained in Section 3.5. there will be a random region $D_t\subset \Z^d$ of possibly large drift. It turns out to
be useful to localize the $b$ on the set $D_t$.
We will iterate 
a representation 
\qq
\label{bloc}
b_t=\sum_{\CD\subset D_t}\sum_{A\subset \Z^{d+1}}b_{t,
\CD, A}
\qqq
 Note also that $\CD=\emptyset$
is included (and is actually the more probable contribution). The representation (\ref{bsum}) is of this form, with $b_{t,
\CD, A}=0$, if $\CD \neq \emptyset$.
We will denote
\qq
\label{sdef}
s_{t, A}:=b_{t, \emptyset, A}
\qqq
and write
\qq
\label{cdec}
b_t=s_t+\ell_t
\qqq
("small" and "large" contributions, the $s$  term being more probable and the $\ell$ one including the large, but rare, contributions). %Here 
%$$\ell_{A}=0\ \ \ \NA\cap D_t=\emptyset.$$
Given $b_t,D_t$, we need to explain how $b'_{t'},D'_{t'}$
are defined in terms of them. This will need several steps and definitions, to which we turn now.

\subsection{Relevant and irrelevant variables}
Let us first discuss the  linear RG (\ref{linearRG}). 
 Thus let $\sigma=\CL s$ i.e.
\qq
\label{linearRGs}
\sigma_{t'}
 = \sum_{t\in I_{t'}}\sum_{A\subset\Z^{d+1}}\CL_i s_{t, A}
\qqq
where 
$$I_{t'}=[L^2t', L^2(t'+1)-1]$$
and $i=t-L^2t'$ as in (\ref{bite}).
Next, we localize $\sigma$. We introduce some notation.
Given $z'\in\Z^{d+1}$ let
$\bf z'$ denote the unit cube in $\R^{d+1}$ with center $z'$. 
Define also "blocking": given $A\subset \Z^{d+1}$ set
\begin{equation}
[A]:={\rm int} \{(L^{-2}t,L^{-1}u)\ | \ (t,u)\in A\}
\label{block}
\end{equation}
where $\rm int $ denotes the integer part.
Given a subset $A'$ of $\Z^{d+1}$ we define a family   $\CA(A')$  of subsets $A$ of $\Z^{d+1}$
as follows. 
Let first $|A'|=1$ i.e.
$A'=z'\in\Z^{d+1}$ (by a slight abuse and in order to simplify notations, we write $z'$ instead of $\{z'\}$). 
Then $A\in\CA(z')$ if either $[A]\subset {\bf z'}$, or
 $[A]\cap  z' \neq\emptyset$, ${\rm diam}(A) \leq \frac{L}{4}$
 and the first point $z$ (in lexicographical order) in $[A]$ belongs to $\Nz^\prime$. 
For $|A'|>1$ we let $A\in\CA(A')$ if
 $[A]=A'$ and ${\rm diam}(A) > \frac{L}{4}$. Then, defining
\qq
\label{sigmaA}
\sigma_{t', A'}
 = \sum_{t\in I_{t'}}\sum_{A\in\CA(A')}\CL_i s_{t, A},
\qqq
we have
\qq
\label{sigdec}
\sigma_{t'}=\sum_{A'}\sigma_{t', A'}.
\qqq

The virtue of this decomposition is that, as we will see in (\ref{contra}) below, the nonlocal terms
$\sigma_A$ with $|A|>1$ will strongly contract.

The term will $|A|=1$ does not contract deterministically, but it will contract with high probability, basically
because its variance contracts, as 
explained in (\ref{line}).  We will come back to this lack of deterministic contraction in Section 4.4.

\subsection{Composite operators} Let us next turn to the remainder term $G_{t'}$ in (\ref{bite}) 
consisting of a polynomial in $b$.  We introduce some notation. For matrices $F_t (x,y)$,
indexed by time, we denote
\qq
%\label{expa}
\nonumber
F_I=F_tF_{t-1}\dots F_s
\qqq
for an interval of times $I=[s,t]$. 
Let $\CP$ be the set of nontrivial partitions $\pi$ of the time interval $I_{t'}$
into intervals $\pi_1,\dots,\pi_k$ arranged in increasing
time order and $\pi_1$, $\pi_k$ possibly empty, with $T_{\pi_k}$, $T_{\pi_1}$ being then the identity
(thus $k\leq L^2+2$). Then,
\qq
\label{pdec}
%\nabla\cdot c'
p'=L^d{\mathcal S}_LT^{L^2}+R'
\qqq
with
\qq
\label{rprime}
R'=L^d
\sum_{\pi\in\CP}{\mathcal S}_LT_{\pi_k}(\nabla_n\cdot b)_{\pi_{k-1}}T_{\pi_{k-2}}\dots  (\nabla_n\cdot b)_{\pi_2}T_{\pi_1}.
\qqq
(remembering that $T_{\pi_k}$ or $T_{\pi_1}$ could be the identity).
Thus, writing $\pi_i=\{\tau_i\}\cup \pi'_i$ with $\tau_i$ the largest time, we get, by integration by parts,
\qq
\label{bprime}
R'=\nabla_{n+1}\cdot Q',
\qqq
with
\qq
Q'=L^{d-1}\sum_{\pi\in\CP} {\mathcal S}_LT_{\pi_k}b_{\tau_{k-1}}(\nabla_n\cdot b)_{\pi'_{k-1}}\nabla_n\cdot T_{\pi_{k-2}}\dots b_{\tau_{2}} (\nabla_n\cdot b)_{\pi'_{2}}T_{\pi_1},
\label{rem1}
\qqq
where the $L^{-1}$ comes from the commutation of $\nabla$ and ${\mathcal S}_L$, as in (\ref{linearRG}).  Note that, by integration by parts, we have
transferred the action of  one $\nabla_n$ from a $b$ to a $T$. Thus, we never encounter factors like $T \nabla_n\cdot b \nabla_n\cdot b $ or $\nabla_n\cdot T\nabla_n\cdot b$.
 It turns out  that there
is nevertheless an obstruction to bounding these nonlinear terms due to the gradients in (\ref{rem1}). Indeed,  we will show (see Proposition \ref{sell})
that, in a suitable norm, 
\qq
\label{sbound}
\|s\|\leq (L^{-n}\epsilon)^{\zeta}
\qqq
with $0<\zeta<1$. This bound is not sufficient to show that the  nonlinear terms
in eq. (\ref{rem1}) (with $b$ replaced by $s$) are smaller than the linear one. Indeed, on the 
lattice $(L^{-n}\Z)^d$  the operator $\nabla_n$ is bounded in sup norm by $CL^n$ and hence
$\|\nabla_n \cdot s\|\leq  CL^{(1-\zeta) n}\epsilon^{\zeta}$, which is large. To show that the  nonlinear terms
are subleading, we need to show that $\|s\nabla_n \cdot s\|$ has a better bound than $\|s\|$ and 
that $\|\nabla_n \cdot s\nabla_n \cdot s\|$ is also small (in Proposition \ref{sell} below we will state bounds on these variables and, in
subsection 4.10, we will show that this is enough to control all the nonlinear terms). In Renormalization Group language, we need to
show that these "composite operators" are irrelevant. Hence, let us look  at the linear RG for them.

%We will denote $B^1_t:=b_t$ and set
Let
\begin{equation}
B_t:=b_{t+1} \nabla_n \cdot b_t
\label{B}
\end{equation}
 Inserting the expansions (\ref{rem1}) 
 %and (\ref{rem2})
and using 
$$({\mathcal S}_Lc)({\mathcal S}_Ld)=L^{-d}{\mathcal S}_L(cd)
$$ 
which follows by a change of variables,
we  observe that $B'_{t'}$ is given by the  expansion (\ref{rem1}) on the time interval 
$I_{t'}\cup I_{t'+1}$
with the additional constraint that there is at least one factor of $b$ on both intervals 
$I_{t'}$ and $I_{t'+1}$. In the terms in  (\ref{rem1}) involving only two consecutive $b$'s,
    the only one involving $\nabla_n \cdot b$ is the linear RG one (because
the last  $b$ factor in    each term in (\ref{rem1}) is  $b_{\tau_{k-1}}$  and not $\nabla_n \cdot b_{\tau_{k-1}}$):   
\begin{equation}
(\CL B)_{t'} := L^{d-1} {\mathcal S}_LT^{L^2-1}   B_{L^2(t'+1)-1} T^{L^2-1},
\label{f54}
\end{equation}
 and it turns out that this  {\it linear} RG will be contractive in our norm.

%The composite operator
%\begin{equation}
%B^3_t:=\nabla b_{t+1} \nabla \cdot b_t
%\label{B}
%\end{equation} 
 %is given by the expansion (\ref{rem1}) on the time interval $I_{t'}\cup I_{t'+1}$
 %with the aforementioned constraint. Observe that there are no quadratic terms with this constraint where both $b$ carry a gradient. Hence there is no linear RG for this composite operator. 

The localization of 
$B$ is inherited from the one of $b$ in (\ref{bloc}). We get% for $\alpha=2,3$
 % $D^-\subset D_t$
%and $D^+\subset D_{t+1}$ set $\ND=(D^-,D^+)$and write
\qq
\label{Bloc}
B_t=\sum_{\CD}\sum_{A\subset \Z^{d+1}}B_{t, \CD, A}.
\qqq
where now $\CD:=(\CD_t,\CD_{t+1})$ with $\CD_t\subset D_t$, $\CD_{t+1} \subset D_{t+1}$
and  
$$
B_{t, \CD, A}=\sum_{A_{t+1}, A_t}
b_{t+1,\CD_{t+1}, A_{t+1}} \nabla_n \cdot b_{t, \CD_{t}, A_{t}},
$$
 with $A_{t}\cup A_{t+1}= A$.

\subsection{Drift region}
The drift region is inductively built out of the regions where $\sigma=\CL s$
fails to contract. Let us denote the local part of $s$ by $s^{loc}$, i.e.
\begin{equation}
s^{loc}_t:=\sum_{u\in\Z^d}s_{t,u}.
\label{srel}
\end{equation} 
and 
\begin{equation}
\sigma^{1}_{t'}:=\sum_{u'\in\Z^d}(\CL s^{loc})_{t',u'}.
\label{srel1}
\end{equation} 
Here and below, when we use the indices ${t,u}$, we mean them as in (\ref{sdef}),  i.e.  with $\CD=\emptyset$, and with $|A| =1$, $A=\{u\}$.

The size of $b_n$ will be measured in terms of a running parameter
\rr
\epsilon_n=L^{-n}\epsilon
\label{endefi}
\rrr
where $\epsilon$ is a bound on  the size of the initial $b$, see (\ref{1norm}). We also need
a parameter $\gamma$ satisfying
\rr
0<\gamma<1/12.
\label{aldefi}
\rrr
Given a space-time point $z'=(t',u')$ define
random variables
\qq
r_{z'}:=(n_0+n) \log L\ \ {\rm if}\ \  \|\sigma^{1}_{t', u'}\|_{\lambda} \geq {_1\over^8} \epsilon_{n+1}^{1-3\gamma} 
\label{f18a}
\qqq
and zero otherwise, where the norm $ \|\cdot\|_{\lambda} $ is defined in (\ref{f24}) below.
% WE NEED log L HERE multiplying $n_0+n$, OR PUT $L^-N$ in the defintion of the norm, see \ref{n0choise}. WHAT TO CHOOSE?
%(with a slight abuse of notations, we identify
%$\sigma^{1}_{t', u'}$ with the sequence $\sigma^{1}_{t'\cdot}\delta_{\cdot,u'}$). 
The number  $n_0$ is
chosen so that:
\qq
L^{-n_0}=\epsilon^{3\gamma}.
\label{jb1}
\qqq
% in eq. (\ref{n0choise}) below. 

$D_t$ is now defined in terms of random integers $N_{n, z}$, $z= (t, u) \in\Z^{d+1}$ and where we usually suppress the index $n$ (and, as before, primes will refer to the scale $n+1$)
\qq
D_t=\{ u\in \mathbb Z^d \ | \ N_{(t,u)} \neq 0 \}
\label{f3a}
\qqq
which, in turn, are given recursively in terms of the random variables $r$
of previous scales as follows. 
Let 
\begin{equation}
\tilde N_{z'}=\sum_{z:[z]=z'} N_z.
\label{a7a}
\end{equation}
where we write $z$ for $\{z\}$ and the blocking operation $[z]$ is defined in (\ref{block}).
Then we set
\begin{equation}
N^\prime_{z^\prime} := \tilde N_{z'} - 1 + r_{z'},\ \ {\rm if}\ \
\tilde N_{z'} \geq \lambda/3 \ \ {\rm or} \ \ r_{z^\prime} \neq 0
\label{a7}
\end{equation}
\begin{equation}
N^\prime_{z^\prime} = 0,\ \ {\rm if}\ \
\tilde N_{z'} < \la/3 \ \ {\rm and} \ \ r_{z^\prime} =0.
\label{a8}
\end{equation}
The subtraction by one means that, if no large fluctuations $r_{z^\prime}$  occur during a sufficient number of scales,
then $N$ will  eventually vanish.

\subsection{Norms} We will iterate bounds for the %drift $b_t$ and
%the composite operators   
$b_t$ and $B_t$. %and  $\NB_t$. 
For
this, we need to introduce suitable norms. For a kernel $b(x,y)$ with $x,y\in 
L^{-n}\Z^d$ let, for $u,v\in\Z^d$
\begin{equation}
  \|b  \|_{u,v}=  \sup_{x\in \Nu, y\in \Nv}  |b (x,y)  |.
 \label{f24a}
\end{equation} 
Strictly speaking, the norm depends on $n$, but we will suppress that in the notation (we 
 always use unprimed variables for objects on scale $n$ and primed ones for objects on scale $n+1$).
Define, for $C\subset \Z^d$, 
$
 \tau(C)$ to be 
 the minimal number of edges of a connected graph whose vertex set is $C$
and whose edges are  nearest neighbor bonds in 
$\mathbb Z^d$, and denote by
$d(A)$
 the diameter of a set $A\subset \Z^{d+1}$.  Set 
 \begin{equation}
 l(\CD,A,t,u,v):=\tau (\CD\cup u\cup v)+d(A\cup (t,u)\cup(t,v)).
 \label{weight}
\end{equation} 
Let us also define, for $\CD\subset\Z^d$,
\begin{equation}
N_{t} (\CD):=\sum_{u\in \CD}N_{(t,u)},
\label{NA}
\end{equation}
and, for $\CD=(\CD_1,\CD_2)$, let analogously $N_{t}(\CD):=N_{t}(\CD_1)+N_{t+1}(\CD_2)$.
Now, define the norm %where for  $D\subset \Z^d$, $\bar\CD:=\CD$ and for $\CD=(\CD_t,\CD_{t+1})$, $\bar\CD:=\CD_t\cup\CD_{t+1}$. We use the norm
\begin{equation}
  \|b_t\|_\lambda = \sup_{v\in \mathbb Z^d} \sum_{\CD,A,u }%\subset \mathbb Z^d \atop
 %v \in \mathbb Z^d} 
    \|b_{t, \CD, A} \|_{u,v}e^{ \lambda l(\CD,A,t,u,v)}e^{-N_{t} (\CD)}.
\label{f24}
\end{equation} 
For $B$ we use the same formula with the difference that the argument of $\tau$ in
(\ref{weight}) is replaced by $\CD_t\cup\CD_{t+1}\cup u\cup v$.
%and for $\gammapha=2,3$ we replace the exponent by $\lambda l(\CD_t\cup\CD_{t+1},A,t,u,v)$
%\begin{equation}
 % \|B_t\|_\lambda = \sup_{u\in \mathbb Z^d} \sum_{\ND,A,v }%\subset \mathbb Z^d \atop
 %v \in \mathbb Z^d} 
 %   \|B_{t\ND A} \|_{u,v}e^{\lambda l(D^-\cup D^+,A,t,u,v)}.
%\label{Bnorm}
%\end{equation} 

\subsection{Definition of $b'$ and $B'$}
We are now ready to give the inductive construction of  $b'_{t', D', A'}$ and $B'_{t', D', A'}$. 
  To have a unified notation,
set $B^1:=b$, $B^2:=B$, $B^3:=\nabla\cdot B$, $B^4:=T$ and $B^5:=\nabla T$, with the convention that $B^4, B^5$
are zero if either $\CD$ or $A$ is not empty.  
Then  (\ref{bdef}), (\ref{pdec}),  (\ref{bprime})  and (\ref{rem1}) give,
using the shorthand $\E^\perp:=1-\E$, 
\begin{equation}
b'=\CL b+\E^\perp \CN^1,
\label{b'dec}
\end{equation}
where $\CL$ is defined in (\ref{linearRG}), and $\CN^1$ is a sum of products
\begin{equation}
\Pi=L^{d-1}{\mathcal S}_LB^{\alpha_1}_{t_1}\dots B_{t_N}^{\alpha_N},
\label{Ndec}
\end{equation}
with $N\leq L^2$,  that are at least quadratic in $b$ (where we count $B$ as quadratic). As a consequence of (\ref{b'dec}), (\ref{bdef}), (\ref{Tdef}), we have
 \begin{equation}
 T'=L^d{\mathcal S}_LT^{L^2}+\E\ \nabla\cdot\CN^1.
\label{51}
\end{equation}

In the same way, we get
\begin{equation}
B'=\CL B+\CN^2
\label{b'dec1}
\end{equation}
 with $\CL$
given by (\ref{f54}), and $\CN^2$ is a sum of products  $\Pi$, as in (\ref{Ndec}),  with
% and for $\alpha=3$  $\CL=0$ 
$N\leq 2 L^2$ and at least one $b$ on each $L^2$ interval. In both $\CN^1$, $\CN^2$, $\alpha_1\neq 3,5$ (because 
the products $\Pi$ do not start with a $\nabla$, see (\ref{rem1})) .

Also, it suffices to give the definitions for $t'=0$, so that the times $t_i$ lie on $[0,L^2-1]$ for $b'$ and
on $[0,2L^2-1]$ for $B'$. For simplicity we  suppress the index $t'$ in the notation.

We need to localize these expansions to get the primed versions
of (\ref{bloc}) and (\ref{Bloc}).
 For this,
insert the decompositions
(\ref{bloc}) and (\ref{Bloc})  to  (\ref{Ndec}). The result is localized
basically  by taking unions of the $\CD_t$ and $A_t$ and blocking the result, except that we need to
take care of the constraints, in (\ref{bloc}), (\ref{Bloc}), $\CD'\subset D'$. Remember that, because of (\ref{a8}), $D'$ can be a smaller than the blocking
of $D$. This means that, at each step, parts of the large fields may become small, due to the fact that $N'_{z'}$
may become zero, see (\ref{a8}), and have therefore to be reabsorded into the small fields (see (\ref{s'}) below).

%To do this recall that (NOT NEEDED)
%$$
%D'_{t'}=([\cup_{t\in I_{t'}}D_t]\
%\setminus \{z'\ |\ \tilde N_{z'}<K\}) \cup \{z'\ |\ r_{z'}\neq 0 \}.
%$$
Write in (\ref{Ndec})
\begin{equation}
b_{t_i}=\sum_{\CD_i\subset \CD_{t_i}}\sum_{A_i}b_{t_i,\CD_i, A_i}=\sum_{\tilde \CD_i\subset D_{t_i}\cap LD'}
\sum_{\tilde A_i}\tilde 
b_{i\tilde \CD_i\tilde A_i}
\label{btdec}
\end{equation}
with
\begin{equation}
\tilde 
b_{i\tilde \CD_i\tilde A_i}:= \sum^{\hspace{6mm}\raisebox{-2mm}{$\scriptstyle \prime$}}_{\CD_i\subset D_{t_i}}\sum^{\hspace{6mm}\raisebox{-2mm}{$\scriptstyle \prime$}}_{A_i}  
b_{t, \CD_i, A_i}
%1(D^t\cap
%LD_{t'}'=\tilde D_t)1(A\cup (D^t\setminus\tilde D_t)=\tilde A_t)
\label{btilde}
\end{equation}
where the sums are constrained by
\begin{equation}
\CD_i\cap
LD'=\tilde \CD_i\ \ {\rm and}\ \  A_i\cup (\CD_i\setminus\tilde \CD_i)=\tilde A_i.
\label{constraint}
\end{equation}
where $LD'=\{x \in \Z^d | [x] \in D'\}$.
For $B$ and $\nabla \cdot B$ we proceed the same way except that, this time,
$\CD_i$ is a pair $(\CD_{t_i},\CD_{t_i+1})$. 
%and in the same way for $B_{t_i}$ and $\NB_{t_i}$. 
This way we end up with
\begin{equation}
\Pi=L^{d-1}\sum {\mathcal S}_L \tilde B^{\alpha_1}_{1\tilde \CD_{1}\tilde A_{1}}\dots\tilde B_{N
\tilde \CD_{N}\tilde A_{N}}^{\alpha_N},
\label{Pisum}
\end{equation}
where the sum runs over $\tilde \CD_{i}$, $\tilde A_{i}$ and, by convention, $\tilde B^{\alpha_i}= \ B^{\alpha_i}$ for $\alpha_i=4,5$.  
Now,  we may localize
\begin{equation}
\Pi=\sum_{\CD'\subset D',A'}\Pi_{\CD'A'}
\label{Piloc}
\end{equation}
 with $\Pi_{\CD'A'}$ being a sum of terms in (\ref{Pisum}) with 
\begin{equation}
\CD'=[\cup_i\tilde \CD_{i}], \ \ \ A'=[\cup_i\tilde A_{i}].
\label{a'}
\end{equation}
Our notation is a bit abusive: if $\al_i$ equals 2 or 3, both sets in the pair
$\tilde \CD_{i}$ occur in the union. Since  $\CN^1$ is a sum of $\Pi's$,
(\ref{Piloc}) yields a similar expansion for $\CN^1$. 

We still need to localize $\CL b$.  For $\CL s=\sigma$ this was done  in Section 4.2.
For $\CL \ell$ we proceed as above with $\CN^1$; now all $\alpha_i=4$ except for
one which equals $1$. We get
\begin{equation}
\CL\ell=\sum_{\CD'\subset D',A'}(\CL\ell)_{\CD'A'}
\label{CLloc}
\end{equation}
with
\rr
\label{linearRGl}
(\CL \ell)_{\CD' {A'}}
 = \sum_{t=0}^{L^2-1}\sum_{\tilde \CD_t \neq\emptyset}
\sum_{\tilde A_t}\CL_t
\tilde \ell_{t, \tilde \CD_t, \tilde A_t}.
\rrr
where $\tilde \ell$ is defined in terms of $\ell$ as  in (\ref{btilde}),   ${\tilde \CD_t} \subset D_{t}\cap LD_{t'}'$,  $\CD'=[\tilde \CD_{t}]$, $A'=[\tilde A_{t}]$ and  $\CL_t$  is given by  (\ref{linearRG}).

Let $\chi_{A'}=1$ if $A'=z'$ and $r_{z'}=0$, and $0$ otherwise. 
 We define:

\begin{equation}
s'_{A'}=\E^\perp(\sigma^{1}_{A'}\chi_{A'}+\sigma^2_{A'}+(\CL\ell)_{\emptyset A'}+\CN^1_{\emptyset A'}),
\label{s'}
\end{equation}
where $\sigma^2$ is defined by
\begin{equation}
\CL s=\sigma^{1}+\sigma^2
\label{lsdeco}
\end{equation}
and $\sigma^{1}$ is the  local part given in (\ref{srel1});
we define also:
\begin{equation}
\ell'_{\CD'A'}=\E^\perp( (\CL\ell)_{\CD' A'}+\sigma^{1}_{\CD'}(1-\chi_{\CD'})\delta_{A',\emptyset}+\CN^1_{\CD' A'}).
\label{ell'}
\end{equation}
%I DELETED $\CB$ in the argument of $\CN^1$; WHY SUBTRACT EXPECTATION VALUE IN $\ell'_{\CD'A'}$? WE NEVER USE THAT.
%so that $\sigma^{1}_{\CD'}=0$ unless $\CD'=z'$.
 It is easy to verify that, with these definitions, 
$$
b'=s'+ \ell'
$$
with $b'$ given by (\ref{b'dec}).

Both $s'$ and $\ell'$ have a linear and a nonlinear part. Besides, when one goes from one scale to the next, there is a new large field arising from the previous scale small field
and a part of the previous scale large field that becomes small. 
The factor $\chi_{A'}$ in (\ref{s'}) ensures that the contribution coming from the local part
of $s$ will be small, and the remaining part $\sigma^{1}_{\CD'}(1-\chi_{\CD'})\delta_{A',\emptyset}$ contributes to the new large field and is thus put into $\ell'$. The term
$(\CL\ell)_{\emptyset A'}$ corresponds to the part of previous scale large fields that have become small (due to the decrease of $N$ in (\ref{a7}-\ref{a8})), and which thus contribute to $s'$.

For $B'$ we need no special treatment of the linear term as
we did for $\CL b$ and both terms  in (\ref{b'dec1}) are localized in the same way by  expressing $B$ in terms of $\tilde B$ as in  (\ref{btilde}) and collecting terms with (\ref{a'}). We get
\begin{equation}
B'_{\CD'A'}=(\CL B)_{\CD' A'}+\CN^2_{\CD' A'}.
\label{B'}
\end{equation}
 %$B^3$ receives only contributions from $\CN(B)^3$.
 
Next, we will state the estimates on the linearized RG, that will be used in Section 6, when we prove Theorem \ref{main}.
 Proposition \ref{sell}, which gives deterministic estimates on the random part $b$ of transition probabilities of our random walk
 will be proven in subsection 4.10; because of the $e^{-N_t(\CD)}$ factor in (\ref{f24}), these bounds
 are useful only if non-zero values of $N$ are improbable, which is the content of  Proposition \ref{prob1}. This Proposition
  will be proven in subsection 4.12; this proof uses Proposition \ref{prob2}, which itself is proven in subsection 4.11. Proposition \ref{T} deals with 
  the nonrandom part of the  transition probabilities, given by (\ref{Tdef}),  and is proven in subsection 4.13.

\subsection{Deterministic bounds}

Let us now state the bounds that will be proven inductively in
the scale $n$ for  $b$ and $B$.
Recalling the definition (\ref{srel}), decompose $s=s^{\rm loc}+s^{\rm nloc}$.
We suppress both the index $n$ and the index $t$ in our notation. The number $\gamma$ has been introduced in
(\ref{aldefi}) and $\la$ is chosen small enough so that  (\ref{1norm}) holds for $n=0$, and  that the bound (\ref{Tnorm})  below also holds.

\prop{sell}
For  $\epsilon$  small enough we have, $\forall n  \in \mathbb N$, $\forall t \in \mathbb N$,
 \begin{eqnarray}
 \| s^{loc}\|_\lambda & \leq& \ep_n^{(1-3\gamma)}  \label{f26}
\\
 \| s^{\rm nloc} \|_\lambda  & \leq &  \ep_n^{(1-2\gamma)} \label{f26a}
\\  \| \ {\ell}\  \|_\lambda & \leq &  \ep_n^{(1-\gamma)}  \label{f27}
\end{eqnarray}
Moreover
\qq
  \| B \|_\lambda  &\leq&  \ep_n^{(1-\gamma/2)} 
\label{f29}\\
   \|\nabla\cdot B   \| _\lambda &\leq &  \ep_n^{1/ 2}.
\label{f31}
\qqq
\eprop
%I CHANGED THE POWER FOR B, BECAUSE THIS IS THE LEADING NONLINEAR TERM FOR b, HENCE IT MUST RUN  FASTER THAN THE POWER FOR ELL.
We note also a simple consequence of (\ref{f26})-(\ref{f27}) and the fact that, since the lattice spacing is
$L^{-n}$, the norm of $\nabla_n$ is bounded by $CL^n$:
\rr
  \|\nabla\cdot b   \| _\lambda \leq  C\ep \ep_n^{-3\gamma}.
\label{grads}
\rrr

\begin{remark}\label{alpha} The different powers introduced in the Proposition are chosen for convenience
and are not optimal.
 \end{remark} 

Proposition \ref{sell} is a deterministic statement holding for
any realization of the noise. For this statement to be useful, we need to show (see the definition (\ref{f24}) of our norm)
that the random variables $N_z$ are  zero with high probability.

\subsection{Probabilistic bounds} Our main estimate is:
\prop{prob1}
% For every  $z\in{\mathbb Z^d}\times\N $, 
The random integers $N_{n, z}$, defined in (\ref{f18a}-\ref{a8}), satisfy
 for any $A \subset \Z^{d+1}$,
any $\{N_z\}_{z\in A}$, with $N_z \neq 0$, %ADDED
\qq
 \mathbb P(\{N_{n, z}=N_z\}_{z\in A}) \leq \exp(- K_n n|A|-2K_n{}\sum_{z\in A} N_z)
 \label{38a}
\qqq
%implies that (\ref{38}) holds also for $N_{n+1}$, 
with
$K_n=K\prod_{\ell=2}^{n-1}(1-\ell^{-2})$, $n\geq 3$, $K_n=K$, $n<3$, and
%\qq
%\mathbb P(N_{n, z}=N>0)\leq \exp(-K(n+N))
%\qqq
where   $K \to \infty$ as $\ep \to 0$.
\eprop
Since $N$ is determined in terms of the random variables $r$ in previous scales
this Proposition follows ultimately from an estimate on the probability that $r_z\neq 0$.
This in turn follows from the control of the variance of deterministically relevant part of $s$, i.e. $s^{loc}$. Given  $u\in\Z^d$ consider
the random variables
 \begin{equation}
 (s_{t, u},f):= \int dxdys_{t, u}(x,y)f(x,y).
\label{fs}
\end{equation}
with $f: \mathbb R^d\times \mathbb R^d\to \mathbb R$ a measurable function such that
 \begin{equation}
 \|f\|_u:= \int dxdy|f(x,y)| \exp (-2\lambda \tau(u,x,y))<\infty.
\label{w0}
\end{equation}
Here, again by an abuse of notation, we write $\tau(u,x,y)$ instead of  $\tau(\{ u,v,w \} )$, with $x\in {\bf v}$ 
 $y\in {\bf w}$ (since $\tau$ is defined on sets).
Then  we
prove the following exponential moment estimate: 
\prop{prob2} 
For all $t \in \mathbb N$, $u\in\Z^d$% and $f$ with $\|f\|_u\leq CL^n$ DO WE NEED C?
\begin{equation}
\mathbb E\  e^{ (s_{t, u},f)}\leq  e^{\delta_n \|f\|_u^2},
\label{w1}
\end{equation}
with $\delta_n  =   L^{-n\gamma}\ep^{2-6\gamma}_n$.
\eprop
Note that (\ref{w1}) implies, by replacing $f$ by $a f$, subtracting $1$ on both sides,
 dividing by $a^2$, letting $a \to 0$ and using 
$\mathbb E \ s_{t, u} (x,y)=0$:
\begin{equation}
\mathbb E\  (s_{t, u},f)^2\leq  \delta_n \|f\|_u^2.
\label{w1a}
\end{equation}

\subsection{Inductive bounds for $T$} The following result describes the
inductive bounds for the average (\ref{Tdef}) of $p$. 
We can view $T$ as being of the form (\ref{bloc}), with the only nonzero term corresponding to $A=\emptyset$, $\CD=\emptyset$, so that
$\|T\|_\lambda$ is bounded from above and below by $C \int T(x) e^{\la |x|}dx$. 
 
\prop{T} $T=T_n$ satisfies uniformly in $n$
\qq
&& \|T\|_\lambda+ \|\nabla T\|_\lambda \leq C
\label{Tnorm}\\
%&&T^t(x)\leq C(k)(t+1)^{-d/2}e^{-\frac{ k}{\sqrt {t+1}}|x|},
% \label{Tnorm1}
%\qqq
%\qq
&&0\leq T^t(x)\leq C(k)(t+1)^{-d/2}e^{-c(t)|x|},
%for  $k\leq \lambda L$ and $t\in [0,2L^2]$.
 \label{Tnorm1}
\qqq
for $\frac{k}{L}\leq \frac{ \lambda}{2}$,  and $t\in [0,2L^2]$,
with $c(t)=\min (\la, \frac{ 4k}{\sqrt {t+1}})$.
Moreover, there exists $D>0$ such that
 \begin{equation}
\lim_{n\to\infty}\|T_n-T^\ast_D\|_1= 0 .
\label{j4a}
\end{equation}
 \eprop
A consequence of  (\ref{Tnorm1}) is:
\begin{corollary}\label{ttilde}
 Let
 $\tilde{T}^t (x):=e^{\frac{k}{L}|x|}T^t(x) $. Then, for $\frac{k}{L}\leq \frac{ \lambda}{2}$ 
 and $t\in [0,2L^2]$ we have $\frac{k}{L}\leq \hf c(t)$, and so,
\begin{equation}
\| \tilde T^t \|_\infty \leq C(k)(t+1)^{-\frac{d}{2}} ,\ \ \| \tilde T^t \|_1\leq C(k).
\label{w9}
\end{equation}
\end{corollary}

\begin{remark}\label{Constants} In the proofs below we shall use the following conventions. $C$ or $c$ denote  constants that may vary
from place to place, even in the same equation, but that do not depend on $L$ or on the scale $n$. $C(L)$ is similar, but depends on $L$ but not on $n$. We assume throughout that $L$ has been chosen large enough so that  inequalities of the form $CL^{-\gamma} \leq 1$, for $\gamma >0$,  can be assumed.
We then choose $\ep$ small enough so that we can assume  inequalities of the form $C(L) \ep^{\gamma} \leq 1$.
 \end{remark} 

\begin{remark}\label{tprime} In the proofs below, whenever the index $t'$ does not appear, it means that it is set equal to $0$.
 Then, in (\ref{bite}), the interval  $I_{t'}=[0, L^2-1]$
 and $i=t$. When considering composite operators we have $t\in [0, 2L^2-1]$.
 \end{remark} 

\subsection{ Proof of Proposition  \ref{sell}} For $n=0$, we have, by assumption (\ref{1norm}), 
\rr  \|b\|_\lambda    \leq\epsilon,\ \  \|B\|_\lambda   \leq C\epsilon^2.
 \label{othstep}
 \rrr 
Thus, since for $n=0$ the norm of $\nabla_0$ is of order one,
the assumptions hold for small $\epsilon$ and  we have $\CD=\emptyset$, $N=0$, $\ell=0$.

Before proceeding to the induction, let us sketch the main steps of the proof. For the nonlinear terms in (\ref{s'}, \ref{ell'}),
we need only to show that the norm of a product of $b$'s, $b\nabla \cdot b$'s and $ \nabla \cdot b\nabla \cdot b$'s is suitably bounded in terms of the product of their norms
(all terms of the form $\Pi$ in (\ref{Ndec}) can be written as a product of those three factors times $T$'s or $\nabla T$'s whose norms are bounded by (\ref{Tnorm})),  that the resummation (\ref{btilde}) does not increase the norm too much,
and then use inductively Proposition  \ref{sell}. This is done in Lemma \ref{T1} below and involves only simple estimates on the exponential
factors in (\ref{f24}). 

For the linear terms in (\ref{s'}), $\sigma^{1}_{A'}\chi_{A'}$ is controlled trivially because of the characteristic function,
the norm of $\sigma^2_{A'}$, being nonlocal, will be small because of the exponential decay factor included in the norm (\ref{f24}) (see (\ref{contra}) below)
and the one of $(\CL\ell)_{\emptyset A'}$ will be small because our inductive bound on $\| \ell \|_\la$ is smaller than our bound on $\| s\|_\la$, see (\ref{f26}-\ref{f27}). 
For  the linear terms in (\ref{ell'}), the norm of $\sigma^{1}_{\CD'}(1-\chi_{\CD'})\delta_{A',\emptyset}$ will be small because our definition
(\ref{f18a})
of $r_{z'}$ (see  (\ref{n0choise}) below) and the one of $\CL\ell$ will be small because the operator $\CL_i$
is a contraction for fixed $i$, see (\ref{cboundl}) below, and because, unlike $s$, $\ell$ occurs only at times where $N_{z}\neq 0$,
and this effectively controls the sum over times in  (\ref{linearRG}) that made $s$ a relevant variable, see (\ref{barcl}) below. This contraction also implies that the linear term in (\ref{b'dec1}) contracts since there is no sum over times in 
(\ref{f54}).

Let us now bound the nonlinear 
term $\CN^i$, $i=1,2$ in (\ref{b'dec}, \ref{b'dec1}).  This is given by a sum of
the products $\Pi$ in  (\ref{Ndec}) localized in  (\ref{Piloc}). 
To simplify the notation, we consider here only $\tilde \CD_{i}$ being a single
set and not a pair of such sets, as they occur for $\alpha=2,3$, but the Lemma below extends to those cases. 
\lem{T1} 
%There exists a $L(\rho,\lambda)$ such that 
$\Pi$ defined in (\ref{Pisum}), (\ref{Piloc}) satisfies: 
\begin{equation}
\| \Pi\|_{2\lambda}\leq C(L)\prod_i\| \tilde B_{t_i}^{\alpha_i} \|_{\frac{\lambda}{3}}
\label{barPi}
\end{equation}
where $\tilde B_{t_i}^{\alpha_i}$ is defined in (\ref{btilde}) (see the discussion leading to (\ref{Pisum})), and satisfies:
\begin{equation}
\|\tilde B_{t_i}^{\alpha_i}\|_{\frac{\lambda}{3}}\leq C\| B_{t_i}^{\alpha_i}\|_\lambda
\label{barC}
\end{equation}
\elem

\no{\bf Proof}.  Recall that, from (\ref{f24}),
\begin{equation}
  \| \Pi\|_{2\lambda} = \sup_{v'} \sum_{\CD' ,A',u' }
      \|\Pi_{t', \CD', A'} \|_{u',v'}e^{-N_{t'}'(\CD')}e^{ 2\lambda l(\CD',A',t',u',v')}.
\label{f241}
\end{equation} 
We have from (\ref{Pisum}-\ref{a'})
\begin{equation}
      \|\Pi_{t', \CD', A'} \|_{u',v'}\leq  L^{d-1}\sup_{u_1\in L\Nu',\ v_N\in L\Nv'}\sum_{v_1,u_2,\dots,u_N}
      \sum_{\{\tilde \CD_i, \tilde A_i\}}
       \prod_{i=1}^N\|\tilde B^{\alpha_i}_{t_i, \tilde \CD_i, \tilde A_i}\|_{u_iv_i}.
\label{f240}
\end{equation} 
 From
 (\ref{a7}) we have $N'_{z'}\geq\tilde N_{z'}-1$. Since, see (\ref{a'}), $\CD'=[\cup_{i}\tilde \CD_{t_i}]$ (leaving out the index ${t'}$ in $D'$)
 we deduce, using (\ref{a7a}) and  (\ref{NA}),
\rr
 -N_{t'}' (D')\leq-\sum_iN_{t_i}({\tilde \CD_{t_i}})+|D'|.
 \label{n'bound}
\end{equation}
For the geometrical factor (\ref{weight}), we have
\begin{equation}
l(D',A',t',u',v')\leq \sum_i(\frac{c}{L}l(\tilde \CD_i,\tilde A_i,t_i,u_i,v_i)+c).
\label{ellbound}
\end{equation}
Indeed, this inequality obviously holds both for the graph length and for the diameter terms in (\ref{weight}),
since the blocking operation (\ref{block}) effectively scales space-time distances by a factor at least $L^{-1}$ (except for small sets, hence the $+c$
term in (\ref{ellbound})) and since $u_1\in L\Nu' $, $ v_N\in L\Nv'$. 
 Since the LHS is also larger than $|D'|$ (because $\tau (D')\geq |D'|$), 
we deduce:
\qq
2\lambda l(D',A',t',u',v')\leq (2{\lambda}+1) l(D',A',t',u',v')- |D'|
\non \\
 \leq\sum_i(2{\lambda}+1)\frac{c}{L}
l(\tilde \CD_i,\tilde A_i,t_i,u_i,v_i)+(2\lambda+1)L^2c-|D'|.
\label{ellbound1}
\qqq
Taking $L(\lambda)=3(2+1/\la)c$ and inserting (\ref{f240}),
(\ref{n'bound}) and (\ref{ellbound1})  in (\ref{f241}), the claim (\ref{barPi})
follows.

For (\ref{barC}), we recall that
 $\tilde \CD_{i}=\CD_{i}\cap LD'$ and that, for $z'\notin D'$, $\tilde N_{z'}<{\la\over 3}$.
 These imply
\rr
-N_{t_i}(\tilde \CD_{i})\leq -N_{t_i}( \CD_{i})+{\la\over 3}|\CD_{i}\setminus LD'|\leq -N_{t_i}( \CD_{i})+{\la\over 3}|\CD_{i}|.
\label{Ni}
\rrr
As for the geometric factors, using (\ref{constraint}), we get:
$$
\tau(\CD_i,u,v)\geq\hf(\tau(\tilde \CD_i,u,v)+d((\tilde A_i\setminus A_i)\cup (u,t)\cup (v,t)),
$$
which we rewrite as
$$
\tau(\tilde \CD_i,u,v) \leq 2\tau(\CD_i,u,v)-d((\tilde A_i\setminus A_i)\cup (u,t)\cup (v,t)),
$$
and combine with
$$
d(\tilde A_i\cup (u,t)\cup (v,t))\leq d(A_i\cup (u,t)\cup (v,t))+d((\tilde A_i\setminus A_i)\cup (u,t)\cup (v,t))
$$
in order to derive, see (\ref{weight}),
\rr
l(\tilde \CD_i,\tilde A_i,t_i,u,v)\leq 2l(\CD_i,A_i,t_i,u,v)\leq 3l(\CD_i,A_i,t_i,u,v)- |\CD_i|.
\label{Di}
\rrr
where in the last inequality we used $l(\CD_i,A_i,t_i,u,v)\geq |\CD_i|$.
%Now, take $K<\frac{\lambda}{3}$ (WE NEED ONE THIRD HERE) and combine (\ref{Ni}) and(\ref{Di}) 
%with (\ref{n'bound}) and (\ref{ellbound}). 
Insert (\ref{Ni}), (\ref{Di})  in (\ref{f24}), (\ref{btilde}), %(\ref{CC}),
 and
the claim (\ref{barC}) follows.
\halmos

This Lemma allows us 
  to bound the nonlinear terms $\CN^i$, $i=1,2$, by inspection
  using the inductive bounds (\ref{f26})-(\ref{f31}), the bound (\ref{Tnorm})
  for $T$ and the fact that $\CN^i$ are sums of $\Pi$'s with an $L$ dependent number
of terms. 
%We need to
%discuss separately 
%$n\geq n_0$ and $n<n_0$ where $n_0>n_0(L)$ will be taken suitably large.

Let us start with the nonlinear term  $\CN^1$ corresponding to $b$, setting $t'=0$ (see Remark \ref{tprime}). The largest contribution
is linear in $B$ i.e. coming from the term
$$
L^{d-1}\sum_t {\mathcal S}_L T^{L^2-t-2} B_t T^{t}.
$$
whose norm is bounded by $C(L) \ep_n^{1-\gamma/2}$ (using (\ref{f29})). Terms involving $b \nabla B$ 
are bounded, using inductively Proposition \ref{sell}, by $C(L) \ep_n^{3/2-3\gamma}$. 
Other terms are smaller and 
 we get from the Lemma:
\rr
\| \CN^1\|_{2\lambda}\leq  \frac{1}{8}  \ep_{n+1}^{1-\gamma}
\label{N3}
\end{equation}
for  $\ep$ small.

For  $\CN^2$,  corresponding to $B$, the largest terms are the quadratic ones
%$$
%L^{d-1}s_LT^{\tau_1}b_{2L^2-\tau_1}\nabla \cdot T^{\tau_2}b_{2L^2-\tau_1-\tau_2-1}T^{\tau_3}
%$$
%I WOULD PUT:
% CHANGED ACCORDING TO OUR DISCUSSION
$$
L^{d-1}{\mathcal S}_LT^{\tau_1}b_{\tau_2+\tau_3+1}\nabla \cdot T^{\tau_2}b_{\tau_3}T^{\tau_3}
$$
with $\tau_1+\tau_2+ \tau_3 +1 = 2L^2  -1$, $\tau_2 \neq0$, $ \tau_3 \leq L^2 -1$ $\tau_2+\tau_3 +1\geq L^2$,
and 
$$
L^{d-1}{\mathcal S}_LT^{L^2-1}b_{L^2}\nabla \cdot B_{L^2-2}T^{L^2-2}
$$
%(in the former, $\tau_1$ equals $L^2-1$ or $L^2-2$ WHY??? SEE SUGGESTION ABOVE). 
 The norms of these are bounded 
by $ C(L) \ep_n^{2(1-3\gamma)}$ and $  C(L) \ep_n^{(1-3\gamma)+1/2}$
respectively; other terms, e.g.
  $L^{d-1}{\mathcal S}_LT^{L^2-2}B_{L^2}\nabla \cdot B_{L^2-2}T^{L^2-2}$ are of order $  C(L) \ep_n^{(1-\gamma/2)+1/2}$, %CHANGED POWER
and so
\rr
\| \CN^2\|_{2\lambda}
\leq \frac{1}{2} \ep_{n+1}^{(1-\gamma/2)}
\label{N4}
\end{equation}
 since
$\gamma<1/12$ and $\epsilon$ is small. 

Finally, to estimate $\nabla\cdot\CN^2$ for (\ref{f31}) the leading term is of the form 
$L^{d-1}{\mathcal S}_L\nabla \cdot b_{2L^2-1}\nabla \cdot T^{2L^2-t-2}b_{t}T^{t}$, for $t\leq L^2-1$; %CHANGED-THE PREVIOUS FORM WAS NOT GENERAL
indeed, if $\nabla$ acts on a $T$, we can use (\ref{Tnorm}), which gives a smaller contribution;
we use the bound (\ref{grads}) for  $\nabla \cdot b_{2L^2}$
%fact that the operator $\nabla$ is bounded by $CL^n$
 to get 
\rr
\|\nabla\cdot \CN^2\|_{2\lambda}
\leq C(L)\ep \ep_n^{1-6\gamma} \leq \hf \ep_{n+1}^{1/2}.
\label{N4'}
\end{equation}
Since $\gamma<\frac{1}{12}$ and $\epsilon$ is small. 

To bound $\nabla\cdot \CL B^2$, we use inductively (\ref{f29}) and the fact that, if we apply $\nabla$ to (\ref{f54}), 
it acts on a $T$ and we can use (\ref{Tnorm}) to bound it.
We get then
\rr
\| \nabla\cdot \CL B^2\|_{2\lambda}
\leq \hf \ep_{n+1}^{1/2}.
\label{N4''}
\end{equation}
This and (\ref{N4'})  finishes the iteration of (\ref{f31}). 
% this is smaller than the  bound (\ref{f31}). 

We will now consider the  linear RG
contributions to (\ref{s'})-(\ref{B'}). To deal with the $\sigma$ of (\ref{sigmaA})
and $\CL\ell$ of (\ref{linearRGl}), we need a fundamental bound on the operator $\CL$:
\lem{lrgnorm} Let $\beta$ denote $s$ or $\ell$.  For $t\in I_{t'}$ define
\begin{equation}
c^t_{D'A'}
: = e^{\sum_{s\neq t}N_s(LD')}\sum_{\tilde \CD, \tilde A}\CL_i \tilde \beta_{t, \tilde \CD, \tilde A}.
\label{linearRGb}
\end{equation}
where $i=t-L^2t'$, $[\tilde \CD]=D'$ and $\tilde A\in\CA(A')$ (for $\beta=s$) or $[\tilde A]=A'$ (for $ \beta=\ell$). Then,
\begin{equation}
\|c^t\|_{2\lambda}\leq CL^{-1}\|\tilde \beta_t\|_{\frac{\lambda}{3}},
\label{cboundl}
\end{equation}
%where, as in (\ref{CC}), $\CC_t=e^{-N_t}\tilde \beta_t$.  
\elem
\no{\bf Proof.} Write, using (\ref{linearRG}) and $ |u'-\frac{u}{L}|\leq |x'-\frac{x}{L}|+2$, $ |v'-\frac{v}{L}| \leq |y'-\frac{y}{L}|+2$,
%I PUT 2 INSTEAD OF c
for $x'\in {\bf u'}$, $x\in {\bf u}$, $y'\in {\bf v'}$, $y\in {\bf v}$,
\rr
\|\CL_i \tilde \beta_{t, \tilde \CD, \tilde A}  \|_{u',v'} \leq
 C L^{d-1} \sum_{u,v} 
\|\beta_{t, \tilde \CD, \tilde A}\|_{u,v}e^{-\frac{k}{L}( |Lu'-u|+|Lv'-v|)}S_i(u,v,u',v'),
 \label{f79}
\rrr
%I ADDED INDEX $S_i$ 
with
\rr
S_i(u,v,u',v')=
 \sup_{x^\prime  \in \Nu^\prime \atop y^\prime \in \Nv^\prime}
\int_{\Nu} dx \tilde{T}^{L^2-i-1} (Lx^\prime-x)  \int_{\Nv} dy \tilde{T}^{i} (y-Ly^\prime) 
 \label{Suv}
\rrr
and $\tilde{T}^t (x)=T^t(x) \exp (\frac{k|x|}{L})$.  As in (\ref{n'bound}), we get
\rr
-N'_{t'}(D')\leq -N_t(\tilde \CD)-\sum_{s\neq t}N_s(LD')+|D'|
 \label{n'b1}
\rrr
As for the geometric factors,
since  $A'\subset [\tilde A]$
for all $\tilde A\in\CA(A')$ we have, as in (\ref{ellbound}), 
\rr
l(D',A^\prime,t',u',v')
 \leq \frac{c}{L}(l(\tilde \CD,\tilde A,t,u,v)+|Lu'-u|+|Lv'-v|)+c .\non
%\label{w7}
\rrr
Since the LHS is also larger than $ \hf(|D'|+|u'-v'|)$ we get as in (\ref{ellbound1})
\qq
2\lambda l(D',A^\prime,t',u',v') &\leq &({\lambda}+1)\frac{2c}{L}( l(\tilde \CD,\tilde A,t,u,v)
+|Lu'-u|+|Lv'-v|)\non\\
&&+(\lambda+1)2c - (|D'|+|u'-v'|).\non
%\label{w7}
\qqq
We take  
\rr
k=({\lambda}+1)2c
\label{kdefi}
\rrr 
and so we have
 $\frac{k}{L}=({\lambda}+1)\frac{2c}{L}\leq \lambda/6$ for $L$ large. Using $ l( \tilde \CD, \tilde A,t,u,v)\geq |u-v|$,
we get:
\qq
2\lambda l(D',A^\prime,t',u',v') &\leq &\frac{\lambda}{3} l(\tilde \CD,\tilde A,t,u,v)
+ \frac{k}{L}(|Lu'-u|+|Lv'-v|)\non\\
&&+k - (|D'|+|u'-v'|) -\frac{\lambda}{6} |u-v|.
\label{w7}
\qqq
Then, inserting (\ref{n'b1}) and (\ref{w7}) in (\ref{linearRGb}), (\ref{f79}), we get:
\rr
\|c^t\|_{2\lambda}\leq C L^{d-1}\| \tilde \beta_t\|_{\frac{\lambda}{3}} \sup_{v'}
\sum_{u', u, v}e^{- |u^\prime-v^\prime|} e^{-\frac{\lambda}{6} |u-v|} S_i(u,v,u',v').
\label{cfinal}
\rrr 
Note that we have $\frac{k}{L}\leq \frac{\la}{2}$ and that $C(k)$ in (\ref{w9}) is $L$-independent, i.e. we may
 write $C(k)=C$ in (\ref{w9}). Then,  use (\ref{w9})  to bound, for any $v'$,
\begin{equation}
\sum_{u, v} e^{-\frac{\lambda}{6} |u-v|}S_i(u,v,u',v')  \leq C \min ( i^{-\frac{d}{2}}, (L^2-i-1)^{-\frac{d}{2}}) \leq C' L^{-d}
\label{f81}
\end{equation}
where the $L^\infty$ bound is used in (\ref{Suv}) for the integral corresponding to the minimum in (\ref{f81}); 
 the $L^1$ bound  and the factor $e^{-\frac{\lambda}{6} |u-v|}$ control
 the remaining integral and the  sum over $u, v$.
 Finally, use $e^{-  |u^\prime-v^\prime|}$ to control the sum over $u'$ in (\ref{cfinal}).

Combining (\ref{cfinal}) and (\ref{f81}) we get the claim (\ref{cboundl}). \halmos

Let us look at the consequences of Lemma \ref{lrgnorm}. The simplest
one is gotten by taking $\beta=s$ in   (\ref{linearRGb}). Then summing
( \ref{cboundl}) over the times  and using (\ref{barC}) yields, for $\sigma=\CL s$,
%since $\|\cdot\|_{\lambda}$ is increasing in $\lambda$, 
%CHANGED IS IT s TILDE IN THE MIDDLE?
\qq
\label{expa}
\|\sigma\|_{2\lambda}\leq CL\|\tilde s\|_{\frac{\lambda}{3}} \leq CL\|s\|_{\lambda},
\qqq
which just expresses the fact that the linear RG has an expanding bound.

Next, recall we have two  decompositions for $\sigma$: in  (\ref{lsdeco})
$\sigma=\sigma^1+\sigma^2$ where  $\sigma^1=\CL s^{\rm loc}$;
%ADDED
 but we can also write, as in 
(\ref{srel}),  $\sigma=\sigma^{\rm loc}+\sigma^{\rm nloc}$. Thus
the local part of $\sigma^2$ equals the local part of $\CL s^{\rm nloc}$, which exists because of the ``blocking"
in 
(\ref{sigmaA}). 
Lemma  \ref{lrgnorm} and (\ref{barC}) then gives
\qq
\label{expa1}
\|(\sigma^2)^{\rm loc}\|_{2\lambda}\leq CL\|s^{\rm nloc}\|_{\lambda} \leq CL \ep_{n}^{1-2\gamma} \leq {_1\over^8} \ep_{n+1}^{1-3\gamma},
\qqq
using (\ref{f26a}) and $\ep$ small in the last two inequalities.

Since from  (\ref{f18a}) and the definition of $\chi$ in (\ref{s'}), (\ref{f18a}), we have 
\begin{equation}
 \|\sigma^{1}\chi\|_{\la}
\leq {_1\over^8} \ep_{n+1}^{1-3\gamma}
\label{s'2}
\end{equation}
we conclude
\qq
\label{expa11}
\|\E^\perp(\sigma^1\chi+(\sigma^2)^{\rm loc})\|_{\lambda}\leq {_1\over^2} \ep_{n+1}^{1-3\gamma} 
\qqq
since the norm of $\E^\perp$ is bounded by $2$.  
%and since $\|\cdot\|_{\lambda}
%\leq \|\cdot\|_{2\lambda}$.

Next,  apply  Lemma \ref{lrgnorm} to  $\beta=s1_{d(A)>  \frac{L}{4}}$. Then, since $l$ in (\ref{weight}) is larger than $d(A)$
$$\|\beta\|_{\frac{\lambda}{3}}\leq e^{-\frac{\lambda L}{6}}\| s^{\rm nloc}\|_{\lambda}.$$
Summing over the times $t$, we get from (\ref{cboundl}), the definition (\ref{sigmaA}) and (\ref{f26a}) used inductively,
\qq
\label{contra}
\|\sigma^{\rm nloc}\|_{2\lambda}\leq CL e^{-\frac{\lambda L}{6}} \|s^{\rm nloc}\|_\lambda \leq {_1\over^4} \ep_{n+1}^{1-2\gamma},
\qqq
which means that  the non local part is irrelevant under the linear RG.

%Let now $n\geq n_0$. 

Next apply (\ref{cboundl}) to $\beta=\ell$. Given $D'$, let 
$$
T(D')=\{s\in I_{t'}\ |\
N_s(LD')\neq 0\}.
$$
 Thus, for $s\in T(D')$, $N_s(LD')\geq \la/3$ and so
$$
\sum_{t\in T(D')}e^{-\sum_{s\neq t}N_s(LD')}\leq |T(D') | e^{-{\la\over 3}(|T(D')|-1)}\leq C.
$$
%BARS ADDED.
 Now, writing 
 (\ref{linearRGb}) with $\beta=\ell$ as $ \sum_{\tilde \CD, \tilde A}\CL_i \tilde l_{t, \tilde \CD, \tilde A} =e^{-\sum_{s\neq t}N_s(LD')} c^t_{D'A'}$,
using (\ref{linearRGl}), and combining the last bound (to control the sum over times) with (\ref{cboundl}), (\ref{barC}) and the inductive bound (\ref{f27}), we get 
\rr
\| {(\CL\ell)}_{t'}\|_{2\lambda}\leq CL^{-1} \ep_n^{ 1-\gamma}\leq \frac{1}{8} \ep_{n+1}^{ 1-\gamma},
\label{barcl}
\rrr
for $L$ large.
Eq. (\ref{barcl}) is also a bound for the $\|\cdot\|_\lambda$ norm of the
%Combining (\ref{cboundl}) and (\ref{barC}), and summing over times, we get that
third term on the RHS of (\ref{s'}). Inserting (\ref{expa11}),  (\ref{barcl}) and
(\ref{N3}) in (\ref{s'}), and bounding the norm of $\E^\perp$ by $2$, the induction step for (\ref{f26}) follows; inserting
 (\ref{contra}),  (\ref{barcl}) and (\ref{N3}) in (\ref{s'}), we get the induction step for (\ref{f26a}). 
 
Finally, to bound  (\ref{ell'}), use (\ref{barcl}) and (\ref{N3}) for the first and third terms on the RHS. For the second one,
we use $r_{z'}=(n_0+n) \log L$ and the definition (\ref{f24})  to bound its norm by
\rr
C(L)L^{-n-n_0}\epsilon_n^{1-3\gamma}\leq {_{1}\over{8}}\epsilon_{n+1}^{1-\gamma}
\label{n0choise}
\rrr
since we have taken, see (\ref{jb1}), $L^{-n_0}=\ep^{3\gamma }$, and we use $\ep^{\gamma }$ to control $8C(L)$. Finally, bound the norm of $E^\perp$ by two.

To finish the inductive step in the proof of Proposition \ref{sell} we need to consider the linear RG  (\ref{f54})
for $B$. The operator $\CL$ is bounded as  in (\ref{cboundl}), and since, in  (\ref{f54}), there is no sum over times,
we get ($B^2=B$)
\rr
 \| {\CL B^2} \|_{2\lambda} \leq CL^{-1} \| B^2 \|_{\lambda}.
\label{lb2}
\rrr
Combining this with  (\ref{b'dec1}), (\ref{N4}), and using inductively (\ref{f29}), we get (\ref{f29}) on scale $n+1$.
% since $\|\cdot\|_{\lambda} \leq \|\cdot\|_{2\lambda}$.
 \halmos

\subsection{Proof of Proposition \ref{prob2}}  It suffices to consider $t'=0$ (see Remark \ref{tprime}) and so we will suppress the argument $t'$. In Section 4.10. we have shown, see   (\ref{s'}, \ref{f18a}),
 \rr
 s'_{u'}=\E^\perp \sigma^1_{u'} \chi +\rho
 \label{P1}
\rrr
with $$\chi=\mathbbm{1}(\|\sigma_{u'}^1\|_{\lambda}\leq  {_{1}\over{8}} \ep_n^{1-3\gamma})$$
%IT MUST BE LAMBDA NORM, NOT $2$ LAMBDA I CHANGE r TO rho AND PUT 1 FOR CHI
where, by  the second inequality in (\ref{expa1}),   (\ref{contra}),  $\sigma^2=(\sigma^2)^{loc}+ \sigma^{nloc}$, (\ref{barcl}) and (\ref{N3}),
 \rr
\|\rho\|_{2\lambda}\leq C L\ep_{n+1}^{1-2\gamma}.
 \label{P2}
\rrr
%BUT HERE WE NEED $2$ LAMBDA.  WE HAVE PUT CL NOT C OF L, AND WE HAVE $n+1$. CHANGED BELOW ALSO.
From (\ref{linearRG}), (\ref{srel1}), we have
 \rr
\sigma^1_{u'}=( \CL s^{loc})_{u'}= \sum^{L^2-1}_{t=0} \sum_{u\in Lu^\prime} \CL_is_{t, u}.
 \label{P3}
\rrr

By Schwarz' inequality,
 \rr
 \E\ e^{(s'_{u'},f)}\leq  (\E\ e^{\E^\perp(\sigma^1_{u'} \chi,2f)})^\hf (\E\ e^{(\rho_{u'},2f)})^\hf
 \label{P4}
\rrr
%I PUT CHI INSIDE THE PARENTHESIS

Let us first discuss the second factor on the RHS of (\ref{P4}). 
Write, using (\ref{fs}),
$(\rho_{u'},2f)= 2\int dx dy \rho_{u'}(x,y) f (x, y) \exp(-2\la \tau (u', x,y))  \exp(2\la \tau (u', x,y))$.
Using, for $x \in {\bf u}$, $y \in {\bf v}$,  $\tau (u', x, y)\leq \tau (u \cup v)+ d(u' \cup u \cup v)$
(since the right hand side is greater than the sum of the length of path joining $u$ and $v$ and
one joining $u'$ and $u$ or $v$),
and using the definition of the norms (\ref{f24}) and (\ref{w0}),  we get:
 \rr
|(\rho_{u'},2f)|\leq C\|\rho\|_{2\lambda}\|f\|_{u'}\leq C L\ep_{n+1}^{1-2\gamma}
\|f\|_{u'},
 \label{P5}
\rrr
using  (\ref{P2}) in the last inequality.
Let first $f$ be such that the right-hand side of (\ref{P5}) is less than $1$.  Using $\E (\rho_{u'},2f)=0$ and the inequality $|e^x-1-x|\leq x^2$
for $|x|<1$ we get, using (\ref{P5}), 
\rr
\E\ e^{ (\rho_{u'},2f)}\leq 1+\E (\rho_{u'},2f)^2\leq \exp( CL^2\ep_{n+1}^{2-4\gamma}
\|f\|^2_{u'}) \leq \exp(  \delta_{n+1}
\|f\|^2_{u'}),
\label{P6}
\rrr
for $\ep$ small, with $\delta_{n+1}=L^{-(n+1) \gamma}\ep_
{n+1}^{2-6\gamma}$. 

On the other hand, if $f$ is such that the right hand side of (\ref{P5}) is larger than $1$, we can bound: 
$$
\mathbb{E} \exp(\rho_{u'},2f)\leq \mathbb{E} \exp |(\rho_{u'},2f)|,
$$
and use (\ref{P5}) and $x<x^2$ for $x>1$ to get  (\ref{P6}) again.

Consider then the first factor on the RHS of (\ref{P4}).
We have, by (\ref{fs}), (\ref{sigmaA}), (\ref{linearRG}),
\rr
(\sigma^1_{u'},f)= \sum^{L^2-1}_{t=0} \sum_{u\in Lu^\prime}(s_{t, u} ,
 f_{t}),
 \label{P7}
\rrr
with
\begin{equation}
f_t (x,y) \mathrel{\mathop:}= L^{d-1} \int  dx^\prime dy^\prime  f (x^\prime, y^\prime)T^{L^2-t-1} (Lx^\prime -x) T^{t} (y-Ly^\prime).
\label{w3}
\end{equation}
\lem{ft} Let $u\in L\bf u'$. Then, $ \forall f: \mathbb R^d\times \mathbb R^d\to \mathbb R$ measurable, %ADDED 
\qq
\| f_{t}\|_{u}&\leq &CL^{-1}(t+1)^{-d/2}\|f\|_{u'}
 \label{P8}\\
 \sum_{u\in L{\bf u'}}\| f_{t}\|_{u}&\leq&CL^{-1}\|f\|_{u'}
  \label{P9}
\qqq
\elem
\no{\bf Proof.}
 By (\ref{w0}) and (\ref{w3}) we get
\begin{eqnarray}
  \|f_t\|_u  \leq L^{d-1} \int dxdydx'dy'|f(x',y')|T^{L^2-t-1} (Lx^\prime -x) T^{t} (y-Ly^\prime)
 e^{-2\lambda\tau(u,x,y)}.
   \non
\end{eqnarray}
Now use the triangle inequality to get:
\rr
\tau(u^\prime,x^\prime,y^\prime)) 
 \leq  \frac{c}{L}(\tau(u,x,y)+|Lx_i^\prime-x|+|y-Ly^\prime|)+c,
 \non
\rrr
which gives, if
$\frac{c}{L}<1/4$ %CHANGED,
\begin{eqnarray}
  \|f_t\|_u \leq C L^{d-1} \int dx^\prime dy^\prime
  |f(x^\prime, y^\prime)| \exp (-2\lambda\tau(u^\prime,x^\prime, y^\prime))
  I(x',y',u)
  \label{w8}
\end{eqnarray}
  with
  \begin{eqnarray}
 I(x',y',u)=  \int dxdy  \tilde T^{L^2-t-1} (Lx^\prime - x) \tilde T^{t} (y - L y^\prime)  \exp (-\lambda\tau(u,x, y)) 
 \label{w88}
\end{eqnarray}
and, as before, $\tilde T^t (x) = T^t (x) \exp (\frac{2c\lambda |x|}{L})$. By symmetry, we may suppose
 $t \leq \frac{L^2}{2}$. Since $k= 2c\lambda$,  and $\frac{c}{L}<1/4$, $\frac{k}{L} \leq \frac{\la}{2} $,  and we may use (\ref{w9}), with   $C(k)$  $L$-independent. 
 The sup norm bound in (\ref{w9}) 
 yields
   \begin{eqnarray}
&& I(x',y',u)\leq  CL^{-d}(t+1)^{-d/2} \int dxdy  \exp (-\lambda\tau(u,x, y)) 
 \\
 \non
&& \leq  CL^{-d}(t+1)^{-d/2}.
 \label{w88a}
\end{eqnarray}
Combined with (\ref{w8}) we obtain the first claim (\ref{P8}). 

 For the second claim, use 
(\ref{Tnorm1}) to get
\begin{equation}
\exp (- \frac{\lambda}{2} \tau(u,x, y))
\tilde T^{t} (y- L y^\prime) \leq C (t+1)^{-d/2}  \exp(-c\la \frac{|u-Ly^\prime|}{\sqrt {t+1}}),
\label{w88b}
\end{equation}
where $C$ and $c$ are $L$-independent. %ADDED
Then,  using the sup norm bound in (\ref{w9}) to get $\tilde T^{L^2-t-1} (Lx^\prime - x)\leq C L^{-d}$, for  $t \leq \frac{L^2}{2}$,
and
the bounds
$$
 \int dxdy  \exp (- \frac{\lambda}{2}\tau(u,x, y)) \leq  C,
 $$
 $$
\sum_{u \in \Z^d} (t+1)^{-d/2}  \exp(-c\la \frac{|u-Ly^\prime|}{\sqrt {t+1}})\leq C,
 $$
  we get:

    \begin{eqnarray}
\sum_{u \in \Z^d} I(x',y',u)\leq  CL^{-d} 
 \label{w89}
\end{eqnarray}
whereby, using  (\ref{w8}),  (\ref{P9}) follows. \halmos
The lemma allows us to prove
\lem{eesf}
  \begin{equation}
  \E\ e^{(\sigma^1_{u'},2f)}
\leq e^{L^{-\gamma}\delta_{n+1} \|f\|_{u'}^2}.
\label{P10}
\end{equation}
\elem
\no {\bf Proof.}
As in (\ref{P5}), we have $|(\sigma^1_{u'},2f)|\leq C \|\sigma^1\|_{2\lambda}  \|f\|_{u'} $.
We obtain, as in (\ref{expa}), taking $\beta=s^{loc} $, instead of $s$ in (\ref{linearRGb}),
and using inductively Proposition \ref{sell}, 
\qq
 \|\sigma^1\|_{2\lambda}\leq CL \ep_n^ {1-3\gamma}.
  \label{jb3}
 \qqq
We then get that
 \rr
|(\sigma^1_{u'},2f)|\leq CL \ep_n^ {1-3\gamma} \|f\|_{u'}.
 \label{P11}
\rrr

Consider first   $f$ such that the RHS of (\ref{P11}) is less than $1$.
Using $\E (\sigma^1_{u'},2f)=0$ and the same argument
as before (\ref{P6}),
we get 
\rr
\E\ e^{(\sigma^1_{u'},2f)}\leq 1+\E (\sigma^1_{u'},2f)^2. 
\label{P12}
\rrr
By (\ref{P7}),
\rr
\E (\sigma^1_{u'},f)^2= \sum^{L^2-1}_{t_1,t_2=0} \sum_{u_1,u_2\in Lu^\prime}\E (s_{t_1, u_1},
 f_{t_1})(s_{t_2, u_2},
 f_{t_2})
 \label{P13}
\rrr
Denote $z_i=(u_i, t_i)$ and  distinguish between the sum over  $|z_1-z_2| \leq L^{\beta}$ and  $|z_1-z_2| \geq L^{\beta}$, where $\beta$ will be chosen below.

For the first sum, we use Schwarz inequality on $\E (s_{t_1, u_1},
 f_{t_1})(s_{t_2, u_2},
 f_{t_2})$ and  then  Schwarz inequality
 again for the sum over $z_1, z_2$ with the constraint $|z_1-z_2| \leq  L^{\beta}$, and (\ref{w1a}), 
to bound that sum  by $C L^{\gamma}\delta_n\sum_{u,t} \|f_t\|_u^2$, if we choose $(d+1)\beta \leq \gamma$. 

For $|z_1-z_2| \geq  L^{\beta}$ we apply (\ref{decay}) with $F_i=(s_{t_i, u_i},
 f_{t_i})$. The supports of these functions are, in the original lattice, at a distance at least $L^{\beta +n}$ and their size
 is bounded by $C L^{(2+d)n}$;  
 we have, $|(s_{t_i, u_i},
 f_{t_i}) |\leq C  \|s_{t_i}\|_{2\lambda}  \|f_{t_i}\|_{u_i} \leq CL \ep_{n-1}^ {1-3\gamma} \sup_{u,t} \|f_t\|_u$, where, in the last inequality, we use  (\ref{P1}) (with prime referring to scale $n$),  (\ref{P2}), (\ref{jb3})
%and Proposition \ref{sell} 
(on scale $n-1$) and  where the supremum is taken over $u,t$ in (\ref{P13}). This gives a bound on  $\|F_i\|_\infty$ and we have $\E( F_i)=0$.
%we use Proposition \ref{sell} to bound  $\|F_i\|_\infty$. 
 By (\ref{decay})  the sum of the absolute value of 
  those terms is bounded by 
  $$C L^{(2+d)n} L^2 \ep_{n-1}^{2-6\gamma} \exp(-cL^{\beta +n})  \sup_{u,t} \|f_t\|_u^2
  .$$
  %ADDED $L^{(2+d)n}$
   Since for $L$ large   $C L^{(2+d)n} L^2 \exp(-cL^{\beta +n}) \leq 
L^{-2} L^{-n\gamma}$ and %ADDED 
$ \ep_{n-1}^{2-6\gamma} L^{-2} L^{-n\gamma}\leq \delta_n$,  we conclude that:
  \begin{equation}
\E (\sigma^1_{u'},2f)^2\leq C L^\gamma \delta_n \sum_{u,t} \|f_t\|_u^2 .
\label{w41}
\end{equation}
By Lemma \ref{ft}
\begin{eqnarray}
\sum_{u,t}  \|f_t\|_u^2 \leq C L^{-2} \|f\|_{u'}^2
\sum_{t\leq L^2}(t+1)^{-d/2}
  \label{w899}
\end{eqnarray}
%PUT ${t\leq L^2}$ (NOT ${t\leq \hf L^2}$)
Since the sum is bounded by
$C\log L \leq L^\gamma$, for $d \geq 2$ we get, using (\ref{P12}), (\ref{w41}), in the case where  $ f $ is such that the RHS of (\ref{P11}) is less than $1$,
  \begin{equation}
  \E\ e^{(\sigma^1_{u'},2f)}
\leq e^{CL^{-2+2\gamma} \delta_{n} \|f\|_{u'}^2}\leq e^{L^{-\gamma} \delta_{n+1} \|f\|_{u'}^2}.
\label{P14}
\end{equation}

Now, consider the case where   $f$ is such that the RHS of (\ref{P11}) is larger than $1$.
 Decompose $\mathbb Z^{d+1}$ into $C L^{\beta}$ sublattices $\mathbb L_\ell$   indexed by $\ell $, so that, if $z_1=(t_1,u_1), z_2=(t_2,u_2)$ belong to the same sublattice, we have $|z_1-z_2| \geq L^{\beta}$, with again $(d+1)\beta \leq \gamma$.
Using H\"older's inequality, we bound
\begin{equation}
\E\ e^{(\sigma^1_{u'},2f)} \leq \prod_{\ell}\mathbb E \Big( \exp (CL^{\gamma}  \sum^{\hspace{6mm}\raisebox{-2mm}{$\scriptstyle \ell$}}_{u,t }
 (s_{t, u} , f_t)
  )\Big)^{(CL^{\gamma})^{-1}},
\label{w4}
\end{equation}
where $\sum^\ell$ means that the sum is restricted to $ (u, t)\in (Lu^\prime \times [0,L^2-1] ) \cap  \mathbb L_\ell$.
Each of the factors in (\ref{w4}) is of the form $\displaystyle \mathbb E ( \prod^k_{i=1} F_i )$ 
where the support of the functions $F_i$ are, in the original lattice, at a distance at least $L^{\beta+n}$
and their size is bounded by $CL^{(2+d)n}$. Hence, by (\ref{decay1}),
\begin{equation}
\E\ e^{(\sigma^1_{u'},2f)} \leq 
\prod_{\ell}  \prod^{\hspace{6mm}\raisebox{-2mm}{$\scriptstyle \ell$}}_{u, t }\Big( \mathbb E \big( \exp (CL^{\gamma}  
 (s_{t, u} , f_t)
  )\big)^{(CL^{\gamma})^{-1}}\exp e^{-cL^{\beta+n}}\Big).
\label{ww4}
\end{equation}
where the product $\prod^\ell$ runs over $ (u, t)\in (Lu^\prime \times [0,L^2-1] ) \cap  \mathbb L_\ell$ and where the factor $CL^{(2+d)n}$
is controlled by the exponential  $e^{-cL^{\beta+n}}$.

Using inductively (\ref{w1}),
\begin{equation}
 \E\ e^{(\sigma^1_{u'},2f)}\leq  \exp [CL^{\gamma} \delta_n \sum_{u,t} \|f_t\|_u^2+L^{2+d}e^{-cL^{\beta+n}}].
\label{w6}
\end{equation}
Use (\ref{w899}) and then use   the fact that if $\|f\|_{u'}$ is such that the RHS of (\ref{P11}) is larger than $1$,
then $\delta_n \|f\|_{u'}$ is larger than an inverse power of $L^n$,
which shows that $L^{2+d} e^{-cL^{\beta+n}}$ is less than the first term, for $L$ large enough and all $n$; we end up again with the bound (\ref{P14}).  
 The claim follows.
 \halmos

Let us now bound the first factor in (\ref{P4}) using Lemma \ref{eesf}. First,
by the exponential Tchebycheff inequality, (\ref{P10}) implies:
\rr
\mathbb P(|(\sigma^1_{u'},2f)| > k\|f\|_{u'}) \leq 2 \exp (- \frac{L^\gamma k^2}{4\delta_{n+1}})
\label{P20}
\rrr
Let $a:=L^{-\hf\gamma}\delta_{n+1}^\hf \|f\|_{u'}$. Then
 \qq
 \mathbb{E} ((|(\sigma^1_{u'},2f)|^m)& \leq& a^m \sum^{\infty}_{r=0} (r+1)^m \ \mathbb{P} (|(\sigma^1_{u'},2f)| \in [a r, a (r+1)])\non\\
 \non
 & \leq&2 a^m \sum^\infty_{r=0} (r+1)^m \exp \big(-\frac{r^2}{4}\big) \leq (Ca)^m (m!)^{1/2} .
 \qqq
 Thus, since $\chi\leq 1$ and $\E\E^\perp=0$, 
\qq
\E\ e^{\E^\perp (\sigma^1_{u'}\chi,2f)}&\leq& 1+ \sum^\infty_{k=2} \frac{2^k}{k!} \max_{m\leq k} ( \mathbb{E}(|(\sigma^1_{u'},2f)|^m)( \mathbb{E}(|(\sigma^1_{u'},2f)|)^{k-m})\non\\
& \leq& 1 + \sum^\infty_{k=2} \frac{(C a)^k}{k!} (k!)^{1/2}\leq e^{C a^2}=e^{CL^{-\gamma}\delta_{n+1} \|f\|^2_{u'}}\leq
=e^{\delta_{n+1} \|f\|^2_{u'}}.
\non
\qqq
Combining this with (\ref{P6}), (\ref{P4}), shows that (\ref{w1}) iterates.
\halmos
%for $n\geq n_0$.

%Finally to get the induction started at $n=n_0$, recall from the previous Section
%that we have then $\|s\|_\lambda\leq L^{n_0(1-2\alpha)}$. Hence proceeding as
%in the case of r above we conclude
%$$
%\E \exp{(s_u,f)}\leq \exp(L^{n_0(2-4\gamma)}\|f\|_u^2)\leq  \exp(\epsilon_n^2\|f\|_u^2).
%$$

\subsection{Proof of Proposition \ref{prob1}}
Let us first deduce the following Corollary from Proposition \ref{prob2}
\begin{corollary}\label{ncoro} For all $z'\in\N\times \mathbb Z^d$, and $n\geq 1$,
\begin{equation}
 \mathbb P (r_{z'} \neq 0) \leq \exp (-c L^{(n+1)\gamma})
\label{w14}
\end{equation}
for some $c>0$.
\end{corollary}
\no {\bf   Proof. }
%It suffices to take $n\geq n_0$ since
%for smaller $n$, $r\equiv 0$. 
From (\ref{f18a}) we infer that if  $r_{z'} \neq 0$
then there exist $x,y$, $\zeta>0$, such that 
\rr
|\sigma^1_{u'}(x,y)|\geq \zeta \ep_
{n+1}^{1-3\gamma} \exp (-3\lambda\tau(u', x,  y)/2)  .
\label{unprob}
\rrr
Indeed, otherwise, we could integrate the opposite bound over $x \in {\bf u}$, $y \in {\bf v}$, multiply the result by
 $\exp (\lambda\tau(u', u,  v))$, sum over $u,v$ and get, for $\zeta$ small enough,
that $\|\sigma^1_{u'}\|_\lambda \leq {_1\over^8} \ep_
{n+1}^{1-3\gamma}$, i.e. $r_{z'} = 0$
%From  (  (\ref{expa1}-\ref{contra})) we 
 %have $\|(\CL s^{loc})\|_{2\lambda}\leq C \ep L
%L^{-n(1-3\gamma)}$ which implies 

The bound (\ref{jb3}) implies $\forall x,y$, 
$$
|\sigma^1_{u'}(x,y)|\leq
CL\ep_n^{1-3\gamma}e^{-2\lambda\tau(u',x,y)} \leq  \zeta \ep_
{n+1}^{1-3\gamma} \exp (-3\lambda\tau(u', x,  y)/2),
$$
if $C e^{-\hf \lambda\tau(u',x,y)} \leq \zeta L^{-2}$.
Thus it suffices to estimate the probability for the event that (\ref{unprob}) happens 
for, say, some $ x,  y\in (L^{-(n+1)}\Z)^d$ %WE ARE ON SCALE $n+1$ 
with $|x-u'|, |y-u'|<L$.  Let
\[
f(\cdot,\cdot) = \hf L^{2nd}\de_{ x,\cdot} \de_{ y,\cdot} \exp (2\la \tau(u',x, y)),
\]
with $\delta$ the Kronecker delta, so that  $\|f\|_{u'}=\hf$ and  $(\sigma^1_{u'}, 2f)=\sigma^1_{u'}(x,y) \exp (2\lambda\tau(u', x,  y))$.  
Using (\ref{P20}) with $k= \zeta \ep_
{n+1}^{1-3\gamma}$, using $\exp (\hf \lambda\tau(u', x,  y)) \geq 1$, and remembering that $\delta_{n+1}=L^{-(n+1) \gamma}\ep_
{n+1}^{2-6\gamma}$, we  infer that the probability in question is bounded from above by
$$
2L^{2(n+2)d} \exp (-L^{(n+1)\gamma}),
$$
where $L^{2(n+2)d}$ bounds the number of pairs $(x,y)$ with $|x-u'|, |y-u'|<L$. This
implies the claim.\halmos

To prove Proposition \ref{prob1}, 
we need 
the following 
\lem{N'Lemma}  There exists a  $C$, where $C \to \infty$ as $\ep \to 0$,  such that
if $K=C\log L$, and  $n\geq m_0$, where 
\qq
L^{\frac{\gamma m_0}{2}}= n_0,
 \label{jb12}
\qqq
with $n_0$ defined in (\ref{jb1}),
then, for any $A \subset \Z^{d+1}$,
any $\{N_z\}_{z\in A}$, with $N_z \neq 0$, %ADDED
\qq
 \mathbb P(\{N_{n, z}=N_z\}_{z\in A}) \leq \exp(- Kn|A|-2K{}\sum_{z\in A} N_z)
 \label{38}
\qqq
implies that (\ref{38}) holds also for $N_{n+1}$, with $K$
%=K\prod_{\ell=1}^n(1-\ell^{-2})$
replaced by
%by $K_{n+1}$.
$K'=K(1-n^{-2})$. 
%I PUT  $n^{-2}$ SINCE WE DONT NEED $n+2$ now that $n\geq m_0$.
\elem
\no {\bf Proof.} 
We set again $t'=0$ (see Remark \ref{tprime}). Using the recursion relation (\ref{a7}), we can write, for any $(N'_{z'})_{z'\in A'}$, $ N'_{z'} \neq 0$,
\qq
&&\mathbb P(\{N_{n+1, z'}=N'_{z'}\}_{z'\in A'}) = \nonumber\\
&& \sum_{B'\subset A'}\mathbb P(\{\tilde N_{z'} = N'_{z'}+1-(n+n_0) \log L,\ r_{z'} \neq 0\}_{z'\in B'}\&
\{\tilde N_{z'} = N'_{z'}+1\}_
{z'\in A'\backslash B'})
\label{40}
\qqq
where we recall that, by (\ref{f18a}))  $r_{z'}\neq  0$ means that $r_{z'}=(n_0+n) \log L$. Using H\"older's inequality, the summand is less than
\qq
&& \mathbb P(\{\tilde N_{z'} = N'_{z'}+1-(n_0+n) \log L\}_{z'\in B'}\&
\{\tilde N_{z'} = N'_{z'}+1\}_
{z'\in A'\backslash B'})^{(1-{1\over n^{2}})}\non\\
&&\cdot\mathbb E (\prod_{z'\in B'} \mathbbm{1} (r_{z'}\neq 0))^{{1\over n^{2}}}.
\label{41}
\qqq
Using the inductive assumption (\ref{38})  the first factor is bounded by
\qq
\prod_{z'\in A'}\Big( \sum_{A_{z'}} \sum_{\{N_z\}_{z\in A_{z'}}} \exp \Big[- K'n|A_{z'}|-2K'\sum_{z\in A_{z'}} N_z\Big] \Big):=\prod_{z'\in A'}P_{z'}
\label{42}
\qqq
where the sum $\sum_{A_{z'}}$ runs, for $z'=(u', 0)$, over all non empty subsets $A_{z'} \subset L {\bf u}' \times [0, L^2 -1] $,
 and the sum over $\{N_z\}$ runs over $N_z\neq 0$, satisfying the constraint 
\rr
\sum_{z\in A_{z'}}N_z=\left\{
\begin{array}{rl}
&N'_{z'}+1-(n_0+n) \log L, \; \text{if }z'\in B'\\
&N'_{z'}+1,\;\;\;\; \;\;\;\;\;\;\; \;\;\;\;\;\;\;\;\;\;\;\;\;\;\; \text{if } z'\in A'\backslash B'.
\end{array} \right.
\label{ntilded}
\rrr
To estimate $P_{z'}$ let us first observe that we have an a priori bound for all $x$ and $n$,
\qq
N_{n, x} \leq L^{(3+d)n}n_0.
\label{42a}
\qqq
Indeed, let $N_n= \sup_z N_{n, z}$. Then, we get from (\ref{a7a}-\ref{a7}) that 
$$
N_{n+1}\leq L^{2+d} N_n+(n_0+n) \log L.
$$
  From this, we get
easily  (\ref{42a}).
%that
%$N^n_x \leq L^{(3+d)(n+n_0)}$, which implies for $n\geq n_0$. 

Let first  $z'\in A'\backslash B'$ and write $A_{z'}=A$. By (\ref{ntilded})
\qq
P_{z'}=\sum_{A}\CN(|A|) \exp [- K'n |A|-2K'(N'_{z'}+1)]  
\non
\qqq
where $\CN(|A|)$ is the number of choices for $\{N_z\}$. For  $|A|=1$ there is only one possible $z$ for which $N_z=N'_{z'}+1$ and thus $\CN(1)=1$. For  $|A|>1$,
 $\CN(|A|)\leq (C L^{(3+d)n} n_0)^{|A|}$ since each $N_z$ 
  takes a discrete set of values, by (\ref{a7a}-\ref{a7}), and is bounded by  (\ref{42a}).
  Since the sum over $A$ runs over subsets of a set of cardinality $L^{2+d}$, and thus contains less than $L^{(2+d) |A|}$ terms for given $|A|$, we infer
  \qq
P_{z'}\leq\sum_{M=1}^{L^{2+d}}\CN_M \exp [-K'n M-2K' (N'_{z'}+1)]  
\non
\qqq
with $\CN_1=L^{2+d} $ and $\CN_M= (CL^{(3+d)(n+1)} n_0)^M$ for $M>1$. 

Taking $K=C\log L$, for $C$ large,  implies $K'>\hf C\log L$ (since $n\geq m_0$, we may assume $n\geq 2$).
We   then use  $\exp [-K'n (M-1)]\leq \exp [-K'(n+1) M/4]  $, for $M>1$, 
the bounds  $ \exp [-K'(n+1)M/8]\leq \hf L^{-2-d} (CL^{(3+d)(n+1)})^{-M}$
and  $ \exp [-K'nM/8] \leq n_0^{-M}$, which hold for $C$ large in $K'>\hf C\log L$, 
and the fact that $n\geq m_0$, see (\ref{jb12}), in order to bound
the summand  by $\hf L^{-2-d} \exp [- K'(n+1)-2K' N'_{z'}]  $, for $M>1$.

For $M=1$, $\CN_1=L^{2+d} $ and we can use 
 a factor $\exp [-K' ]  $  to bound that term by
 $\hf L^{-2-d} \exp [- K'(n+1)-2K' N'_{z'}]  $. 
 %I REORDERED THE ARGUMENT

Thus, since the sum over $M$ contains $L^{2+d}$ terms,
  \qq
P_{z'}\leq \hf  \exp [- K'(n+1)-2K' N'_{z'}]  
\label{43}
\qqq
for  $z'\in A'\backslash B'$.

Now consider $z'\in B'$. Proceeding exactly as above, but using the first
equality in (\ref{ntilded}), 
%$\widetilde N_{z'}=\sum_{z\in A} N_z=N'_{z'}+1-n-n_0$, 
we get that
\qq
P_{z'}\leq
{1\over 2} \exp [- K'(n+1)-2K' N'_{z'}+2K'(n_0+n) \log L].
\label{45}
\qqq

\par\noindent
We will prove below  that
\qq
\mathbb E(\prod_{z'\in B'} \mathbbm{1} (r_{z'}\neq 0))\leq \exp (-c'L^{(n+1)\gamma}|B'|).
\label{46}
\qqq
Now, insert  (\ref{43}) and (\ref{45}) in (\ref{42}); then,  insert the result and (\ref{46}) in (\ref{41})  to obtain
\qq
(\ref{41}) \leq  2^{-|A' |} \exp[(-\frac{c'L^{(n+1)\gamma}}{n^2}+2K' (n_0+n) \log L)|B'|
- K'(n+1)|A'|-2K'\sum_{z' \in A'}N'_{z'}]   .
\non
\qqq 
Substitute this result into the  sum  (\ref{40}). Since

\rr
\exp[-\frac{c'L^{(n+1)\gamma}}{n^2}+2K' (n_0+n) \log L]%|B'|] 
\leq  2^{-1}%|B'|},
 \label{problem}
 \rrr
 which follows from $n\geq m_0$ and $K'\leq K=C\log L$, since we have both
 $L^{(n+1)\gamma/2} \geq C (\log L)^2{n^3}$, for $L$ large,  and $L^{(n+1)\gamma/2} \geq n_0$, see (\ref{jb12}). Note that we can let $C\to \infty$, when $\ep \to 0$, since $n_0, m_0 \to \infty$ as  $\ep \to 0$,  see (\ref{jb1}),
 (\ref{jb12}).
 
 % and $L^{n_0}=\ep^{-3\gamma}$ (see (\ref{jb1})),
 
 We may use
 $$
 2^{-|A' |}\sum_{B' \subset A'}  2^{-|B'|} \leq 1,
 $$
 to conclude the iteration of (\ref{38}).

   To prove (\ref{46}),   decompose $\mathbb Z^{d+1}$ into $2^{d+1}$ sublattices $\mathbb L_\ell$   indexed by $\ell=1,\ldots,2^{d+1}$, such that if $z_1=(u_1,t_1), z_2=(u_2,t_2)$ belong to the same sublattice, we have $|z_1-z_2| \geq 2$.
Use the  H\"older inequality, as in the derivation of (\ref{w4}),
 to reduce the proof of (\ref{46}) to the case where  $B'$ is  included in one of those sublattices.
 Now apply (\ref{decay1}):  the support of the functions $F_i$ are, in the original lattice, at a distance at least $L^{n+1}$ and $|A_i| \leq L^{(n+1)(d+2)}$. 
 Since   $\exp(2|A_i|e^{-cL^{n+1}})\leq C$ we
 conclude by using  Corollary \ref{ncoro}  to estimate (\ref{46}) for $B'$ reduced to a point (and absorbing thus the constant $C$).
\halmos

To finish the proof of Proposition \ref{prob1}, 
observe that we have   $r_{z'}=0$ $\forall z'$, for  $n<m_0$, which implies that $N_n=0$, $D_n=\emptyset$ and $\ell_n=0$ for
$n<m_0$ (see (\ref{f3a}-\ref{a8}) and (\ref{bloc}-\ref{cdec})).
 To see this, note that this holds
  for $n=0$ and that we get inductively, from (\ref{expa}) that the linear term in the iteration (\ref{b'dec})  of $b$ is bounded by:
  \qq
  \|\CL b_{n}\|_{\la}\leq CL \|b_{n}\|_{\la}
  %+C(L)\epsilon_n^{1-2\gamma}.
   \label{small n}
\qqq
  and, as long as this remains smaller than $\epsilon_n^{1-3\gamma}$, one can show, as in the proof of (\ref{N3}), that the nonlinear contributions are smaller, of order $\epsilon_n^{1-\gamma}$.
%  CHANGED
%  NOT CLEAR: THE BOUND ON THE NONLINEAR TERM IS OF THIS FORM ONLY IF THE b CONTRACTS;
  This means that $\|b_{n}\|_{\la}$ and in particular $\|\sigma^1\|_{\la}$
  is bounded by
  $(CL)^n\epsilon$. Thus, $r_{z'}=0$ $\forall z'$ as long as this is smaller
  than  ${_1\over^8}\epsilon_n^{1-3\gamma}$ i.e. as long as
  $C^n(L^{n}\epsilon)^{3\gamma}\leq {_1\over^8}$, which is true for $\ep$ small and   $n<m_0$, since then $L^n\leq n_0^{2/\gamma}$,  see (\ref{jb12}),
 $ C^n$ is also bounded by a small power of $n_0$,
  and (\ref{jb1}) means that $n_0\leq | \log \epsilon|$.
  
Thus, Proposition \ref{prob1} holds trivially for  $n<m_0$. For $n\geq m_0$, where we can assume that $m_0$ is  larger than $3$, Lemma \ref{N'Lemma} implies inductively  Proposition \ref{prob1}.
  %for any fixed $n_0$, provided that $\ep$
  %is small enough.
  \halmos
 % $(CL)^n\epsilon_n^{1-2\gamma}$. Thus, $r_{z'}=0$ $\forall z'$ as long as this is smaller
  %than  ${_1\over^8}\epsilon_n^{1-3\gamma}$ i.e. as long as
  %$C^nL^{(1-\gamma)n}\epsilon^\gamma\leq {_1\over^8}$, which is true for $\ep$ small and   $n<m_0$, since then $L^n\leq n_0^{2/\gamma}$,  see (\ref{jb12}),
  %and (\ref{jb1}) means that $n_0\leq | \log \epsilon|$.

\subsection{Proof of Proposition \ref{T}} 
We start with the study of the RG iteration for $T$: %THE r HERE IS DETERMINED BY OUR m IN OUR ASSUMPTIONS
\lem{CTlemma} There exist
$r,c>0$ such that for all $n\geq 0$, $\hat \CT_n$, defined in (\ref{bite11}), (\ref{defT}), 
is analytic in $|\ima k|< r^2L^{n\over 4}$   and 
for such $k$
\begin{equation}
\hat \CT_n(k)=(1+\CO(L^{-2n}|k|^4))e^{-{D_0\over 2d}k^2}
\label{z1}
\end{equation}
if $|k|\leq rL^{n\over 4}$ and
\begin{equation}
| \hat \CT_n(k)|\leq e^{-cL^{n\over 2}}
\label{z2}
\end{equation}
otherwise. 
\elem
\no {\bf Proof.}  Our assumption (iv) implies that 
$\hat T(k)$ is analytic in a neighborhood of
$\T^d$.   Combining with Assumption (vi), we conclude 
 that, for $r$ small enough, $|\hat T(k)|\leq \rho(r)<1$ for $|\rea k|>r$,
$|\ima k|\leq r^2$.  This implies (\ref{z2}) for $|k|> r L^n$, $|\ima k| \leq r^2 L^{n \over 4} $. 

Assumption (iii) in turn gives the representation (\ref{bite111})
near the origin, which means that
 $\hat T(k)=e^{-{D_0\over 2d}k^2} (1+ \CO(|k|^4))$
for $|k| \leq r$. This implies (\ref{z1}) for $|k| \leq rL^n$, in particular for $|k| \leq rL^{n \over 4}$, $|\ima k| \leq r^2 L^{n \over 4} $.
Since 
$$|e^{-{D_0\over 2d}k^2}|= e^{-{D_0\over 2d}((\rea k)^2-(\ima k)^2)} \leq e^{-{D_0\over 4d}|k|^2}$$
for $\hf |k| > |\ima k|$, the claim (\ref{z2}) holds also for  $rL^{n \over 4} <|k| \leq  r L^n$, $|\ima k| \leq r^2 L^{n \over 4}$, if   $ r$
is taken small enough.
\halmos

The Lemma implies that
$k{\hat \CT_n}(k)$ has similar bounds in the strip, hence it is
integrable there  (recall that
$\rea k$ is on the $L^n$ torus) and thus  it
is exponentially decaying and we deduce (for $\la$ small)
\begin{equation}
\|\CT_n\|_{2\lambda}+\|\nabla\CT_n\|_{2\lambda} \leq C  .
\label{5333}
\end{equation}

Next, write again $T$ for $T_n$ and $T'$ for $T_{n+1}$.  
Recall   (\ref{51}): 
 \begin{equation}
 T'=L^d{\mathcal S}_LT^{L^2}+\E\ \nabla\cdot\CN^1.
\label{51b}
\end{equation}
We need
\lem{deltaTlemma} There is a $\eta>0$ s.t.  $\beta:=\E \nabla\cdot\CN^1$ satisfies
\begin{equation}
\|\beta\|_{2\lambda}+\|\nabla \beta\|_{2\lambda} \leq \epsilon_{n+1}^\eta% L^{-\eta n}  .
\label{53}
\end{equation}
\elem
\no{\bf Proof.} Recall that $ \CN^1$ is a sum of products $\Pi$ (see (\ref{Ndec})).
We use two simple bounds for such products. 

Note first that $\E\ \Pi$ is a function of $(x,y)$ only.
So, $\|\E\ \Pi\|_{2\lambda}\leq
\sup_{v\in \mathbb Z^d} \E (\sum_{u } 
    \|  \Pi \|_{u,v}e^{2 \lambda |u-v|})$, and we get, using (\ref{weight}):
\qq
&&\|\E\ \Pi \|_{2\lambda}\leq 
\sup_{v\in \mathbb Z^d} \E(  \sum_{A,u } \sum_{\CD} \| \Pi _{t, \CD, A} \|_{u,v}
e^{2 \lambda l(\CD,A,t,u,v)} 
e^{-2 \lambda \tau (\CD \cup u \cup v)})
%e^{-N_{t} (\CD)}e^{+N_{t} (\CD)}).
\label{fix1}
\qqq
%Using Schwarz inequality on $\sum_{\CD} $, and 
Writing $1= e^{-N_{t} (\CD)}e^{+N_{t} (\CD)}$, and taking the sup over  $u, \CD$, we get:
\qq
&&\sum_{A, u, \CD}  \| \Pi _{t, \CD, A} \|_{u,v}e^{2 \lambda l(\CD,A,t,u,v)} 
e^{-2 \lambda \tau (\CD \cup u \cup v)} 
%e^{-N_{t} (\CD)} e^{+N_{t} (\CD)}
\non \\
 &&\leq
 (\sum_{A, u, \CD} \| \Pi _{t, \CD, A} \|_{u,v}
 e^{2 \lambda l(\CD,A,t,u,v)} 
e^{-N_{t} (\CD)})
(\sup_{u\in \Z^d, \CD}
e^{-2 \lambda \tau (\CD \cup u \cup v)} 
e^{N_{t} (\CD)})
\label{fix2}
\qqq
Next, insert this in (\ref{fix1}),
%taking the supremum over $u \in \Z^d$ of the second factor, and
 take the supremum over the random variables $\om$ for the sum $ \sum_{A, u, \CD }$,
 and replace the $\sup_{u\in \Z^d, \CD}$ by a sum;
we get:
\qq
&&\|\E\ \Pi \|_{2\lambda}\leq 
\non \\
 &&
\sup_{v\in \mathbb Z^d} \Big( \big[ \sup_\om  \sum_{A,u, \CD }%\subset \mathbb Z^d \atop
 %v \in \mathbb Z^d} 
 \| \Pi _{t, \CD, A} \|_{u,v}e^{2 \lambda l(\CD,A,t,u,v)}e^{-N_{t} (\CD)}  \big]
 \big[\sum_{u\in \Z^d, \CD}\E(
e^{-2\lambda \tau (\CD \cup u \cup v)} 
e^{N_{t} (\CD)})  \big] \Big)
\label{fix3} 
\qqq
By Proposition \ref{prob1} $ \E(
e^{N_{t} (\CD)})$ is bounded by $1+  \exp(-c\la K)$, where $K$ is as in Proposition \ref{prob1}, and, for fixed $v$,
$$
\sum_{u \in \Z^d, \CD} 
e^{-2 \lambda \tau (\CD \cup u \cup v)} \leq C.
$$
 So, combining these last two bounds,
\qq
&&\|\E\ \Pi \|_{2\lambda}\leq C
\sup_\om  \sup_{v\in \mathbb Z^d}    \sum_{\CD,A,u }%\subset \mathbb Z^d \atop
 %v \in \mathbb Z^d} 
 \| \Pi _{t, \CD, A} \|_{u,v}e^{2 \lambda l(\CD,A,t,u,v)}e^{-N_{t} (\CD)}.
 \non \\
 &&=C \sup_\om \|\Pi\|_{2\lambda},
\label{EPi1}
\qqq
and it suffices
to bound $\|\Pi\|_{2\lambda}$ uniformly in $\om$.
%$ \|\Pi\|_{2\lambda}$.

The second bound uses independence. Consider first a product $\Pi$ such that there exists $i$ with
 $\alpha_i=1$ and  no $j\in \{1,\dots,N\}$ with $|i-j|=1$ 
and $\alpha_{j}\in \{1,2,3\}$ i.e. there is $b$ separated by two time units
from other $B$'s.  This means that, in the original lattice, the distance between that 
$b$ and other $B$'s is at least $L^{2n}$; thus, we can use  $\E b=0$ and the bound
(\ref{decay}) which gives
\begin{equation}
\|\E\ \Pi\|_{2\lambda}\leq \|\Pi\|_{2\lambda}e^{-cL^{2n}},
\label{EPi}
\end{equation}
since the support of the functions in $\Pi$ is a power of $L^n$. The norm $ \|\Pi\|_{2\lambda}$ is bounded
as in (\ref{N3}).
Since the gradient is bounded by $CL^n$ (\ref{EPi}) we may use $e^{-cL^{2n}}$ to control the two gradients in $\nabla\nabla_x \cdot \Pi$, 
for the bound on $\nabla \beta$. 
 
Finally, to control the two gradients in other cases,  i.e. when there are no isolated $b$'s, which we shall assume from now on,
we use the translation
invariance of the expectation values, which implies:
\rr
\E\nabla_x \cdot \Pi(x,y)=-\E\nabla_y \cdot \Pi(x,y).
\label{symme}
\rrr
Let us bound now $\nabla\nabla_x \cdot \Pi$ using (\ref{EPi1}) and (\ref{symme}). 

%I REWROTE A BIT THE NEXT 2 PARAGRAPHS

If ${\alpha_1}=4$ and  ${\alpha_N }=4$ in (\ref{Ndec}) , we can use  (\ref{symme}) 
so that both gradients act on  $T$'s and,   using inductively (\ref{Tnorm}), we
arrive at the bound (\ref{N3})
for $\|\nabla\nabla \cdot \Pi\|_{2\lambda} $. Note that these $\alpha$'s cannot take the value 5, because there are no
$\nabla T$ at the beginning or at the end of the products in (\ref{Ndec}), see (\ref{rem1}).

Consider the other cases:  if ${\alpha_1} \neq 4$ and  ${\alpha_N }=4$, since there are no isolated $b$'s,  ${\alpha_1} =2$, i.e.
the product starts with $b \nabla \cdot b$ (by (\ref{rem1}), there is no $\nabla$ in the beginning). Then, we use (\ref{symme})
for one $\nabla$ and (\ref{f31}) to control $\nabla \cdot b \nabla \cdot b$. This is bounded by $\epsilon_{n+1}^\eta$
for $\eta$ small. If the product starts with $b \nabla \cdot b  \nabla \cdot b$, we use (\ref{symme})
for one $\nabla$,
  (\ref{grads}) for the first $\nabla \cdot b$ and (\ref{f31}) for $\nabla \cdot b  \nabla \cdot b$. This is 
again bounded by $\epsilon_{n+1}^\eta$
for $\eta$ small. Other terms are even smaller. We can proceed similarly if 
 ${\alpha_1} = 4$ and  ${\alpha_N } \neq 4$, by using  (\ref{symme}) for
one of the $\nabla$ acting on $T_{\pi_k}$ ( see (\ref{rem1})), so that it acts on a $b$.

Finally, consider  ${\alpha_1} \neq 4$ and  ${\alpha_N } \neq 4$. If we have $\Pi= B^2T^{L^2-4}B^2$,
we bound the norm of one  $\nabla$ by $CL^n$ in $\nabla\nabla \cdot \Pi$, and use (\ref{f31}) for $\nabla \cdot  B^2$ and (\ref{f29}) for $B^2$.
The result is bounded by $\epsilon_{n+1}^\eta$
for $\eta$ small.
For $\Pi= B^2\nabla \cdot bT^{L^2-5}B^2$, we get $\nabla\nabla \cdot \Pi=\nabla \nabla \cdot  B^2\nabla \cdot bT^{L^2-5}  B^2$. 
%(NO NEED TO INTEGRATE BY  PARTS THE NABLA SINCE IT IS BOUNDED IN NORM ANYWAY)
%A moment's reflection then shows that  the largest contribution to $\nabla\nabla \cdot \Pi$ 
%using (\ref{EPi1})-(\ref{symme}) comes from
%$\Pi= B^2\nabla \cdot bT^{L^2-5}B^2$ in the form $B^3\nabla \cdot bT^{L^2-5}\nabla_y \cdot B^2$ where $\nabla_y$
%acts from the right, as in (\ref{symme}).
 Its norm is bounded by $$C(L)L^{n} \epsilon \epsilon_{n}^{3/2-4\gamma},$$
 %I GET FOUR GAMMA NOT FIVE
 using (\ref{f31})  for $\nabla \cdot  B^2$,   (\ref{grads}) for $\nabla \cdot b$, (\ref{f29})  for $ B^2$ and 
%L^{-n(\hf-3\gamma-1+(1-\gamma))}$$
bounding the norm of the remaining $\nabla$ by $CL^n$.
Since $\gamma<1/12$ this is bounded  by $\epsilon_{n+1}^\eta$
for $\eta$ small. \halmos

Estimate (\ref{53}) implies that the Fourier transform $\hat\beta(k)$  is analytic
in $|\ima k|<2\lambda$ %I PUT 2 LAMBDA AND TAKE A SMALLER STRIP BELOW 
and bounded there by $ \ep_{n+1}^{\eta}$.
By the symmetry assumption (iii) in Section 2.5 and the fact that $\hat\beta(0)=0$, which follows from (\ref{51})
 and $\hat T(0)=1=\hat T'(0)$, the Taylor expansion
reads
\rr
\hat\beta(k)=\zeta k^2 +\CO(|k|^4).
 \label{54a}  
 \rrr
 By a Cauchy's estimate
%   $(u$  is a unit vector)  and (48),(52)
\rr
|\zeta |\leq C \ep_{n+1}^{\eta},
   \label{j26}
   \rrr
and 
\rr
|\hat \beta(k)-\zeta k^2 | \leq  C \ep_{n+1}^{\eta} |k|^4,
   \label{j27}  
\rrr
for  $|k| \leq 1$, in the strip $|\ima k|<\frac{3\lambda}{2}$.
 $\zeta$   will ``renormalize" the effective
diffusion constant. A convenient way to keep track of this "marginal" variable
in the RG flow is the following inductive representation:
\lem{Tdeco} $T_n$ can be decomposed as
\rr
  T_n(x)=\rho_n^{-d}\ \CT_n({x}/{\rho_n})+t_n(x)
   \label{Tndeco}  
\rrr
where  $\CT_n$ is defined in (\ref{bite1}), $\rho_n$ is a convergent sequence 
and  there exists a constant $A$ so that 
\qq
 \|t_n\|_\lambda+ \|\nabla t_n\|_\lambda \leq A \ep_n^{\eta}
\label{tnnorm}
\qqq
with $\hat t_n(k)=\CO(|k|^4)$ at origin.
\elem
\no{\bf Proof.} We proceed by induction, with primes referring to scale $n+1$. Set
\qq
\rho'^2 = \rho^2- {2d\zeta  D_0^{-1}}.
   \label{j28}
    \qqq
    with $\rho_0=1$.
The first term in (\ref{51b}) can be written as
\begin{equation}
 L^d{\mathcal S}_LT^{L^2}=L^d{\mathcal S}_L(\tilde{{\mathcal T}}+t)^{L^2}=\rho_n^{-d}\ \CT'({\cdot}/{\rho_n})+\tau
\label{51a}
\end{equation}
where $\tilde{{\mathcal T}_n}({\cdot})=\rho_n^{-d}{\mathcal T}({\cdot}/{\rho_n})$, and
\begin{equation}
{\tau} =L^d \sum^{L^2}_{m=1} ( ^{L^2}_m ){\mathcal S}_L {t}^m  \tilde{{\mathcal T}} ^{L^2-m}
 \label{001}
 \end{equation}
 
 Letting
 \rr
{ r}:=\rho^{-d}{\mathcal T'}({\cdot}/{\rho})-{\rho'}^{-d}{\mathcal T}'({\cdot}/{\rho'}),
   \label{j30}
    \rrr
  %  I CHANGED SIGNS AND ADDED A LINE IN (\ref{j30a}) 
    whose Fourier transform is, using (\ref{z1}):
    \rr
{\hat r}(k) := \hat{\mathcal T}'(\rho k)- \hat{\mathcal T}'(\rho'k) =-\frac{D_0}{2d} k^2 ({\rho}^{2}-{\rho'}^{2})+\CO(|k|^4)= -\zeta k^2+\CO(
|k|^4),
   \label{j30a}
    \rrr
  we get from  (\ref{51b})  
and  (\ref{51a})  that $T'= {\rho'}^{-d}{\mathcal T}'({\cdot}/{\rho'}) +t'$, with 
 \rr
{t}' = {r}+  \tau + \beta.
\label{j32}
\rrr
Using (\ref{j30a}),   (\ref{54a}) for $r+\beta$, and using $\hat t (k)=\CO(|k|^4)$ in 
(\ref{001}), we get that $\hat t'=\CO(|k|^4)$.

By (\ref{j26}), (\ref{j28}), 
\rr
|\rho'-\rho|\leq C \ep_{n+1}^{\eta},  
\label{j30aa}
    \rrr
so that the sequence $\rho_n$ is convergent.

In order to iterate (\ref{tnnorm}), consider first $ r$.     As in the derivation of (\ref{5333}) and using  (\ref{j30a}), (\ref{j30aa}),
    we get
\qq
 \|r\|_\lambda+ \|\nabla  r\|_\lambda \leq C \ep_{n+1}^{\eta}.
\label{rnorm}
\qqq
As for $\tau$, separate the linear part in $t$ in  (\ref{001}):
\begin{equation}
{\tau} =L^d {L^2}{\mathcal S}_L {t}  \tilde{{\mathcal T}} ^{L^2-1}+\tilde\tau := \tau_0 +\tilde\tau,
 \label{taudeco}
 \end{equation}
 with
 $$
 \tilde\tau=L^d \sum^{L^2}_{m=2} ( ^{L^2}_m ){\mathcal S}_L {t}^m  \tilde{{\mathcal T}} ^{L^2-m}.
 $$
 This last term is of the same form as  $\Pi$ in (\ref{Ndec}) and is bounded using the
 inductive hypotheses (\ref{tnnorm}) and (\ref{5333}),
 \qq
 \|\tilde\tau\|_\lambda+ \|\nabla \cdot \tilde\tau\|_\lambda \leq C(L) \ep_n^{2\eta} \leq   \ep_{n+1}^{\eta}.
\label{tilnorm}
\qqq
for $\ep$ small. 
%The bounds (\ref{53}),   (\ref{rnorm})  and  (\ref{tilnorm})
%imply that (\ref{tnnorm}) will iterate provided 
We will now show that the term $\tau_0$ in (\ref{taudeco}), linear 
in $t$,  contracts. %REWROTE THIS
 This happens since
$\hat t=\CO(|k|^4)$. Indeed, in Fourier space, the first term $\tau_0$ in (\ref{taudeco})
equals
\begin{equation}
\hat\tau_0(k)={L^2}\hat{t}  ({_k \over^L}) \hat{{\mathcal T}}  (\rho{_k \over^L})^{L^2-1}
\non% \label{001}
 \end{equation}
 (\ref{tnnorm}) implies that $\hat t$ is analytic in $|\ima k|l<\la$ and
 $$
  \sup_{|\ima k|l \leq \la}(1+|k|)|\hat{t}  (k)|\leq C (\|t\|_\lambda +\|\nabla t\|_\lambda).
 $$
 %WE NEED $(\|t\|_\lambda +\|\nabla t\|_\lambda)$ HERE
 Hence,
 by a Cauchy estimate, and the fact that $\hat t=\CO(|k|^4)$, 
 $$
  \sup_{|k|\leq {L\la\over 2}}(1+|k|)|\hat{t}  ({_k \over^L})|\leq  CL^{-3}
 (\|t\|_\lambda +\|\nabla t\|_\lambda).
 $$
By Lemma \ref{CTlemma},
$$
 \sup_{|k|\geq {L\la\over 2}} \hat{{\mathcal T}}  (\rho{_k \over^L})^{L^2-1}\leq e^{-cL^2}.
$$
We conclude, using the obvious bound $|\hat{{\mathcal T}}  (\rho{_k \over^L})| \leq 1$ for  $|k|\leq {L\la\over 2}$,
%and  $(1+|k|)|\hat{t}  ({_k \over^L})|\leq \|t\|_\lambda$ for  $|k|\geq {L\la\over 2}$,
 that:
 $$
  \sup_{|\ima k |  leq  {L\la\over 2}}(1+|k|)|\hat{\tau}_0  (k)|\leq CL^{-1} (\|t\|_\lambda +\|\nabla t\|_\lambda),
 $$
which implies the desired contraction, using (\ref{tnnorm}) inductively:
\qq
 \|\tau_0\|_\lambda+ \|\nabla  \tau_0\|_\lambda \leq  CL^{-1}(\|t\|_\lambda +\|\nabla t\|_\lambda) \leq \frac{1}{2} A\ep_{n+1}^\eta.
\label{tilnorm2}
\qqq
Using (\ref{rnorm}), (\ref{taudeco}), (\ref{tilnorm}), (\ref{tilnorm2}), (\ref{53}) to bound (\ref{j32}), and choosing $A$ so that
 $$
 C+2 \leq \hf A,
 $$
 for $C$ coming from (\ref{rnorm}),
we get the iteration of  (\ref{tnnorm}).  \halmos

%Finally, for $n\leq n_0$ the previous estimates change as follows.   (\ref{53}), (\ref{j26}) and
%(\ref{j30aa})
%become $C(L)\epsilon^2$ and  (\ref{tnnorm}) $A(L)\epsilon^2$. As $n_0$ was $L$
%dependent for small $\epsilon$ we may iterate up to $n_0$ and satisfy the
%inductive assumption there. 
To finish the proof of Proposition \ref{T} we need to check (\ref{Tnorm1}). Write, using  (\ref{Tndeco}), $T_n^t= 
(\rho_n^{-d}\ \CT_n)^t+\sum_{i=1}^t t^i_n (\rho_n^{-d}\ \CT_n)^{t-i}$
It is readily verified, using  (\ref{z1}), (\ref{z2}), that $ (\rho_n^{-d}\ \CT_n)^t$ satisfies (\ref{Tnorm1}).  Now, use (\ref{5333}), (\ref{tnnorm}),
to bound $\| t^i_n (\rho_n^{-d}\ \CT_n)^{t-i}\|_\la \leq  A \ep_n^{ \eta } C^{t}$, for any $i \geq 1$. Since $t\leq 2L^2$, 
we may bound $(t-1)  A \ep_n^\eta C^t \leq (t+1)^{-d/2}$, which proves the bound (\ref{Tnorm1}) also for $\sum_{i=1}^t t^i_n (\rho_n^{-d}\ \CT_n)^{t-i}$, since a bound on  the $\|\cdot\|_\la$
norm implies the exponential decay in  (\ref{Tnorm1}).
%THE BOUND ON THE SUM WORKS ONLY FOR $ k \leq \la \sqrt {t+1}$ because the t part does not have better decay.
  \halmos

\section{The nonlinear analysis.}

In this Section we will bound the nonlinear part of the map $f_t$. Recall that
we wrote in Sect. 3  (\ref{defbeta}),
\begin{equation}
 Df_t (E) = T + \nabla b_t (E) 
\label{f1}
\end{equation}
and in Section 4 we have studied $b_t(0)$ and the composite $b_t(0)\nabla b_t(0)$.
 Now we need to extend that analysis to
the $E$ dependence. Let 
$B^1_t(E)=b_t(E)$ and $B^2_t(E)=b_{t+1}(f_t(E))\nabla b_t(E)$, and $B^3=\nabla \cdot B^2$, $B^4=T$, $B^5=\nabla T$ as before.
It turns out that for the nonlinear part we only need to keep track of the localization of $B_t$ in the region $D_t$:
we will establish inductively a decomposition
\rr
B^\alpha_t(E)=\sum_{\CD\subset D_t}B^\alpha_{t, \CD}(E).
\label{bEdef}
\rrr
Of course we have $B^\alpha_{t, \CD}(0)=\sum_{A}B^\alpha_{t, \CD A}$.
%We will continue writing $b_t=s_t+\ell_t$ where  $s_T(E)=b_{t\emptyset}(E)$.

\subsection{Inductive definition of $B$} We wish to show that the nonlinearity in
$B^{\alpha}$ is irrelevant under the RG iteration. For $\alpha=2$ the reason for
this will be the same as for the $E=0$ part. However,  for $\alpha=1$ i.e. for $b$,
we need to show the irrelevancy of its derivative $Db$ as explained in Sect. 3.6. 
To study its iteration consider the iterative formula (\ref{LRG1}) for the
 second derivative $D^2f$. We have outlined in Sect. 3.6. the argument that on
 the linear level  $Db$ contracts. However, we face again the problem that
 in the nonlinear terms $\nabla b$ and $\nabla D b$ will enter and they are not small in our norm. 
 Hence as for the iteration of $b$ we need to iterate bounds for composites involving
 $Db$. We denote $Db$ by $\CC^1$. The composites that are analogous to $B^2$ are
  \qq
 (\CC_t^2,\CC_t^3,\CC_t^4) :=
 %(b\nabla\CC, \nabla b\nabla\CC, \CC(\nabla b\otimes 1),\CC(\nabla b\otimes \nabla b), \nabla%\CC(\nabla b\otimes 1), \nabla\CC(\nabla b\otimes \nabla b))
 (b_{t+1}\nabla \cdot Db_t,  Db_{t+1}(\nabla \cdot b_t\otimes 1),Db_{t+1}(\nabla \cdot b_t\otimes \nabla \cdot b_t)),
 \label{compo}
 \qqq
 %WE NEED A CONVENTION ON THE INDEX t FOR  (\ref{f25yy}) below.
 where we use the notation (\ref{f10a}),
 %which we denote by $(\CC^2,\dots,\CC^7)$.
 and the ones analogous to $B^3$ are their divergences $\nabla\cdot\CC^\al:=\CC^{\al+3}$, for $\al = 2,3,4$.
 
Let us introduce some notation:  
$I$ stands for a set $\{(\alpha_i,t_i)\}$ with $t_i\in[0,2L^2-1]$ and $\alpha_i\in\{1,\dots,5\}$.
Denote
\begin{equation}
B_I:=B^{\alpha_1}_{t_1}(E_{t_1})\dots B_{t_N}^{\alpha_N}(E_{t_N})
\label{Ndec1}
\end{equation}
with, for $t_i\geq 1$,
 \rr
 E_{t_{i}}=f_{t_{i}-1}\circ\dots\circ f_0 (E)\ \ \  {\rm with} \ \ \ E=L^{-d}E'(\cdot/L).
 \label{eti}
 \rrr
 
With this notation
\rr
Df_{t'}'(E')=L^{d}{\mathcal S}_LT^{L^d}+\nabla\cdot\CM^1(E')
\label{dfprime}
\rrr
where $\CM^1$ is a sum 
\rr
\CM^1=L^{d-1}\sum_I{\mathcal S}_LB_I
\label{CMsum}
\rrr
 Thus,  see (\ref{f1}), (\ref{Tdef}),
\rr
b'(E')=\CM^1(E')-\E\CM^1(0).
\label{dfprime1}
\rrr
Similarly we have
\begin{equation}
B'_{t'}=(\CL B)_{t'}+\CM^2(B)_{t'},
\label{BB'}
\end{equation}
where we write explicitly the linear term, with  $\CL$ given by the analogue of (\ref{f54}), 
\begin{equation}
(\CL B)_{t'}(E') := L^{d-1} {\mathcal S}_LT^{L^2-1}   B_{L^2(t'+1)-1}(E_ {L^2(t'+1)-1})T^{L^2-1}.
\label{f544}
\end{equation}
%and $\CL_3=0$ . 

Let us next derive the recursion relation for the $\CC^\al$. Consider first
$\CC^1$  i.e. $Db$. 
Inserting to  (\ref{LRG1}) the decomposition (\ref{f1}) and expand as
in (\ref{rprime}) and (\ref{rem1}).  This way ${\CC'}^1$ becomes a sum
of terms of the form
$$
L^{d-1} {\mathcal S}_LQ_{\pi} \CC_t(\nabla Q_{\pi'}\otimes \nabla Q_{\pi''})
$$
where 
$$
Q_{\pi}=
T_{\pi_k}b_{\tau_{k-1}}(\nabla\cdot b)_{\pi'_{k-1}}\nabla\cdot T_{\pi_{k-2}}\dots b_{\tau_{2}} (\nabla\cdot b)_{\pi'_{2}}T_{\pi_1},
$$
as in (\ref{rem1}). We may express this again in terms of the products (\ref{Ndec1}) and the
composites invoving $Db$ (\ref{compo}) and their derivatives:
\qq
{\CC'}^1=\tilde\CL_1 \CC^1+L^{d-1}\sum_{\{I_i\}}{\mathcal S}_LB_{I_1} \CC_{I_2}(B_{I_3}\otimes B_{I_4}).
 \label{compo1}
 \qqq
 where $\tilde\CL_1 =\tilde\CL$  defined in (\ref{f41}) and where
 $I_2=\{(\alpha,t)\}$, $\alpha\in\{1,\dots,7\}$. It is easy to see that all seven of them
 can occur!

%As in the case of
 %the $b$ iteration there are some obvious constraints for $I_1$
 %and $I_2$. WHERE DO WE USE THIS? Thus, if  $I_1\neq \emptyset$ then  
 %$\alpha_1\neq 3, 5$ (NOT EQUAL TO 5 ALSO) for $(\al_1,t_1)\in I_1$ with the largest time $t_1$, because
 %formula (\ref{rem1}) does not start with a $\nabla$. 
 %Similarly 
 %if $t_k$ is the smallest time
 %with $(\al_k,t_k)\in I_1$ and if $\al_k=4$ then
  %$\al\in \{1,\dots,4\}$ for $I_2=\{\alpha,t\}$, otherwise the $\nabla$ in $\CC^\al$, $\al>4$, would be integrated by parts and we would have
  %$\al_k=5$. The same holds if $I_1=\emptyset$.

 As in the case of $B^2$ we obtain for $\CC^\al$ with $\al>1$ 
an expansion as in eq.  (\ref{compo1}) on the time interval  $[0,2L^2-1]$.
For $\al=2,5$ the times in $I_1$ lie on $[L^2, 2L^2-1]$ and those
on $I_k $ $k>1$ lie on $[0,L^2-1]$ and, for  $\al=3,4,6,7$, we have $I_1, I_2 \subset [L^2, 2L^2-1]$, $I_3, I_4 \subset [0, L^2-1]$.

 We have, since $\nabla$ acts on a $T$, as in (\ref{N4''}), that:
\qq
\tilde\CL_\alpha=0,\ \ \  \alpha>4.
\label{nolin}
\qqq
%For $\alpha\in \{2,3,4\}$ the $I_1$
 %and $I_2$ satisfy the same constraints as in the case  $\alpha=1$.

We also need to localize
\rr
\CC^1_t(E)=\sum_{\CD\subset D_t}\CC^1_{t, \CD}(E)
\label{cEdef}
\rrr
and similarly for the $\alpha>1$ with $\CD=(\CD_t, \CD_{t+1})$ as in  (\ref{Bloc}).
We will now explain how (\ref{bEdef}) and (\ref{cEdef}) are carried through the
induction.

First write, as in (\ref{btdec}) and (\ref{btilde})
\begin{equation}
B^\alpha_{t_1}=\sum_{\CD_i}B^\alpha_{t_i, \CD_i}=\sum_{\tilde \CD_i}
\tilde 
B^\alpha_{t_i, \tilde \CD_i }
\label{btdec1}
\end{equation}
%with
%\begin{equation}
%\tilde 
%B^\alpha_{i\tilde \CD_i}:={\sum_{\CD_i\subset D_{t_i}}}\hspace{-2mm}'\hspace{2mm}
%B^\alpha_{tD_iA_i}
%1(D^t\cap
%LD_{t'}'=\tilde D_t)1(A\cup (D^t\setminus\tilde D_t)=\tilde A_t)
%\label{btilde}
%\end{equation} with the constraints 
%\begin{equation}
%\bar\CD_i\cap
%LD_{t'}'=\tilde \CD_i
%\label{constraint}
%\end{equation}
%and in the same way for $B_{t_i}$ and $\NB_{t_i}$. 
This way we end up with
\rr
\CM^1(E')=L^{d-1}\sum_J{\mathcal S}_L\tilde B_J
\label{cme'}
\rrr
with $\tilde B_J
=\tilde B^{\alpha_1}_{t_1, \tilde \CD_{1}}(E_{t_1})\dots \tilde B_{t_N, \tilde \CD_{N}}^{\alpha_N}(E_{t_N})$
and $J$ stands for the set $\{(\alpha_i,t_i,\tilde \CD_{i})\}$. 

Localize (\ref{cme'}) by collecting all terms with $[\cup \tilde \CD_i]=\CD'$
\rr
\CM^1(E')=\sum_{\CD'}\CM^1_{\CD'}(E'),
\label{cme''}
\rrr
which we write below as $\CM^1_{\emptyset}(E')+\sum_{\CD'\neq \emptyset}\CM^1_{\CD'}(E')$.
In the same way we localize
\rr
{\CC'}^\alpha=L^{d-1}\sum {\mathcal S}_L\tilde B_{J_1} \tilde\CC_{J_2}(\tilde B_{J_3}\otimes
\tilde B_{J_4})=\sum_{\CD'}{\CC'}^\alpha_{\CD'}
\label{db'e'}
\rrr
where again  $J_2$ is the singleton $\{(t_2,\alpha_2,\tilde\CD_2)\}$ and we included the
linear term in (\ref{compo1}) to the sum as well.

Localization of $b'$ is less obvious. Basically the idea is that $b'(E')-b'(0)$
contracts (unlike $b'(0)$!) because it is controlled
by $Db'$ which contracts. However, from (\ref{compo1}) we
see that $Db'$ has terms that are {\it quadratic} in $B_t^\al$
for $t\in J_3,J_4$. Hence the bound for $Db'$ will involve
$e^{2N_{t'\CD'}}$ instead of the $e^{N_{t'\CD'}}$ factor in the
bound of $b'$ (see eq. (\ref{f25yy}) below). Thus we need to be
very careful not to propagate such factors from $Db'$ to $b'$.

 Let $\CJ_0$ be the set of $J$
with all $\tilde\CD_i=\emptyset$.  Then
\rr
D\CM^1_\emptyset=L^{d-1}\sum_{J_1, J_2,J_3\in \CJ_0} {\mathcal S}_L\tilde B_{J_1} \tilde\CC_{J_2}(\tilde B_{J_3}\otimes
\tilde B_{J_4})=\sum_{\CD'}{\CN}_{\CD'}.
\label{CN}
\rrr
 We set $b'_{t', \emptyset}(E')={\CN}_{\emptyset}( E')$, which we write as:
\begin{equation}
b'_{t', \emptyset}(E')-b'_{t', \emptyset}(0)=\int^1_0 d\la {\CN}_{\emptyset}(\la E')E'
\label{bploc1}
\end{equation}
and for $\CD'\neq\emptyset$,
\begin{equation}
b'_{t', \CD'}(E')-b'_{t'\CD'}(0)=\int^1_0 d\la {\CN}_{\CD'}(\la E')E'+\CM^1_{\CD'}(E')-\CM^1_{\CD'}(0).
\label{bploc2}
\end{equation}
Denote the $\CD=\emptyset$ and $\CD\neq\emptyset$ parts of $b(E)$ by
$$
m_t(E)= b_{t, \emptyset}(E)-b_{t, \emptyset}(0)
$$
 and 
 $$
 M_t(E)= \sum_{\CD \neq \emptyset} b_{t, \CD}(E)-b_{t, \CD}(0),
 $$
 so that we have:
 \qq
 b_t(E)-b_t(0)= m_t(E) + M_t(E),
 \label{jb14}
 \qqq
and
\qq
 b_t(E)= s_t+\ell_t + m_t(E) + M_t(E).
 \label{jb141}
 \qqq

Note that these definitions avoid the problem with the $N$ factors since
the only contribution to $b'$ from $Db'$ is from (\ref{CN}) where only the $J_4$
term
can contribute a $e^{N_{t, \CD}}$ factor. $\CN$ will contract for reasons stated above and
$\CM^1_{\CD'}(E')$ for $\CD'\neq\emptyset$ will contract 
for the same reasons that  $\ell= \CM^1(0) -\E (\CM^1(0))$ contracted.

\subsection{Inductive bounds}
The analysis is now similar to  the $E=0$ case, if only it is simpler due to less
localization. 
 We use the norm, analogous to (\ref{f24}), but without the $A$
sum:
\begin{equation}
  \|B^\alpha_t\|_\lambda = \sup_{E\in B^+_\delta}
  \sup_{v\in \mathbb Z^d} \sum_{\CD,u }%\subset \mathbb Z^d \atop
 %v \in \mathbb Z^d} 
    \|B^\alpha_{t, \CD} \|_{u,v}e^{ \lambda \tau(\CD,u,v)}  e^{-N_{t, \CD}},
\label{f24new}
\end{equation} 
where, for $\al=2$, $\CD$ is a pair $(\CD_1, \CD_2)$, with the same conventions as in (\ref{f24}).

Note that since $f_t$ preserves $\int E dx$  the arguments (\ref{eti}) of the functions $B$ in (\ref{Ndec}) 
are in  $B^+_\delta$ if $E$ is. For convenience we also choose $\delta=\epsilon$.

%\begin{equation}
% \bar{B}_{t\CD} = B_{t\CD} e^{-N_{t\CD}}.
%\label{f25x}
%\end{equation}
Then we have:
\prop{sell3}
Let  $\gamma$ be as in Proposition \ref{sell}. Then, for all $t, n$,  
 \begin{eqnarray}
 \| m \|_\lambda & \leq & \ep_n^{(1-2\gamma)}  \label{f26c}
\\  \| \ {M}\  \|_\lambda& \leq & \ep_n^{(1-2\gamma)}.  \label{f27aaa}
\end{eqnarray}
\qq
  \| B \|_\lambda &\leq& \ep_n^{(1-\gamma/2)} 
\label{f29a}\\
\| B^3   \|_\lambda =   \|\nabla\cdot B   \|_\lambda &\leq &\ep^{1-2\ga}L^{-n/2}.
\label{f31a}
\qqq
\eprop
As indicated above for the $\CC^\alpha$ we have to use a slightly different definition of the norm:
\rr
\| {\CC}^\alpha_t\|_\la = \sup_{E\in B^+_1} \sup_{v, w\in \mathbb Z^d} \sum_{\CD, u }
%\subset \mathbb Z^d \atop
 %v \in \mathbb Z^d} 
    \|{\CC}^\alpha_{t, \CD} \|_{u,v, w}e^{ \lambda \tau(\CD,u,v,w)} e^{-N^\alpha_{t, \CD}}
\label{f25y}
\rrr
with $ \|C \|_{u,v, w}= \sup_{x \in {\bf u}, y \in {\bf v}, z \in {\bf w}}|C(x,y,z)|$, and 
\qq
N^1_{t, \CD}&=&2N_{t, \CD_t},\non\\
N^2_{t, \CD}&=&2N_{t, \CD_t}+N_{t+1,\CD_{t+1}},%\ \alpha\in\{2,5\}
\label{f25yy} 
\\ 
N^\alpha_{t, \CD}&=&N_{t, \CD_t}+2N_{t+1,\CD_{t+1}},\ \alpha\in\{3,4\}.
\non
\qqq
Note that we introduce a $2N_t$ instead of $N_t$ for the times $t$ where  a $Db$ occurs.
%where $N_{tD}$ is defined in (\ref{}).
The bounds for the $\CC$ are:
\prop{sell4}
For all $t, n$,
 \begin{eqnarray}
 \| Dm \|_\lambda & \leq & CL^{-(1-\gamma)n}  \label{ff}
\\  \|  DM\  \|_\lambda& \leq &  CL^{-(1-\gamma)n}.  \label{ff27}
\end{eqnarray}
Moreover, for $\beta=2,3,4$
\qq
  \| \CC^\beta \|_\lambda &\leq&    \ep^{(1-2\gamma)} L^{-(1-\gamma)n} ,%\ \ \beta=2,3,4
\label{fff}\\
 \| \CC^{3+ \beta } \|_\lambda =  \| \nabla\cdot \CC^\beta   \|_\lambda &\leq &  \ep_n^{1/2}.%\ \ \beta>4
\label{ffff}
\qqq
\eprop
\begin{remark} From  (\ref{f26c}-\ref{f27aaa})  and (\ref{ff}-\ref{fff}) we get, as in (\ref{grads})),
 \qq
\|\nabla \cdot b\|_\la  \leq C \epsilon \ep_n^{-2\gamma}.
  \label{jb15}
\qqq
and
\qq
 \|\nabla \cdot Db\|_\la \leq C L^{\gamma n}.
  \label{jb150}
\qqq
Note also that we cannot deduce (\ref{f26c}-\ref{f27aaa})  from (\ref{ff}-\ref{ff27}) by integration, because of the difference between the
norms (\ref{f24new})  and (\ref{f25y}).
 \end{remark} 
\subsection{Proof of the Propositions}  For $n=0$, the bounds in Proposition \ref{sell3} follow by writing, see  (\ref{f1}),
$$
m(E)= 
Df(E)-DF(0)= \int_0^1 d \mu D^2F (\mu E) E
$$
  using assumption (iv), $\delta=\epsilon$ and the fact that, for $n=0$, the norm of $\nabla$ is of order one.
  We have, for $n=0$, $D = \emptyset$, $M=0$.  The bounds in Proposition \ref{sell4},
  follow similarly, from assumption (iv) for $Dm$, and by combining bounds on $m$ and $Dm$ for $\CC^\beta$.
 %the fact that $\|E\|_1 \leq \delta$ and choosing
%$\delta^2 \leq \ep$. 

The iteration of the bounds in the propositions
goes as in Section 4.10. except that it is simpler due to the lack of the $A$ sums.
Hence, we will  be brief.

The bound (\ref{f29a})  follows as in Section 4.10, see  (\ref{lb2}). The
supremum over $E$ in the norms comes for free since for all $t$ we have $E_t\in B_\delta^+$.
Thus, e.g. the operator (\ref{f544}) has the same bound (\ref{lb2}) as in the $E=0$ case.
The proof of (\ref{f31a}) follows the one of (\ref{f31}), see (\ref{N4'}), (\ref{N4''}). Since the bounds here are different from the ones in Proposition \ref{sell}, we have, instead of (\ref{N4'}),
using  (\ref{jb15}),  (\ref{f26c}),  (\ref{f27aaa}),
 \rr
\|\nabla\cdot \CN^2\|_{2\lambda}
\leq C(L)\ep 
 \ep_n^{1-4\gamma}= C(L) \ep^{2-4\ga} L^{-(1-4\ga) n} \leq \hf \ep^{1-2\ga}L^{-(n+1)/2},
\label{N41}
\end{equation}
since $\gamma<\frac{1}{12}$ and $\epsilon$ is small. 
The bound on $\nabla\cdot \CL B^2$ is similar to  (\ref{N4''}).

Let us consider first  the nonlinear terms in the iteration of (\ref{f26c}-\ref{f27aaa}) and (\ref{ff}-\ref{fff}).

Note that, with the norms (\ref{f24new}), (\ref{f25y}), and with (\ref{f10a}), we have:
\qq
&&\sup_{v,w}\sum_{\CD,u }
    \|{\CC^\al (B_1 \otimes B_2})_{\CD} \|_{u,v, w}e^{ \lambda \tau(\CD,u,v,w)}  \leq 
\non\\
&&    C\sup_{z_1, z_2}\sum_{\CD ,u}
    \|{\CC}^\alpha_{\CD} \|_{u,z_1, z_2}e^{ \lambda \tau(\CD, u, z_1,z_2)}\cdot
    \non\\
&&  
\sup_{v} \sum_{\CD_1,z_1 }
    \|B_{1\CD_1} \|_{z_1, v}e^{ \lambda \tau(\CD_1,z_1,v)} \sup_{w}\sum_{\CD_2,z_2 }
     \|B_{2\CD_2} \|_{z_2, w}e^{ \lambda \tau(\CD_1,z_2,w)}.
\qqq
With this, one can prove the analogue of Lemma \ref{T1}: first, the analogue of (\ref{barPi}), to bound products of the form 
 $B_{I_1} \CC_{I_2}(B_{I_3}\otimes B_{I_4})$ in (\ref{compo1}), with the norm (\ref{f25y}) on the left hand side
 and a factor given by the norm (\ref{f25y}) of $\CC^\al$ on the right hand side.
 The $\exp (N)$ factors in the norms are dealt with  eq. (\ref{n'bound}), which holds
for the new definition (\ref{f25yy}) as well; since we have defined the new $N_{t', \CD}^{'\al}$ so that we have $2N_{t'}$
whenever there is a derivative, we can control both the $2N_{t}$ associated with $\CC_{I_2}$ and the two $N_t$'s
coming from $B_{I_3}\otimes B_{I_4}$. Secondly,
the analogue of the bound (\ref{barC})
 can be proven for  the norm (\ref{f25y}) of $\tilde \CC$ and the one of $\CC$, as in the proof of Lemma \ref{T1}, since
 the bound
 (\ref{Ni}) holds also
for  (\ref{f25yy}).

The nonlinear terms in the iteration of $Dm$ or $DM$, are given by  (\ref{db'e'}), see
 (\ref{dfprime1}), (\ref{compo1}), (\ref{CN}-\ref{jb141}). The leading terms are when all $B_{I_k}$, $k=1, 3, 4$, 
 %CHANGED PUT $3,4$
  are products of $T$'s and $\nabla T$'s. Then, we have, in $J_2$ $\al_2 \neq 1$ (otherwise, it would be the linear term) and $\al_2 \leq 4$
  (otherwise the $\nabla$ would be integrated by parts to a $T$) and 
  we can use inductively (\ref{fff}) and $C \ep^{(1-2\gamma)}L^{-(1-\ga)n} \leq \hf L^{-(1-\gamma)(n+1)}$, for $\ep$ small. 
    All the other terms give rise to smaller contributions.
  
  The same holds for the nonlinear terms in the iteration of $m$ or $M$, that are also of the form of 
   (\ref{db'e'}), see
 (\ref{CN}-\ref{jb141}), except for the nonlinear terms in $\CM^1$, which are bounded in the same way as the nonlinear terms
 for $E=0$, see (\ref{N3}). We use (\ref{fff}) and $C \ep^{(1-2\gamma)}L^{-(1-\ga)n} \leq \hf \ep_n^{1-2\gamma}$ to bound the 
contributions
 of $\CC^\beta$, $\beta>1$ to $m$ and $M$ (we have $C$ here, not $C(L)$, since there is no sum over times in those contributions). 
 The crucial observation, for the nonlinear terms in $\CN$, is that
 the $\exp (N)$ factors work out: we have only one $N_t$
 in the exponentials in the bound for $M'$ (see (\ref{f24new})) but also in the bound
  for $\CN$, since in (\ref{CN}) only $J_4$ can have $\tilde\CD_i\neq\emptyset$, see (\ref{bploc1}). %CHANGED
 
Consider next the nonlinear contributions to the iteration of
 (\ref{fff})  and start with  $\beta=2$.  Since  one cannot have a $\nabla$ in the leftmost
matrix in the recursion (\ref{compo1}), the leading
nonlinear terms are proportional to 
$$
T^{L^2-2}b_{L^2+1}\nabla\cdot {\CC}^\alpha_{L^2-1} (T^{L^2-1}\otimes T^{L^2-1})
$$
 which, using recursively  (\ref{f26c}, \ref{f27aaa}), \ref{ffff}), is $\CO(\epsilon_n^{\frac{3}{2}-2\gamma})$,  and, for $\gamma<1/12$, this is smaller than  $\hf \ep^{(1-2\gamma)} L^{-(1-\gamma)n}$. We also have terms with $b$ and $Db$ separated by powers of $T$ or $\nabla T$ (the $\nabla$   in $\nabla Db$ being integrated by parts). These terms are bounded by combining
 (\ref{f26c}, \ref{f27aaa}) and  (\ref{ff}, \ref{ff27}) and are much smaller than $\ep^{(1-2\gamma)} L^{-(1-\gamma)n}$. 
 
Similarly, for $\beta=3,4$, the leading nonlinear terms are  $T^{L^2-3}\CC^1_{L^2+1}(B^3_{L^2-1}T^{L^2-1}\otimes 1)$ and $T^{L^2-3}\CC^1_{L^2+1}(B^3_{L^2-1} T^{L^2-1}\otimes B^3_{L^2-1}T^{L^2-1})$. The first one
is, since $\CC^1=Db$, using recursively  (\ref{ff}, \ref{ff27}, \ref{f31a}), 
$\CO(\epsilon^{1-2\gamma} L^{-(\frac{3}{2}-\gamma)n})$ which, for $\gamma<1/12$, is again smaller than  $\hf \ep^{(1-2\gamma)} L^{-(1-\gamma)n}$. 
The second term is smaller.

Next we turn to the linear RG for $\CC^\beta$.  For 
$\beta=1$, i.e. for $Dm$ and $DM$, we get, as in (\ref{f41}):
\begin{eqnarray}
&&(\tilde\CL_1 \CC^1)_{t'}( x^\prime,y^\prime,z^\prime,E') = \nonumber \\
&& L^{d-1}\sum_{t\in I_{t'}}\int dxdydz T^{L^2-i-1} (Lx^\prime-x) \CC^1_{t}( x,y,z,E_t)T^{i} (y-Ly^\prime) T^{i} (z-Lz^\prime).
\label{f41a}
\end{eqnarray}
We can now proceed as in (\ref{linearRGb})- (\ref{f81}) with the difference that in (\ref{f79})
we have also a $w$ sum; in  (\ref{w7}), we have $\frac{\lambda}{12}(|u-v|+|u-w|)$ instead of $\frac{\lambda}{6}|u-v|$, and, here 
\rr
S_i(u,v,w,u',v',w'):=
 \sup_{x^\prime  \in \Nu^\prime, y^\prime \in \Nv^\prime, z^\prime  \in \Nw^\prime}
 \int_{\Nu} dx \tilde{T}^{L^2-i-1} (Lx^\prime-x)  \int_{\Nv} dy \tilde{T}^{i} (y-Ly^\prime) 
 \int_{\Nw} dz \tilde{T}^{i} (z-Lz^\prime), 
 \label{Suvw}
\rrr
with $\tilde T$ and $k$ as in (\ref{Suv}), (\ref{kdefi}).
%where the sup is over ${x^\prime  \in \Nu^\prime, y^\prime \in \Nv^\prime, z^\prime  \in \Nw^\prime}$.
Eq. (\ref{f81}) is then replaced by
\begin{equation}
\sum_{u, v, w} e^{-\frac{\lambda}{12}(|u-v|+|u-w|)}S_i(u,v,w.u',v',w')  \leq C \min (i^{-d},i^{-\frac{d}{2}}(L^2-i-1)^{-\frac{d}{2}})\leq CL^{-d}i^{-\frac{d}{2}},
\non%\label{f81aaa}
\end{equation}
since we can use twice the $L^\infty$ bound in (\ref{w9}) and use one $L^1$ bound and the factor $e^{-\frac{\lambda}{12}(|u-v|+|u-w|)}$
to control the sums.
Summing $i^{-\frac{d}{2}}$ over $i$  leads to the bound, for $d\geq 2$,
\begin{eqnarray}
\|\tilde\CL_1\CC \|_\lambda \leq CL^{-1} \log L \|\CC\|_\lambda,
\label{f41aa}
\end{eqnarray}
The other $\tilde \CL_\al$, $\al= 2, 3,4$, have similar bounds, in fact better ones since there is no sum over times. Combined with the
bounds on the nonlinear terms the iteration of (\ref{ff}) , (\ref{ff27}) and (\ref{fff})  follows.

Consider then (\ref{ffff}). We have again a  term  $\nabla \tilde\CL_\beta \cdot
\CC^\beta$, which can be bounded, using (\ref{fff}), as in (\ref{N4''}).
%THERE ARE NONE, BY FORMULA BEFORE cEdef
 %$\tilde\CL_\beta\nabla\cdot
%\CC^\beta$ which contract
%by $CL^{-1}$. 
Among the nonlinear contributions, 
for $\beta=2$, the largest one is    $$\nabla \cdot b_{2L^2-1}\ \nabla T^{L^2-2} \CC^1_{L^2-1}(T^{L^2-1}\otimes
T^{L^2-1})$$ 
i.e. %$\CO(\ep \epsilon_n^{1-4\gamma})$, I HAVE FOUR GAMMA 
$\CO(\ep^{1-2\gamma} L^{-(1-3\gamma)n})$  
using (\ref{jb15}) and (\ref{ff}, \ref{ff27}); this is smaller than $\hf \epsilon_{n+1}^{1/2}$
% (where as in Section 4.10 the discrete gradient $\nabla$ is bounded by $\CO(L^n)$) which
for  $\gamma<1/12$,  which fits to (\ref{ffff}). For $\beta=3$ 
the leading nonlinear
term is proportional to  
$$\nabla \CC^1_{2L^2-1}(\nabla T^{L^2-1}\otimes
T^{L^2-1})(b_{L^2-1}\ T^{L^2-1}\otimes 1)$$
i.e. again, using (\ref{jb150}) and (\ref{ff}, \ref{ff27}),
$\CO(\ep^{1-2\gamma} L^{-(1-3\gamma)n})$, which is less than $\hf \epsilon_{n+1}^{1/2}$. 
%$\CO(\ep \epsilon_n^{1-4\gamma})$.  
For $\beta=4$ the nonlinear terms are smaller.

To finish the proofs,we need to bound the linear contributions to (\ref{f26c}) and (\ref{f27aaa}). Thus consider eqs.
 (\ref{bploc1}) and  (\ref{bploc2}). 
The linear term comes from   (\ref{CN}) with all the $\tilde B$ being powers of $T$
and $\CC^{\al_2}= \CC^1$; we use the bound (\ref{f41aa}) 
 % $$
% D\CM_\emptyset=\tilde\CL_1Dm+\ \ {\rm nonlinear\  terms}
% $$
%Combining this with the contraction of $\CL_1Dm$ and
together with the estimate
 \begin{equation}
\| {\CN}(\la E')E'\|_\lambda \leq \| {\CN}(\la E')\|_\lambda \|E'\|_1
\label{b'loc1}
\end{equation}
which is obtained by writing 
$$
({\CN}(\la E')E')(x,y)= \int {\CN}(\la E')(x, y, z)E'(z) dz,
$$
$\int dz = \displaystyle \sum_{w\in \mathbb Z^d} \int_{\Nw} dz$ and $ \int_{\Nw}  E'(z) dz \leq \|E'\|_1$. We then
use in (\ref{b'loc1}) $ \|E'\|_1=  \|E\|_1 \leq \delta=\epsilon$, 
where
the first equality follows inductively from assumption (ii), and the bounds (\ref{ff}) on $\CC^1$ for $\CN$. 
This controls the linear contribution of  the
 $\CN$ terms in  (\ref{bploc1}) and  (\ref{bploc2}) and, with the previous bounds on 
 the nonlinear terms in $\CN$ and the $\CM^1$ terms in  (\ref{bploc2}), proves (\ref{f26c}) and (\ref{f27aaa})
 \halmos
%in (\ref{f26c}) and (\ref{f27aaa}). The crucial observation is that
% the $N$ factors work out since in (\ref{CN}) only $J_4$ can have $\tilde\CD_i\neq\emptyset$.

%The $\CM$ terms in  (\ref{bploc2}) are bounded as in the $E'=0$ case. \halmos

\section{Proof of the main results.}

We will prove (\ref{result}) for
\qq
D=\lim_{n\to \infty}\rho_n D_0
\label{defD}
\qqq
where $\rho_n$ is the sequence in Lemma \ref{Tdeco}.

 \par\noindent
 {\bf Proof  of Theorem \ref{main}}
%Consider first $t=1$ in (\ref{result}), the subsequence  $\Lambda_n=L^{n}$ and write  $\Lambda= \Lambda_n$.
%NEED TO GENERALIZE LATER.
Using (\ref{jb4}) and (\ref{jb5}) with $t=0$, we have, $\forall x \in (L^{-n}\Z)^d$,
 \qq
L^{nd} E (L^{2n}, L^{n} x) =E_n(1,x)= \int_0^1 d\lambda  \int dy D f_n (\lambda E_n) (x,y) E_n (y),
\label{Df2}
 \qqq
 with $E_n(y)=: E_n (0, y)= L^{nd} E(0, L^ny)$ and $D f_n=: D f_{n,0}$, where, see
 (\ref{jb7}), 
 \qq
 Df_n(\lambda E_n) (x,y) = T_n(x-y)+ \nabla \cdot b_n (\lambda E_n) (x,y),
  \label{jb11}
 \qqq
with $b_n$ at $t=0$.
 Let 
 \rr
 \overline \Omega = \{ \om | \exists m,\mbox{such that } \forall \CD \subset \Z^d, \forall n \geq m, N_n (\CD) \leq  \la \tau( \CD \cup 0 ) \}
\label{p13}
\rrr
By Proposition \ref{prob1}, and writing $K_n\geq cK$, $\forall n$, we get, for any set $\CD  \subset \Z^d$:
%since $N^n_z$ is defined only for $|z|\leq L^n$, TO BE SAID EARLIER, we get
\qq
&&\mathbb{P} (N_n (\CD) \geq \la \tau( \CD \cup 0))
\non \\
&&  \leq \sum_{(N_x)_{x\in \CD}, \sum_{x\in \CD} N_x \geq \la \tau( \CD \cup 0)} e^{-cKn}\prod_{x\in \CD} e^{-cKN_x },
\qqq
since at least one $N_x \neq 0$. So, writing $e^{-cKN_x}= e^{-N_x}e^{-(cK-1)N_x}$, we get,  for $K$ large enough (i.e. for $\ep$ small enough),
\qq
&&\mathbb{P} (N_n (\CD) \geq \la \tau( \CD \cup 0))
\non \\
&&  \leq \exp (-cKn) \sum_{(N_x)_{x\in \CD}}\prod_{x\in \CD} e^{-(cK-1)N_x} \exp(-\la \tau( \CD \cup 0))
\non \\
&&   \leq \exp (-cKn) (1+e^{- cK/2})^{|\CD|} \exp(-\la \tau( \CD \cup 0))
\non \\
&&   \leq \exp (-cKn)  \exp(-\la \tau( \CD \cup 0)/2)
\qqq
 %for $K$ large enough.
  Since $\sum_{\CD \subset \Z^d}  \exp(-\la \tau( \CD \cup 0)/2) \leq C$,  $\sum_n  \exp (-cKn) \leq C$,
we get from  the first Borel-Cantelli lemma, that
 \rr
 \mathbb{P} (\overline \Omega) = 1.
\label{p14}
\rrr

Since $\|E\|_1$ is bounded and $E_n(y)=L^{nd} E(0, L^ny)$ we have
 \qq
 \lim_{n \to \infty} \int dy  \mathbbm{1} (|y| \geq L^{-n/2})E_n (y)=0,
    \label{jbb}
 \qqq
 
Let $\om \in \overline \Omega$.  Then, for $|y| \leq L^{-n/2}$, we get $y \in {\bf 0}$, and, $\forall u$ and $n$ large enough,
 $$
 -N_n (\CD) + \la \tau( \CD \cup u \cup 0) \geq 0.
 $$
 
 Then, from
 (\ref{jb141}, \ref{f24},  \ref{f24new}) and the bounds (\ref{f26}-\ref{f27}, \ref{f26c},\ref{f27aaa}), we  get that,
  for  $|y| \leq L^{-n/2}$,
 $t=0$,
\qq
\int dx |b_n(x,y)|\leq C \epsilon_n^{(1-3\gamma)}.
\label{p16}
\qqq 
 This   implies
 \qq
&& 
 | \int dx dy G(x) \nabla \cdot b_n (x,y) \mathbbm{1} (|y| \leq L^{-n/2}) E_n (y) | \non
 \\
&&
 \leq \int dx dy |\nabla \cdot G(x)| | b_n (x,y | \mathbbm{1} (|y| \leq L^{-n/2}) E_n (y)\non
 \\
&& \leq  C \epsilon_n^{(1-3\gamma)}
\|\nabla \cdot  G(x) \|_\infty
\label{p17}
 \qqq

For any  bounded function $G$, (\ref{jbb}) implies
 \qq
 \lim_{n \to \infty}    \int dx \int_0^1 d\lambda  \int dy G(x) T_n  (x-y) \mathbbm{1} (|y| \geq L^{-n/2})E_n (y)=0,
   \label{jb7a} 
 \qqq
 so that, using  (\ref{Df2}), (\ref{jb11}), (\ref{p17}), (\ref{jb7a}),
 \qq
  \label{jb7}
 && \lim_{n \to \infty}   \int dx G(x)  L^{nd} E (L^{2n}, L^{n} x)=
 \non \\
&&  \lim_{n \to \infty}  \int dx \int_0^1 d\lambda  \int dy G(x) T_n  (x-y) \mathbbm{1} (|y| \leq L^{-n/2})E_n (y).
 \qqq

By Proposition \ref{T}, we get, since $G$ is bounded, 
%and $T_D^* (x)=E_D^* (x)$ (NEED TO DEFINE THINGS SO THAT THIS IS TRUE)
 \rr
\lim_{n \to \infty} \int dx |G(x)||T_n(x)-T_D^* (x)| = 0.
\label{p15}
\rrr
where $D$ is defined in (\ref{defD}). 
One may rewrite (\ref{jbb}) as 
$$
 \lim_{n \to \infty}  \int dy \mathbbm{1} (|y| \leq L^{-n/2}) E_n(y)= \int dy E(y)= \|E\|_{1},
$$
and, using the bound  (\ref{Tnorm}) on $\nabla T$, we get:
$$ 
 \int  dy \mathbbm{1} (|y| \leq L^{-n/2})(T_n(x-y)-T_n(x)) \leq C L^{-n}.
$$
Combining the last three equations, we get 
%since $T_n$  is independent of $E_n$ and is a function of $x-y$,
 \rr
\lim_{n \to \infty} \int dx\int_0^1  d \la \int  dy G(x)  (\mathbbm{1} (|y| \leq L^{-n/2})T_n(x-y) E_n(y)-\|E\|_{1}T_D^* (x)) = 0.
\label{p151}
\rrr
 \\
 Combining  this with
(\ref{jb7})
 concludes the proof.
\qquad $\Box$
 \par\noindent

\bigskip

{\bf Proof  of Corollary \ref{corol_main}.}

Since the equation for $E$ (\ref{Edyn}) is of the same form as (\ref{Edyn}), we need only
to prove that the  assumptions of the Corollary imply those of  Theorem \ref{main}. We prove in the Appendix that
there is a map $\Gamma^*$ conjugating the SRB measure $\nu$  for the random field $\theta$ and  a Gibbs measure $\mu$ for  the random field $\omega$ (see (\ref{hass1})).
We also prove in the Appendix that
the Gibbs measure $\mu$ satisfies the assumptions (\ref{decay}) and (\ref{decay1}) that were assumed for  the random field $\omega$ in  Theorem \ref{main}.

So, we have only to show that assumption (vi') implies the representation   (\ref{bsum}) and the bound
(\ref{1norm}) for the $b$ defined in (\ref{defbeta1}). Since (vi') assumes (\ref{decay2}) that itself implies (as is also shown in the Appendix)
  (\ref{locality} ), (\ref{1norm1}), we can write: $b(0,x,0,\theta)= b(0,x,0,\Gamma (\omega))
  =\sum_{A\subset {\BbbZ^{d+1}}} b_{A}(x,\omega),
$
with:
$$
\sum_{A\subset\Z^{d+1}}| b_{A}(x,\omega)|e^{\lambda d(A\cup \{0\})}<\ep' e^{-m|x|},
$$
since we assumed that $C(w)=\ep' e^{-m|x|}$. Now use $b(x,y,0, \theta(t))= b(0,y-x, 0, \tau_x \theta(t))$, which follow from assumption (iii') and, see (\ref{covar}),
$ \theta(t)= h^t(\theta)$,
 $\tau_xh^t(\theta)= \Gamma( \tau_x \circ\tau^t (\omega))$
 to obtain:
$b(x,y,0,\theta(t))
  =\sum_{A\subset {\BbbZ^{d+1}}} b_{t, A}(x, y, \omega),
$
with 
%$ b_{t, A}(x, y, \omega)=b_{A}(y-x, \tau_x \theta(t))$ and:
$$
\sum_{A\subset\Z^{d+1}}|b_{t, A}(x, y, \omega)|e^{\lambda d(A\cup (x, t))}<\ep' e^{-m|x-y|},
$$
since, in $\tau_x \circ\tau^t (\omega)$, the origin is shifted to $(x, t)$.

Now, use
 the inequality 
 $$
  d(A\cup  (x, t))+|x-y| \geq d(A\cup (x, t) \cup (y, t)),
  $$
to get that  $b(x,y,0, \theta (t))$ has  the representation  (\ref{bsum}),  with (\ref{1norm}) holding  for $\la \leq \hf m$,
and $\ep=C\ep'$. 
%NEED TO CLAIM THE CONVERGENCE WHEN INTEGRATED WITH F so that $F$ and $ |\nabla F(x)|$ are exponentially bounded.
\bigskip

\section{Appendix: Gibbs States}
\label{se:Appendix 1}

\subsection { Infinite Volume Gibbs States.} We will prove in this
Appendix the estimates (a) and (b) of Section 2.2., for the class of Gibbs states
corresponding to SRB measures and we will explain this correspondence in the second section of this Appendix.
 We start by briefly
recalling some definitions pertaining to Gibbs states. For a more thorough
discussion, see \cite{BK5} and the general  discussions  \cite{Ru,S,VFS}. 

%We consider the framework introduced in Section 2.5 above.
 Given $X \subset
{\Z}^{d+1}$, a spin configuration $\om$ in $X$ is an element
$\om\in \Omega_X$. For $Y\subset X$ denote by 
$\omega_Y$  the restriction of $\omega$ to $Y$. An interaction is defined
by a family $\Phi = \{\Phi_X\}$ of  functions indexed by finite
subsets $X$ of ${\Z }^{d+1}$: \qq \Phi_X : \Omega_X \to {\R} \qqq

 %We consider translation invariant
%$\Phi's: \Phi_{X+(x,t)} = \tau_{x,t}\Phi_X$ where $\tau_{x,t}$, $ (x,t) \in {\Z}^{d+1}$ is 
Our interactions are continuous invariant under the natural
action of  translations by ${\Z}^{d+1}$. % on functions defined on $\Omega_X$ to functions defined on $\Omega_{X+(x,t)}$. 
We let $\|\Phi_X\|_\infty$ denote the sup norm of $\Phi_X$ (which is finite since $\Omega_X$ is a finite set).

Given $\Lambda \subset {\Z}^{d+1}$, $|\Lambda| < \infty$, and a
configuration $ \om'\in \Omega_{\Lambda^c}$, the
Hamiltonian in $\Lambda$ with boundary conditions $ \om'$ is defined as 
\qq
{\mathcal  H}_\Lambda( \om| \om') = -\sum_{X \cap \Lambda \neq \emptyset} \Phi_X ( \om \vee \om')
\label{g8}
 \qqq 
 where $ \om \vee \om' $ is the obvious configuration in $ {\Z}^{d+1}$.
 
  We let
 \qq \nu_\Lambda ( \om|\om') :=  \exp (-{\mathcal  H}_\Lambda ( \om|\om')).
\label {Gibbs} 
\qqq 
Then the associated (finite volume) Gibbs measure $\mu_\Lambda$ is the probability
distribution on $\Omega_\Lambda$: 
\qq 
\mu_\Lambda ( \om|\om')=Z_\Lambda(\om')^{-1} \nu_\Lambda ( \om|\om') 
%\om'_{\Lambda^c}) \exp (-{\mathcal  H} ( \om_\Lambda| \om'_{\Lambda^c}))
\label {Gibbs1} 
\qqq 
 with the partition function
 \qq 
 Z_\Lambda(\om') = \sum_{ \om} \nu_\Lambda ( \om|\om') . 
\label{g9}
\qqq
 We also need these objects 
 for open boundary conditions, i.e. when the sum in (\ref{g8}) is restricted to $X\subset \Lambda$. Then we
 write ${\mathcal  H}_\Lambda( \om)$, $\nu_\Lambda ( \om)$, $\mu_\Lambda ( \om)$  and
 $Z_\Lambda$.

We consider interactions of the following type:
\qq
\Phi = \Phi^0 + \Phi^1,
\label {Pot}
\qqq
 where $\Phi^0$ is completely analytic, in the sense of Dobrushin and Shlosman \cite{DS1,DS2}
(see Olivieri and Picco \cite{O,OP} for another approach, used here, to complete analyticity). In our  coupled map lattices case,  $\Phi^0$ will be a finite range interaction
of a one dimensional system (which is easily seen to be completely analytic). 
Let the range of $\Phi^0$ be $r$, i.e. $\Phi^0_X=0$ if $d(X)>r$.

For $\Phi^1$, we will assume:
\qq 
\| \Phi^1 \| = \sum_{0\in X}e^{\la d(X)}
\|\Phi^1_X\|_\infty \leq \epsilon,
\label {norm-phi}
\qqq 
\\
with $\la>0$ . For such $\Phi$'s and $\epsilon=\epsilon(\la, \Phi^0)$
small enough, it is shown in \cite{BK5} that the finite volume Gibbs state
(\ref{Gibbs1}) has a limit, as $\Lambda \to {\Z}^{d+1}$ (in the sense of subsets ordered by inclusion)
independent of $\om'$, which defines a measure $\mu$. We let
$\E$ denote the expectation with respect to $\mu$ and $\E_\Lambda$ the one with respect to $\mu_\Lambda$
with free boundary conditions.

To prove that the conditions (\ref{decay}, \ref{decay1}) hold for $\mu$, we shall use 
the approach of \cite{BK5} to prove: 
\lem{A1}
For $\Phi$ as in (\ref{Pot}), there exists $\epsilon(\la, \Phi^0)>0$ such that for $\epsilon\leq \epsilon(\la, \Phi^0)$ in 
(\ref{norm-phi}), there are constants $m>0$, $C<\infty$, such that the following
holds:

\no a) 
For all $\Lambda$,  $A \subset \Lambda$ and $F:\Omega_A\to\R_+$
\rr
 \exp(-C \epsilon |A| e^{-mR})\leq \frac{\mathbb{E}(F)}{\mathbb{E}_\Lambda (F)} \leq \exp(C\epsilon |A| e^{-mR}).
\label{p3a}
\rrr
where  $R={\rm dist}(A,\Lambda^c)$.

\no b) If $F_i:\Omega_{A_i}\to\R_+$, $i=1,\ldots ,k,$ 
%with $\mbox{supp}\;  F_i = A_i$, 
\rr
\mathbb{E} (\prod^k_{i=1} F_i) \leq \prod^k_{i=1}( \mathbb{E} (F_i) \exp ( C \epsilon |A_i| e^{-mR}))
\label{p4b}
\rrr
where $R=\min_{i\neq j}{\rm dist}(A_i,A_j)$.

\no c)  If $F_i:\Omega_{A_i}\to\R$, $i=1,2$, and $R= {\rm dist}(A_1, A_2)$ then
\rr
|\mathbb{E} ( F_1 F_2) - \mathbb{E} ( F_1) \mathbb{E} ( F_2)| \leq C \min(|A_1|, |A_2|) \|F_1\|_\infty  \|F_2\|_\infty \exp (-mR)
\label{p4a}
\rrr
\elem
\no \textbf{Proof.} a) Writing $\mathbb{E}(F)= \lim_{\Lambda' \to  \mathbb{Z}^{d+1} } \mathbb{E}_{\Lambda'}(F)$ %(fix scale $n=0$ here)
, we get:
\rr
\frac{\mathbb{E}(F)}{\mathbb{E}_\Lambda (F)}= \lim_{\Lambda' \to  \mathbb{Z}^{d+1} }
\frac{\mathbb{E}_{\Lambda'}(F)}{\mathbb{E}_\Lambda (F)}
=\lim_{\Lambda' \to  \mathbb{Z}^{d+1} }
\frac{\sum_{\omega\in\Omega_{\Lambda},\omega'\in\Omega_{\Lambda'}} F(\omega')\nu_{\Lambda}(\omega)\nu_{\Lambda'}(\omega')}
{\sum_{\omega\in\Omega_{\Lambda}\omega'\in\Omega_{\Lambda'}} F(\omega)\nu_{\Lambda}(\omega)\nu_{\Lambda'}(\omega')}
\label{p23aa}
\rrr
We may also assume $\Lambda \subset \Lambda'$.

First we perform a high-temperature expansion to the $e^{-\CH^1}$ part of the Hamiltonian
 in the numerator and the denominator of (\ref{p23aa}). 
Note that we may replace $\Phi^1_X$ by $\Phi^1_X - \inf_{ \om} \Phi^1_X(\omega)$, by adding a constant to
the Hamiltonian. Thus, we may, without loss of generality, assume that 
\rr
\Phi^1_X \geq 0,\ \ \ \| \Phi^1 \|\leq 2 \epsilon.
\label{newphi}
\rrr
%for all $X$, and $\| \Phi^1 \|\leq 2 \epsilon$. 
Now, write
\qq
\exp
(\sum_{X \subset \Lambda} \Phi^1_X ( \om)+\sum_{X \subset \Lambda'} \Phi^1_X ( {\om'}))
 = \sum_{\CX\subset \CS(\Lambda')} \prod_{X \in {\CX}} f_X (\om, \om'):= \sum_{\CX \subset \CS(\Lambda')} f_\CX
 \label{g30}
\qqq
 where $\CS(\Lambda')$ is the set of subsets of $\Lambda'$, and 
\qq 
f_X (\om, \om')= \exp (\Phi^1_X (
\om) \mathbbm{1} (X \subset \Lambda)+ \Phi^1_X ( \om') ) - 1 
\label{fXdef}
\qqq
satisfies by  (\ref{newphi}) 
\qq 
0 \leq f_X, 
\qqq
and 
\qq
 \sum_{0\in
X}e^{\la d(X)} \| f_X \|_\infty \leq C \epsilon.
\label{g2} 
\qqq
Define, for $\CS\subseteq  \CS(\Lambda')$
\rr
 \CN_{\CS}=\sum_{\CX \subset \CS} \sum_{\omega\in\Omega_{\Lambda},\omega'\in\Omega_{\Lambda'}}F(\om')f_{\CX}(\om,\om')
\exp (-{\CH}^0_\Lambda (\om)-{\CH}^0_{\Lambda'} (\om') ) 
\label{p23c}
\rrr 
and $ \CD_{\CS}$ is the same expression with $F(\om')$ replaced by
$F(\om)$. Then
 \rr
 \frac{\mathbb{E}_{ \Lambda'}(F)}{\mathbb{E}_\Lambda (F)}
 =\frac{\CN_{ \CS(\Lambda')}}{\CD_{ \CS(\Lambda')}}.
\label{p2333}
\rrr

%Next we perform a partial expansion for $\CH^0$. 
Let now $\rho>2r$
where $r$ is the range of $\Phi^0$. We cover $\Z^d$ by disjoint cubes of side $\rho$, called $\rho$-cubes.
Two cubes are adjacent if their distance is less than $\rho$, and that implies the notion of connected family
of $\rho$-cubes in the usual way. 
Given $X\subset\Z^d$
we will denote by $\underline X$ the set of $\rho$-cubes intersected by $X$
and by  $\underline{\CX}=\cup_{X \in {\CX} } \underline{X} $.
Let  $\CP=\CP(\CX)=\{P_1,\dots,P_n\}$
be the family of connected components of $\underline{\CX}$ intersecting both
$\underline A$ and $\Lambda^c$ and write $\CX= \CX_1\cup\CX_2$
with $\CX_1$ collecting the $X\subset\cup P_i$.
We have
 \rr
f_{\CX}\leq \bar f_{\CX_1} f_{\CX_2}\non
%\label{p23}
\rrr
where we set $\bar f_X:=\|f_X\|_\infty$. Given a connected union of $\rho$-cubes $P$
set
 \rr
\phi_P:=\sum_{\CX:\underline\CX=P}\bar f_{\CX}.
%\non
\label{xP}
\rrr
With these preliminaries we may now estimate the numerator by
\rr
 \CN_{\CS(\Lambda')}\leq\sum_{\CP\in\Pi_A} \phi_\CP\CN_{\CS_\CP}
 %\sum_{\CX\in \CS(\Lambda'\setminus\CP_A)}  \sum_{ \om, \om'}F(\om')f_{\CX}(\om,\om')\exp (-{\CH}^0_\Lambda (\om)-{\CH}^0_{\Lambda'} (\om') ) 
\label{p231}
\rrr
where $\Pi_A$ is the set of families  $\CP=\{P_1,\dots,P_n\}$ where $P_i$ is a union of $\rho$-cubes,
with $\dist(P_i,P_j)\geq\rho$ and where each $P_i$  intersects both
$\underline A$ and $\Lambda^c$, and $\phi_\CP=\prod_i \phi_{P_i}$.  $\CS_\CP$ in turn consists of families $\CX \subset \CS(\Lambda')$ s.t. 
$ \underline{\CX}$ is disconnected  from $\cup_{P\in\CP}P$ and  no connected
component of $ \underline{\CX}$ intersects both $\underline A$
and $\Lambda^c$.
 Since $ \CS_\CP\subset  \CS(\Lambda')$ we have $\CD_{ \CS(\Lambda')}\geq \CD_{\CS_\CP}$ and so
\rr
\frac{\mathbb{E}_{\Lambda'}(F)}{\mathbb{E}_\Lambda (F)}\leq 
\sum_{\CP\in\Pi_A} \phi_\CP\ \frac{\CN_{\CS_\CP}}{\CD_{\CS_\CP}}
%\non
\label{1stred}
\rrr
To understand our next step, let us suppose for a moment that $\CH^0=0$. Then,
the ratio on the RHS equals 1. Indeed, let, for $\CX \subset \CS_\CP$, $\CX= \CX_A\cup\CX_{A^c}$
with $\CX_A$ collecting the $X$ lying in connected components
of $\underline\CX$ that intersect $\underline{A}$ (and that are contained in $\Lambda$, by definition of  $\CS_\CP$).
 Then, $\CX_A$, $\CX_{A^c}$ are disjoint
and, for $X\in\CX_A$, $f_X(\om,\om')=f_X(\om',\om)$ since $X\subset\Lambda$ (recall
(\ref{fXdef})). Thus
\rr
 \sum_{ \om, \om'}F(\om')f_{\CX}(\om,\om')= \sum_{ \om, \om'}F(\om')f_{\CX_A}(\om,\om')
 \sum_{ \om, \om'}f_{\CX_{A^c}}(\om,\om')= \sum_{ \om, \om'}F(\om)f_{\CX}(\om,\om')
 \label{xx}
\rrr
where in the second step we used in the first sum the above symmetry of $f$ and then interchanged $\om$ and $\om'$. This
renders the numerator equal to the denominator.

To be able to use this trick, we need to decouple the correlations
induced by $\CH^0$.
Given $\CX \subset \CS_\CP$, let 
\qq
V =V({\CX}) =\underline{\CX}\cup
\underline{A}  %\cup \underline{{\CX}} 
\label{Vdefi}
\qqq
Then $Ff_\CX$ depends only on $\om_{V\cap\Lambda},\om'_V$
and
summing
over the other variables we get
\rr
\CN_{ \CS_\CP}=\sum_{\CX \subset  \CS_\CP} 
 \sum_{ \om\in\Omega_{V\cap \Lambda}, \om'\in\Omega_V,}F(\om')f_{\CX}
(\om,\om')e^{-{\CH}^0_{V\cap \Lambda} (\om)-{\CH}^0_V (\om') }Z^0_{\Lambda \setminus V}(  \om) Z^0_{\Lambda' \setminus V}( \om') 
\label{p23a}
\rrr
where,  with a slight abuse of notation, $ Z^0_{\Lambda \setminus V }( \om)$ is the partition function with interaction $\Phi^0$, $
\om$ boundary condition in $V\cap\Lambda$, and open boundary conditions in $\Lambda^{c}$ and similarly for $Z^0_{\Lambda' \setminus V}( \om')$. 
$\CD_{ \CS_\CP}$ has an analogous representation.

Observe that, since the range of  $\Phi^0$ is less than $\rho/2$,
the exponential in (\ref{p23a}) factors over the connected components 
of $V_\al$ of $V$:
\rr
\zeta_V(\om,\om'):=\exp (-{\CH}^0_{V\cap \Lambda} (\om)-{\CH}^0_V (\om') )=
\prod_\al\zeta_{V_\al}(\om,\om'),
\label{p23b}
\rrr
where $\zeta_{V_\al}(\om,\om')$ is a function of $\om_{V_\al\cap \Lambda}$, $\om'_{V_\al}$. 
However, the partition functions do not factor and we need to expand them.
For this, we will use the
\lem{factor} %Let $\La\subset \Z^{d+1}$ and $V\subset\La$
%with $V=\cup_\al V_\al\subset\Lambda$ with $V_\al$  unions of $\rho$
%cubes and $V_\al\cap V_\beta=\emptyset$.  
There exists a constant $z_{\Lambda'\setminus V}$ and functions
$W_A$ on $\Om_{A}$ and $\phi_Y$
on $\Om_{Y\cap V}$ such that 
\qq
 Z^0_{\Lambda'\setminus V} ( \om' ) =z_{\Lambda'\setminus V}
%e^{f|\Lambda\setminus V|}
\prod_{\alpha} W_{V_\al} ( \om') \exp \sum_{Y\in \CU_V} \phi_Y( \om') 
\label{g5}
\qqq
where $\CU_V$ is the set of connected sets of $\rho$-cubes $Y\subset \Lambda'$, so that 
 $Y\cap
V \neq \emptyset$ and $Y\cap
V^c \neq \emptyset$. %$\phi_Y=0$  if $Y \cap V = \emptyset$
%and  
Moreover, for any
$\epsilon > 0$, there exist $\rho < \infty$, $\la >0$ such that
\qq 
\sum_{0 \in Y} \exp(\frac{\la d(Y)}{\rho}) \| \phi_Y \|_\infty \leq \epsilon
\label{g4}
\qqq
A similar representation holds for $ Z^0_{\Lambda\setminus V} ( \om ) $. %ADDED tHIS AND ADDED PRIMES BEFORE.
%Moreover $\phi_Y$ is a constant if $Y \cap \Lambda^c = \emptyset$.
\elem
\no {\bf Proof.} The Lemma goes back to Olivieri and Picco \cite{O,OP}
and is discussed in  \cite{BK5}, see formula (13) (note that there we had 
$\phi_Y$ constant  if $Y \cap V = \emptyset$, so that here
$z_{\Lambda'\setminus V}=\exp(f|\Lambda'\setminus V|+\sum_{Y\subset 
\Lambda'\setminus V}\phi_Y)$).
 Note also that there is no loss of generality in assuming that $\la $ here is the same as in
(\ref{g2}).   \halmos

Define now 
 \qq
\bar \phi_Y = \min_{ \om%\in\Om_{Y\cap V}
} \phi_Y ( \om) 
 \label{g19}
\qqq
and 
\qq
  \Psi_Y ( \om):=\phi_Y ( \om) -\bar \phi_Y
  \label{g18} 
 \qqq
  whereby we have
  \rr
\Psi_Y \geq 0.
%\non
\label{psiy}
\rrr

 We will now  apply (\ref{g5}) to each of the partition functions in eq. 
  (\ref{p23a}) and the corresponding expression for the denominator $\CD_{\CS_\CP}$.
% Let $\CU_V$ be the set of $Y\subset \Lambda'$ which intersect both  $V$ and $V^c$.
 %:=\{Y\subset \Lambda'\ |\ Y \cap  V \neq \emptyset,\  Y \cap    V^c  \neq \emptyset$. 
 We get:
 \qq
 e^{-{\CH}^0_{V\cap \Lambda} (\om)-{\CH}^0_V (\om') }Z^0_{\Lambda \setminus V}(  \om) Z^0_{\Lambda' \setminus V}( \om') 
=
 {\CW_{V}} {\prod_{Y\in\CU_V}} (1+g_Y)
  \label{g20}
 \qqq
 where 
 \qq
 {\CW_{V}}=z_{\Lambda\setminus V}z_{\Lambda'\setminus V}{\prod_{Y\in\CU_V}} 
 e^{\bar \phi_Y (\mathbbm{1}(Y\subset \Lambda)+1)) }  \prod_{\alpha} W_{V_\alpha \cap \Lambda}(\om) W_{V_\alpha}(\om')\zeta_{V_\al}(\om,\om')
 \label{g15}
\qqq
%with the  products  running over $Y \subset\La'$ s.t. $Y \cap  V \neq \emptyset$ and $Y \cap    V^c  \neq \emptyset$. 
We also denoted
\qq
g_Y=\exp ( \Psi_Y
(\om) \mathbbm{1}(Y\subset \Lambda)+ \Psi_Y(\om' ))-1.
\label{gYdef}
\qqq
By (\ref{psiy})
 \qq 
 0 \leq g_Y,
  \qqq
  and by  (\ref{g4})
  \qq \sum_{0
\in Y} \exp(\frac{\la d(Y)}{L})  \| g_Y \|_\infty \leq C \epsilon.
\label{g12}
\qqq

 If we use (\ref{g20}), (\ref{g15}), we see that the obstruction to the factorization, as in (\ref{xx}), of the sums in (\ref{p23a}), 
 comes only from the product of $1+g_Y$. Thus, we  expand:
  \qq
  {\prod_{{Y\in\CU_V}}} (1+g_Y)= \sum_{\CY \subset\CS(\CU_V)} \prod_{Y\in {\CY}} g_Y := \sum_{\CY \subset\CS(\CU_V)} g_ {\CY}
\label{g14}
\qqq
 where  $\CS(\CU_V)$ is the set of subsets of $\CU_V$. 

Combining  (\ref{p23a}),   (\ref{g20})   and  (\ref{g14}) we get
\rr
\CN_{ \CS_\CP}=\sum_{\CX \subset  \CS_\CP} \sum_{\CY \subset\CS(\CU_V)}
 \sum_{ \om\in\Omega_{V\cap \Lambda}, \om'\in\Omega_V,}F(\om')f_{\CX}g_ {\CY}
  {\CW_{V}}
\label{p23abc}
\rrr

We repeat now what we did with $\CX$. Let  $\CQ=\CQ(\CX,\CY)=\{Q_1,\dots,Q_m\}$
be the family of connected components of $\underline{\CX}\cup\underline{\CY}$ intersecting both
$\underline A$ and $\Lambda^c$ and write $\CY= \CY_A\cup\CY_{A^c}$
with $\CY_A$ collecting the $Y\subset\cup Q_i$ and similarly for $\CX$.
We have again
 \rr
f_{\CX}g_{\CY}\leq  \bar f_{\CX_A} f_{\CX_{A^c}}\bar g_{\CY_A} g_{\CY_{A^c}}.\non
%\label{p23}
\rrr
 Given a union of $\rho$-cubes $Q$,
set
 \rr
\psi^{\CP}_Q:=\sum_{\CX \subset  \CS_\CP,\CY\subset\CS(\CU_V)}\bar f_{\CX}\bar g_{\CY}1_{\CQ(\CX,\CY)=\{Q\}}.
%\non
\label{xP1}
\rrr
We get
\rr
 \CN_{\CS_\CP}\leq\sum_{\CQ\in\Pi_A} \psi^{\CP}_\CQ\CN_{\CS_{\CP,\CQ}},
 %\sum_{\CX\in \CS(\Lambda'\setminus\CP_A)}  \sum_{ \om, \om'}F(\om')f_{\CX}(\om,\om')\exp (-{\CH}^0_\Lambda (\om)-{\CH}^0_{\Lambda'} (\om') ) 
\label{p23}
\rrr
with $\psi^{\CP}_\CQ =\prod_{Q\in\CQ}\psi^{\CP}_Q$, and
\rr
\CN_{\CS_{\CP,\CQ}}
=\sum_{\{\CX,\CY\}\in  \CS_{\CP,\CQ}} 
 \sum_{ \om\in\Omega_{V\cap \Lambda}, \om'\in\Omega_V,}F(\om')f_{\CX}g_ {\CY}
  {\CW_{V}},
\non
%\label{}
\rrr
where $\CS_{\CP,\CQ}$ consist of pairs $(\CX, \CY)$, $\CX \subset \CS_\CP$, $\CY \subset\CS(\CU_V)$ such that 
$ \underline{\CX}\cup\underline{\CY}$ is disconnected from $\cup_{Q\in\CQ}Q$ and  no connected
component of $ \underline{\CX}\cup\underline{\CY}$ 
intersects both $\underline A$
and $\Lambda^c$.

There is a similar expansion in the denominator and, since the $\CX$ part of $\CS_{\CP,\CQ}$ belongs to $ \CS_{\CP}$, we have $\CD_{\CS_{\CP}}
\geq\CD_{\CS_{\CP,\CQ}}$
and then 
\rr
\frac{\mathbb{E}_{\Lambda'}(F)}{\mathbb{E}_\Lambda (F)}\leq 
\sum_{\CP\in\Pi_A} \sum_{\CQ\in\Pi_A} \phi_\CP \psi^\CP_\CQ\ \frac{\CN_{\CS_{\CP,\CQ}}}{\CD_{\CS_{\CP,\CQ}}}=\sum_{\CP\in\Pi_A} \sum_{\CQ\in\Pi_A} \phi_\CP \psi^\CP_\CQ
%\non
\label{finall}
\rrr
where in the last step we applied the symmetry argument as in the special case (\ref{xx})
above, which holds here by definition of $\CS_{\CP,\CQ}$.

Let us bound $\phi_P$ in (\ref{xP}). Let $\underline \CX=P$ with
$\CX=\{X_1,\dots,X_n\}$. Then $d(P)\leq \sum_i d(X_i) +2n\rho$. 
%Fix $(t,x)\in\Z^{d+1}$. 
Let $\CA$ be the the set of connected $P$ %with $(t,x)\in P$ and 
s.t. $P$ intersects both $\underline A$ and $\Lambda^c$. Then,
$$
\sum_{P\in\CA}%e^{\hf\la d(P)} 
\phi_P\leq  e^{-\hf\la d(A,\Lambda^c)}\sum_{\CX:\underline\CX\in\CA}\prod_i e^{\hf\la d(X_i)+2\rho}\bar f_{X_i}.
$$
Using (\ref{g2}) the sum is readily bounded, since $P$ is connected and must contain a point in $A$,
 by  $C(\rho)\epsilon |A|$
 and
  \rr
\sum_{P\in\CA}%e^{\hf\la d(P)} 
\phi_P\leq  C(\rho)\epsilon |A|e^{-\hf \la d(A,\Lambda^c)}
\non
%\label{}
\rrr
Similar estimate holds for  $\psi^\CP_Q$, uniformly in $\CP$, using  (\ref{g12}). Therefore,
 since $\rho$ is only 
 constrained to be larger that $2r$, and $r$ is given, we can write $C(\rho)=C$, and get
\qq
(\ref{finall})\leq  (\sum_{k=0}^\infty k!^{-1}
( C\epsilon |A|e^{-\hf  \la d(A,\Lambda^c)})^k)^2\leq  \exp (C \epsilon|A|e^{- \hf \la R}).
\label{g3}
\qqq
since $d(A,\Lambda^c) \geq R$. This is the upper bound in (\ref{p3a}). 

For the lower bound, we simply apply the argument for the upper bound  to the ratio ${\mathbb{E}_\Lambda (F)}/{\mathbb{E}_{\Lambda'}(F)}$.

To prove (\ref{p4b}) 
let $\Lambda_i = \{z | \mbox{dist} \ (z,A_i) \leq \frac{R}{4}\}$, and let $ \Lambda = \displaystyle \bigcup^k_{i=1} \Lambda_i$.
\\
Then, $\Lambda_i \cap \Lambda_j = \emptyset$ for $i\neq j$ and
\[
 \mathbb{E}_\Lambda (\prod^k_{i=1} F_i) = \prod^k_{i=1} E_{\Lambda_i} (F_i).
\]
 Then using (\ref{p3a}) for $F =\displaystyle \prod^k_{i=1} F_i$, with $A=\cup_{i=1}^k A_i$, and using 
 \[
  \frac{\mathbb{E}_{\Lambda_i}(F_i)}{\mathbb{E}(F_i)}  \leq  \exp (C \epsilon |A_i| e^{-cR}),
 \]
 which follows from (\ref{p3a}), for each
$ \frac{\mathbb{E}_{\Lambda_i}(F_i)}{\mathbb{E}(F_i)}$,
  we get
 \[
 \mathbb{E} (\prod^k_{i=1} F_i) \leq \prod^k_{i=1} (\mathbb{E} (F_i)  \exp (C \epsilon |A_i| e^{-cR}))
 \]
i.e. (\ref{p4b}).
 \\
 c) This is proven in \cite{BK5}. \qquad $\Box$

\subsection{Coupled map lattices.} 
We will now recall briefly how the SRB measure of our CML fits into the framework of the
previous section. For definiteness  fix $N=\T^2$. 
Let $\CN=N^{\Z^d}$ which is
a compact metric space with the metric
$$
d(\theta,\theta')=\sum_{x\in\Z^d}2^{-|x|}\rho(\theta(x), \theta'(x))
$$
where $\rho$ is the standard metric on $\T^2$.

Let $h:\CN\to\CN$ in (\ref{dyn}) be of the form
\qq
h(x,\theta)=A\theta(x)+k(x,\theta)
%\label{}
\non
\qqq 
where $A:\T^2\to\T^2$ is a linear hyperbolic torus automorphism and $k$ satisfies the bound (\ref{hass}) (we describe $\T^2$ here by $\theta\in [0,1]^2$, $A$ is in $SL(2,\Z)$ and
$k$ is periodic over $\Z^2$).
Let us recall that $(A,\T^2)$ is conjugate to a subshift of finite type $(\tau,\Sigma)$
where $\Sigma\subset \{1,2,\dots,p\}^{\Z}$ consists of sequences $\omega$
with $M_{\omega(t),\omega(t+1)}=1$ for all $t$ where $M$ is a $p\times p$ matrix
with entries in the set $\{0,1\}$.

The construction of the SRB measure goes through finite volume approximations. Let $V_R$ denote the box of side $2R+1$ in $\Z^d$
with opposite faces identified and set $\CN_R=N^{V_R}$. 
  Let $h_R:\CN_R\to\CN_R$
denote the periodization of $h$ i.e. given a $\theta\in\CN_R$ let $\theta_R\in\CN$
be its periodic extension and set  $
h_R(x,\theta) = h(x,\theta_R)$. For $\theta\in \CN_R$ let $W_R(\theta)$ be the
leaf of the unstable foliation passing through $\theta$ (which exists for $\kappa$ in  (\ref{hass})  small enough) and $\tilde h_R:W_R(\theta)\to
W_R(\theta)$ the restriction of $h_R$ to $W_R(\theta)$. Define the function
$$
\Lambda_R(\theta)=\log\det D\tilde h_R(\theta).
$$
Using translation invariance one shows
$$
\Lambda_R(\theta)=\sum_{x\in V_R}\lambda_R(\tau_x\theta)
$$
where $\tau_x$  is the translation by $x$ and
where the functions $\lambda_R$ (extended periodically to $\Z^d$) 
are uniformly H\"older continuous with
\qq
|\lambda_R(\theta)-\lambda_R(\theta')|\leq \eta d(\theta,\theta')^\alpha
\label{hold}
%\non
\qqq
where $\alpha>0$ and $\eta\to 0$ as $\kappa\to 0$ in (\ref{hass}). One has a good
control of their $R$ dependence and, as $R\to\infty$, they
 converge to $\lambda:\CN\to\CN$ 
satisfying (\ref{hold}).

Next, let $\Sigma_R=\Sigma^{V_R}$ and $\Sigma_\infty=\Sigma^{ \Z^d}$. Equip these with the metric
$$
d(\omega,\omega')=\sum_{(t,x)\in\Z^{d+1}}2^{-|t|-|x|}|\omega(t,x)-\omega'(t,x)|
$$
The map $h_R$ is conjugate  to the time shift $\tau:\Sigma_R\to \Sigma_R$
with $\tau\omega(t,x)=\omega(t+1,x)$ i.e. there exists
a H\"older continuous map $\Gamma_R:\Sigma_R\to\CN_R$
with
\qq
d(\Gamma_R(\omega),\Gamma_R(\omega'))\leq Cd(\omega,\omega')^\beta
\label{ghold}
%\non
\qqq 
for some $\beta>0$ uniformly in $R$ and
\qq
h_R=\Gamma_R^{-1}\circ\tau\circ\Gamma_R.
%\label{}
\non
\qqq 
We can now define a Hamiltonian
\qq
H_{T,R}(\omega)=\sum_{t=-T}^{T}\sum_{x\in V_R}(\lambda_R\circ\Gamma_R)(\tau^t\tau_{x}\omega),
%\label{}
\non
\qqq
and the probability measure
\qq
\mu_{T,R}(d\omega)=Z_{T,R}^{-1}e^{-H_{T,R}(\omega)}\mu^0(d\omega)
%\label{}
\non
\qqq 
where $\mu^0$ is the maximum entropy measure on $\Sigma_R$. 

Let $m_R$ denote the normalized Lebesgue measure on $\CN_R$.
One then has, see \cite{BKL},
\prop{conjug} The weak limits
\qq
\lim_{n\to\infty}h_R^nm_R=\nu_R\ \ {\rm and}\ \ \lim_{T\to\infty}\mu_{T,R}=\mu_R
%\label{}
\non
\qqq 
exist and are related by
\qq
\nu_R=\Gamma^*_R\mu_R
%\label{}
\non
\qqq
\eprop
In particular $\nu_R$ is the SRB measure of the dynamical system $(h_R,\CN_R)$.
 
We will now realize $\mu_R$ as a Gibbs measure with a potential satisfying
the conditions of Section 7.1. We will use telescoping sums to express the
function $\ell_R:=\lambda_R\circ\Gamma_R$ as a sum of localized functions.
Let us order the set $\Z^{d+1}=\{z_1,z_2,\dots\}$ s.t. $|z_{i+1}|\geq |z_i|$.
Given $\omega\in \Sigma_R$, let
$
\omega_n(z_i)=\omega(z_i)
$
if $i\leq n$ and $
\omega_n(z_i)=\tilde\omega(z_i)
$
where for each $n$ we choose an arbitrary extension $\tilde\omega$ of $\Omega_n:=\{\omega(z_i)|
\ i\leq n\}$ to $\Sigma_R$ (it is not hard to see that for each $\Omega_n$
such an extension can be chosen). We can then write
\qq
\ell_R(\omega)=\ell_R(\omega_1)+\sum_{n=1}^\infty(\ell_R(\omega_{n+1})-
\ell_R(\omega_{n})):=\sum_{n=0}^\infty\phi_{R,A_n}(\omega),
\label{tel}
\qqq
where $\phi_{R,A_n}$ is localized in a region $A_n$ with $d(A_n)$
comparable to $(1+n)^{1/d}$. Moreover, since  $\omega_n$ and 
$\omega_{n+1}$ agree in a space time region of radius comparable
to $d(A_n)$  we get from
(\ref{hold}) and (\ref{ghold}),
\qq
|\phi_{R,A_n}(\omega)|\leq C\eta 2^{-\alpha \beta d(A_n)}.
%\label{}
\non
\qqq
Given $A\subset\Z^{d+1}$ let now
\qq
\Phi_{R,A}(\omega)=\sum_{t,x}\phi_{R,A_n}(\tau^{t} \tau_{x}\omega) \mathbbm{1}_{A_n=\tau^{t}\tau_{x}A}
%\label{}
\non
\qqq
Since the number of sets $A_n$ with radius between $r$ and $r+1$
is bounded by $Cr^d$ one easily concludes that
\qq
\sum_{A:0\in A}\|\Phi_{R,A}(\omega)\|_\infty e^{\la d(A)}\leq C\eta
%\label{}
\non
\qqq
for some $\la>0$, uniformly in $R$.
We are hence in the framework of Section 7.1. and 
conclude that  $\mu_R$ has a weak limit  $\mu$
which is the unique Gibbs measure of the potential
$\lim_{R\to\infty}\Phi_{R,\cdot}$. In particular 
the assumptions (a) and (b) in section 2.2 hold for $\mu$.
As a consequence $\nu_R$  also has a weak limit $\nu$,  and it is the SRB
measure of $(h,\CN)$.  These measures are related by the conjugation
 $\nu=\Gamma^*\mu$. 
 
 Finally, consider a local function $w(\theta)$ as in (\ref{decay2}). By
 the above telescoping argument (\ref{tel}) we obtain the claims (\ref{locality}) 
 and (\ref{1norm1}).
 
 If our original system is a perturbation of uncoupled expanding
 circle maps, i.e. $N=S^1$, the Jacobian of the map restricted to the
 unstable manifold, which is used here, is then replaced by the Jacobian of the original map.
 This time the subshift is replaced by the full shift on $\{1,\dots, p\}^\N$
 and the  space time symbolic representation is on  $\{1,\dots, p\}^{\Z^{d+1}}$ .

\newpage

 \section*{Acknowledgments} This work is partially funded by the Academy of Finland
 the European Research Council and the Belgian Interuniversity Attraction Pole, P6/02.

\end{document}